\documentclass{aastex61}
\usepackage{amsmath}

\begin{document}

\title{Early X-ray flares in GRBs}

\author{R.~Ruffini}
\affiliation{ICRA and Dipartimento di Fisica, Sapienza Universit\`a di Roma, P.le Aldo Moro 5, 00185 Rome, Italy}
\affiliation{ICRANet, P.zza della Repubblica 10, 65122 Pescara, Italy. \href{mailto:yu.wang@icranet.org}{yu.wang@icranet.org}}
\affiliation{Universit\'e de Nice Sophia Antipolis, CEDEX 2, Grand Ch\^{a}teau Parc Valrose, Nice, France}
\affiliation{ICRANet-Rio, Centro Brasileiro de Pesquisas F\'isicas, Rua Dr. Xavier Sigaud 150, 22290--180 Rio de Janeiro, Brazil}

\author{Y.~Wang}
\affiliation{ICRA and Dipartimento di Fisica, Sapienza Universit\`a di Roma, P.le Aldo Moro 5, 00185 Rome, Italy}
\affiliation{ICRANet, P.zza della Repubblica 10, 65122 Pescara, Italy. \href{mailto:yu.wang@icranet.org}{yu.wang@icranet.org}}

\author{Y.~Aimuratov}
\affiliation{ICRA and Dipartimento di Fisica, Sapienza Universit\`a di Roma, P.le Aldo Moro 5, 00185 Rome, Italy}
\affiliation{ICRANet, P.zza della Repubblica 10, 65122 Pescara, Italy. \href{mailto:yu.wang@icranet.org}{yu.wang@icranet.org}}

\author{U.~Barres de Almeida}
\affiliation{ICRANet-Rio, Centro Brasileiro de Pesquisas F\'isicas, Rua Dr. Xavier Sigaud 150, 22290--180 Rio de Janeiro, Brazil}

\author{L.~Becerra}
\affiliation{ICRA and Dipartimento di Fisica, Sapienza Universit\`a di Roma, P.le Aldo Moro 5, 00185 Rome, Italy}
\affiliation{ICRANet, P.zza della Repubblica 10, 65122 Pescara, Italy. \href{mailto:yu.wang@icranet.org}{yu.wang@icranet.org}}

\author{C.~L.~Bianco}
\affiliation{ICRA and Dipartimento di Fisica, Sapienza Universit\`a di Roma, P.le Aldo Moro 5, 00185 Rome, Italy}
\affiliation{ICRANet, P.zza della Repubblica 10, 65122 Pescara, Italy. \href{mailto:yu.wang@icranet.org}{yu.wang@icranet.org}}

\author{Y.~C.~Chen}
\affiliation{ICRA and Dipartimento di Fisica, Sapienza Universit\`a di Roma, P.le Aldo Moro 5, 00185 Rome, Italy}
\affiliation{ICRANet, P.zza della Repubblica 10, 65122 Pescara, Italy. \href{mailto:yu.wang@icranet.org}{yu.wang@icranet.org}}

\author{M.~Karlica}
\affiliation{ICRA and Dipartimento di Fisica, Sapienza Universit\`a di Roma, P.le Aldo Moro 5, 00185 Rome, Italy}
\affiliation{ICRANet, P.zza della Repubblica 10, 65122 Pescara, Italy. \href{mailto:yu.wang@icranet.org}{yu.wang@icranet.org}}
\affiliation{Universit\'e de Nice Sophia Antipolis, CEDEX 2, Grand Ch\^{a}teau Parc Valrose, Nice, France}

\author{M.~Kovacevic}
\affiliation{ICRA and Dipartimento di Fisica, Sapienza Universit\`a di Roma, P.le Aldo Moro 5, 00185 Rome, Italy}
\affiliation{ICRANet, P.zza della Repubblica 10, 65122 Pescara, Italy. \href{mailto:yu.wang@icranet.org}{yu.wang@icranet.org}}
\affiliation{Universit\'e de Nice Sophia Antipolis, CEDEX 2, Grand Ch\^{a}teau Parc Valrose, Nice, France}

\author{L.~Li}
\affiliation{ICRANet, P.zza della Repubblica 10, 65122 Pescara, Italy. \href{mailto:yu.wang@icranet.org}{yu.wang@icranet.org}}
\affiliation{Department of Physics, Stockholm University, SE-106 91 Stockholm, Sweden}

\author{J.~D.~Melon~Fuksman}
\affiliation{ICRA and Dipartimento di Fisica, Sapienza Universit\`a di Roma, P.le Aldo Moro 5, 00185 Rome, Italy}
\affiliation{ICRANet, P.zza della Repubblica 10, 65122 Pescara, Italy. \href{mailto:yu.wang@icranet.org}{yu.wang@icranet.org}}

\author{R.~Moradi}
\affiliation{ICRA and Dipartimento di Fisica, Sapienza Universit\`a di Roma, P.le Aldo Moro 5, 00185 Rome, Italy}
\affiliation{ICRANet, P.zza della Repubblica 10, 65122 Pescara, Italy. \href{mailto:yu.wang@icranet.org}{yu.wang@icranet.org}}

\author{M.~Muccino}
\affiliation{ICRA and Dipartimento di Fisica, Sapienza Universit\`a di Roma, P.le Aldo Moro 5, 00185 Rome, Italy}
\affiliation{ICRANet, P.zza della Repubblica 10, 65122 Pescara, Italy. \href{mailto:yu.wang@icranet.org}{yu.wang@icranet.org}}

\author{A.~V.~Penacchioni}
\affiliation{ASI Science Data Center, Via del Politecnico s.n.c., 00133 Rome, Italy}
\affiliation{Dept. of Physical Sciences, Earth and Environment, University of Siena, Via Roma 56, 53100 Siena, Italy}
\affiliation{ICRANet, P.zza della Repubblica 10, 65122 Pescara, Italy. \href{mailto:yu.wang@icranet.org}{yu.wang@icranet.org}}

\author{G.~B.~Pisani}
\affiliation{ICRA and Dipartimento di Fisica, Sapienza Universit\`a di Roma, P.le Aldo Moro 5, 00185 Rome, Italy}
\affiliation{ICRANet, P.zza della Repubblica 10, 65122 Pescara, Italy. \href{mailto:yu.wang@icranet.org}{yu.wang@icranet.org}}

\author{D.~Primorac}
\affiliation{ICRA and Dipartimento di Fisica, Sapienza Universit\`a di Roma, P.le Aldo Moro 5, 00185 Rome, Italy}
\affiliation{ICRANet, P.zza della Repubblica 10, 65122 Pescara, Italy. \href{mailto:yu.wang@icranet.org}{yu.wang@icranet.org}}

\author{J.~A.~Rueda}
\affiliation{ICRA and Dipartimento di Fisica, Sapienza Universit\`a di Roma, P.le Aldo Moro 5, 00185 Rome, Italy}
\affiliation{ICRANet, P.zza della Repubblica 10, 65122 Pescara, Italy. \href{mailto:yu.wang@icranet.org}{yu.wang@icranet.org}}
\affiliation{ICRANet-Rio, Centro Brasileiro de Pesquisas F\'isicas, Rua Dr. Xavier Sigaud 150, 22290--180 Rio de Janeiro, Brazil}

\author{S.~Shakeri}
\affiliation{Department of Physics, Isfahan University of Technology, 84156-83111, Iran}
\affiliation{ICRANet, P.zza della Repubblica 10, 65122 Pescara, Italy. \href{mailto:yu.wang@icranet.org}{yu.wang@icranet.org}}

\author{G.~V.~Vereshchagin}
\affiliation{ICRA and Dipartimento di Fisica, Sapienza Universit\`a di Roma, P.le Aldo Moro 5, 00185 Rome, Italy}
\affiliation{ICRANet, P.zza della Repubblica 10, 65122 Pescara, Italy. \href{mailto:yu.wang@icranet.org}{yu.wang@icranet.org}}

\author{S.-S.~Xue}
\affiliation{ICRA and Dipartimento di Fisica, Sapienza Universit\`a di Roma, P.le Aldo Moro 5, 00185 Rome, Italy}
\affiliation{ICRANet, P.zza della Repubblica 10, 65122 Pescara, Italy. \href{mailto:yu.wang@icranet.org}{yu.wang@icranet.org}}

\begin{abstract}
We analyze the early X-ray flares in the GRB ``flare-plateau-afterglow" (FPA) phase observed by Swift-XRT. The FPA occurs only in one of the seven GRB subclasses: the binary-driven hypernovae (BdHNe). This subclass consists of long GRBs with a carbon-oxygen core and a neutron star (NS) binary companion as progenitors. The hypercritical accretion of the supernova (SN) ejecta onto the NS can lead to the gravitational collapse of the NS into a black hole. Consequently, one can observe a GRB emission with isotropic energy $E_{iso}\gtrsim10^{52}$~erg, as well as the associated GeV emission and the FPA phase. Previous work had shown that gamma-ray spikes in the prompt emission occur at $\sim 10^{15}$--$10^{17}$~cm with Lorentz gamma factor $\Gamma\sim10^{2}$--$10^{3}$. Using a novel data analysis we show that the time of occurrence, duration, luminosity and total energy of the X-ray flares correlate with $E_{iso}$. A crucial feature is the observation of thermal emission in the X-ray flares that we show occurs at radii $\sim10^{12}$~cm with $\Gamma\lesssim 4$.  These model independent observations cannot be explained by the ``fireball'' model, which postulates synchrotron and inverse Compton radiation from a single ultra relativistic jetted emission extending from the prompt to the late afterglow and GeV emission phases. We show that in BdHNe a collision between the GRB and the SN ejecta occurs at $\simeq10^{10}$~cm reaching transparency at $\sim10^{12}$~cm with $\Gamma\lesssim4$. The agreement between the thermal emission observations and these theoretically derived values validates our model and opens the possibility of testing each BdHN episode with the corresponding Lorentz gamma factor.
\end{abstract}


\section{Introduction}
\label{sec:intro}

Following the discovery of the gamma-ray bursts (GRBs) by the Vela satellites \citep{1973ApJ...182L..85K} and the observations by the BATSE detectors on board the Compton Gamma-Ray Observatory \citep[CGRO,][]{1993A&AS...97....5G}, a theoretical framework for the interpretation of GRBs was established. This materialized into the ``traditional" model of GRBs developed in a large number of papers by various groups. They all agree in their general aspects: short GRBs are assumed to originate from the merging of binary NSs \citep[see, e.g.,][]{1986ApJ...308L..47G,1986ApJ...308L..43P,Eichler:1989jb,1991ApJ...379L..17N,1992ApJ...395L..83N,1538-4357-482-1-L29}, and long GRBs are assumed to originate from a ``collapsar'' \citep{1993ApJ...405..273W,Paczynski:1998ey,MacFadyen:1999wc,2013ApJ...764..179B} which, in turn, originates from the collapse of the core of a single  massive star to a black hole (BH) surrounded by a thick massive accretion disk \citep{2004RvMP...76.1143P}. In this traditional picture the  GRB dynamics follows the ``fireball'' model, which assumes the existence of an ultra-relativistic collimated jet \citep[see e.g.][]{1990ApJ...365L..55S,1993MNRAS.263..861P,1993ApJ...415..181M,1994ApJ...424L.131M}. The structures of long GRBs were described either by internal or external shocks \citep[see][]{1992MNRAS.258P..41R,1994ApJ...430L..93R}. The emission processes were linked to the occurrence of synchrotron and/or inverse-Compton radiation coming from the jetted structure, characterized by Lorentz factors $\Gamma \sim 10^2$--$10^3$, in what later become known as the ``prompt emission" phase (see Sec.~\ref{sec:Theory}).

The joint X-ray, gamma ray and optical observations heralded by Beppo-SAX and later extended by Swift have discovered the X-ray ``afterglow'', which allowed the optical identification and the determination of the GRBs cosmological distance. The first evidence for the coincidence of a GRB and a supernova (SN) (GRB 980425/SN 1998bw) was also announced, as well as  the first observation of an early X-ray flare (XRT), later greatly extended in number and spectral data by the Swift satellite, the subjects of this paper. The launch of the Fermi and AGILE satellites led to the equally fundamental discovery of the GeV emission both in long and short GRBs (see Sec.~\ref{sec:obsbackground}).

The traditional model reacted to these new basic informations  by extending the description of the ``collapsar'' model, adopted for the prompt emission, both to the afterglow and to the GeV emission. This approach based on the gravitational collapse of a single massive star, was initially inspired by analogies with the astrophysics of active galactic nuclei, has been adopted with the aim to identify a ``standard model'' for all long GRBs and vastly accepted by concordance \citep[see, e.g.,][]{Piran1999,2004RvMP...76.1143P,Meszaros2002,Meszaros2006,2009ARA&A..47..567G,2014ARA&A..52...43B,2015PhR...561....1K}. Attempts to incorporate the occurrence of a SN in the collapsar by considering nickel production in the accretion process around the BH were also proposed \citep{1999ApJ...524..262M}. In 1999, a pioneering work by \citet{1999ApJ...526..152F} introduced considerations based on population synthesis computations and emphasized the possible relevance of binary progenitors in GRBs.

Since 2001 we have developed an alternative GRB model based on the concept of induced gravitational collapse (IGC) paradigm which involves, as progenitors, a binary system with standard components: an evolved carbon-oxygen core (CO$_{\rm core}$) and a binary companion neutron star (NS). The CO$_{\rm core}$ undergoes a traditional Ic SN explosion, which produces a new NS ($\nu$NS) and a large amount of ejecta. There is a multitude of new physical processes, occurring in selected episodes, associated with this process. The ``first episode'' (see Sec.~\ref{sec:Theory}) of the binary-driven hypernova (BdHN) is dominated by the hypercritical accretion process of the SN ejecta on the companion NS. This topic has been developed in \citep[see, e.g.,][]{Ruffini2001c,Rueda2012,2014ApJ...793L..36F,2015ApJ...812..100B,2016ApJ...833..107B}. These processes are not considered in the collapsar model. Our SN is a traditional type Ic, the creation of the $\nu$NS follows standard procedure occurring in pulsar physics \citep[see e.g.][]{2012A&A...540A..12N}, the companion NS is a standard one regularly observed in binaries \citep[see e.g.][]{2017IJMPD..2630016R,Rueda2012} and the physics of hypercritical accretion has been developed by us in a series of recent articles (see Sec.~\ref{sec:accretion}).

In BdHN the BH and a vast amount of $e^+e^-$ plasma are formed only after the accreting NS reaches the critical mass and the  ``second episode'' starts  (see Sec.~\ref{sec:dynamicse+e-}). The main new aspect of our model addresses the interaction of the $e^+e^-$ plasma with the SN ejecta. We apply the fireshell model which make use of a general relativistic correct space-time parametrisation of the GRBs, as well as a new set of relativistic hydrodynamics equation for the dynamics of the $e^+e^-$ plasma. Selected values of the baryon loads are adopted in correspondence to the different time varying density distribution of the SN ejecta.

In the ``third episode'' (see Sec.~\ref{sec:3.6}), we  mention also the perspectives, utilising the experience gained  both in data analysis and in the theory for the specific understanding of X-ray flares,  to further address in forthcoming publications the more comprehensive  case of the gamma-ray flares, the consistent  treatment of the Afterglow and finally the implication of the GeV radiation.

As the model evolved we soon realized that the discovery of new sources was not leading to a ``standard model'' of long GRBs but, on the contrary, they were revealing a number of new GRB subclasses with distinct properties characterizing their light-curves, spectra and energetics \citep[see][]{2016ApJ...832..136R}. Moreover these seven subclasses  did not necessarily contain a BH. We soon came to the conclusion that only in the subclass of BdHN, with an $E_{iso}$ larger than $10^{52}$ erg, the hypercritical accretion from the SN into the NS leads to the creation of a newly-born BH with the associated signatures in the long GRB emission \citep[see e.g.][]{2015ApJ...812..100B,2016ApJ...833..107B}.

While our alternative model was progressing, we were supported by new astrophysical observations: the great majority of GRBs are related to type Ic SNe, which have no trace of hydrogen and helium in their optical spectra, and are spatially correlated with bright star-forming regions in their host galaxies \citep{2006Natur.441..463F,2010MNRAS.405...57S}. Most massive stars are found in binary systems \citep{2014ARA&A..52..487S} where most type Ic SNe occur and which favor the deployment of hydrogen and helium from the SN progenitors \citep{2011MNRAS.412.1522S}, and the SNe associated with long GRBs are indeed of type Ic \citep{2011IJMPD..20.1745D}. In addition, these SNe associated with long bursts are broad-lined Ic SNe (hypernovae) showing the occurrence of some energy injection leading to a kinetic energy larger than that of the traditional SNe Ic \citep{2016MNRAS.457..328L}.

The present paper addresses the fundamental role of X-ray flares as a separatrix between the two alternative GRB models and leads to the following main results, two obtained by the data analysis, and one obtained from the comparison of the alternative models:

1) The discovery of precise correlations between the X-ray flares and the GRB $E_{iso}$.

2) The radius of the occurrence  X-ray flares ($\sim 10^{12}$~cm) and Lorentz Gamma factor $\sim 2$.

3) The occurrence of a sharp brake between the prompt emission phase and the FPA phase, not envisaged in the current GRB literature. This transition evidence a contradiction in the ultra-relativistic jetted emission for explaining the X-ray flares, the plateau and the afterglow.

In Sec.~\ref{sec:obsbackground} we recall, following the gamma-ray observations by the Vela satellites and the Compton Gamma-Ray Observatory (CGRO), the essential role of Beppo-SAX and the Swift satellite. They provided the X-ray observations specifically on the X-ray flares, to which our new data-analysis techniques and paradigms have been applied. We also recall the Fermi and the AGILE satellites were the existence of the GeV emission was announced, which has become essential for establishing the division of GRBs into different subclasses.

In Sec.~\ref{sec:Theory} we update our classification of GRBs with known redshift into seven different subclasses (see Table~\ref{tab:rates}). For each subclass we indicate the progenitor ``in-states" and the corresponding ``out-states". We update the list of BdHNe (see Appendix~\ref{sec:bdhne}): long GRBs with $E_{iso}\gtrsim10^{52}$~erg, with an associated GeV emission and with the occurrence of the FPA phase. We also recall the role of an appropriate time parametrization for GRBs, properly distinguishing the four time variables which enter into their analysis. Finally we recall the essential theoretical background needed for the description of the dynamics of BdHNe, the role of the neutrino emission in the process of hypercritical accretion of the SN ejecta onto the binary companion NS, the description of the dynamics of the $e^+e^-$-baryon plasma, and the prompt emission phase endowed with gamma-ray spikes.We then shortly address the new perspectives open by the present work, to be further extended to the analysis of gamma-ray flares, to the afterglow and the essential role of each BdHN component, including the $\nu$NS. Having established the essential observational and theoretical background in Secs.~\ref{sec:obsbackground} and \ref{sec:Theory}, we proceed to the data analysis of the X-ray flares.

In Sec.~\ref{sec:sample} we address the procedure used to compare and contrast GRBs at different redshifts, including the description in their cosmological rest frame as well as the consequent K-corrections. This procedure has been neglected in the current GRB literature \citep[see, e.g.,][and references therein, as well as Sec.~\ref{sec:conclusions}]{2010MNRAS.406.2113C}. We then identify the BdHNe as the only sources where early time X-ray flares are identifiable. We recall that no X-ray flares have been found in X-ray flashes, nor in short GRBs. We also show  that a claim of the existence of X-ray flares in short bursts has been superseded. We recall our classified $345$ BdHNe (through the end of 2016). Their $T_{90}$\footnote{The $T_{90}$ is the duration defined as starting (ending) when the $5$\% ($95$\%) of the total energy of the event in gamma-rays has been emitted.}, properly evaluated in the source rest-frame, corresponds to the duration of their prompt emission phase, mostly shorter than $100$~s. Particular attention has been given to distinguishing the X-ray flares from the gamma-ray flares and spikes, each characterised by distinct spectral distributions and specific Lorentz gamma factors. The gamma ray flares are generally more energetic and with specific spectral signatures  (see e.g. the significant example of GRB 140206A in Sec.~\ref{sec:XRTlum} below). In this article we focus  on the methodology of studying X-ray flares: we plan to apply this knowledge to the  case of the early gamma-ray flares. Out of the $345$ BdHNe, there are $211$ which have complete Swift-\textit{XRT} observations, and among them, there are $16$ BdHNe with a well-determined early X-ray flare structure. They cover a wide range of redshifts, as well as the typical range of BdHN isotropic energies ($\sim 10^{52}$--$10^{54}$~erg). The sample includes all the identifiable X-ray flares.

In Sec.~\ref{sec:XRTlum}, we give the X-ray luminosity light curves of the $16$ BdHNe of our sample and, when available, the corresponding optical observations. As usual, these quantities have been K-corrected to their rest frame (see Figs.~\ref{Samplein}--\ref{Sample}, and Sec.~\ref{sec:sample}). In order to estimate the global properties of these sources, we also examine data from the Swift, Konus-Wind and Fermi satellites. The global results of this large statistical analysis are given in Tab.~\ref{tab:grbList} where the cosmological redshift $z$, the GRB isotropic energy $E_{iso}$, the flare peak time $t_p$, peak luminosity $L_p$, duration $\Delta t$, and the corresponding $E_{f}$ are reproduced. This lengthy analysis has been carried out over the past years, and only the final results are summarized in Tab.~\ref{tab:grbList}.

In Sec.~\ref{sec:correlation} we present the correlations between $t_p$, $L_p$, $\Delta t$, $E_{f}$ and $E_{iso}$ and give the corresponding parameters in Tab.~\ref{tab:correlation}. In this analysis we applied the Markov chain Monte Carlo (MCMC) method, and we also have made public the corresponding numerical codes in \href{https://github.com/YWangScience/AstroNeuron}{https://github.com/YWangScience/AstroNeuron} and \href{https://github.com/YWangScience/MCCC}{https://github.com/YWangScience/MCCC}.

In Sec.~\ref{sec:partition} we discuss the correlation between the energy of the prompt emission, the energy of the FPA phase, and $E_{iso}$ (see Tabs.~\ref{tab:grbList2}-\ref{tab:correlation2} and Figs.~\ref{fig:EisoEflareEnd}--\ref{fig:percentage}).

In Sec.~\ref{sec:thermalflare} we analyze the thermal emission observed during the X-ray flares (see Tab.~\ref{tab:grbTemperature}). We derive, in an appropriate relativistic formalism, the relations between the observed temperature and flux and the corresponding temperature and radius of the thermal emitter in its comoving frame.

In Sec.~\ref{sec:thermalflare2} we use the results of Sec.~\ref{sec:thermalflare} to infer the expansion speed of the thermal emitter associated with the thermal components observed during the flares (see Fig.~\ref{fig:081008Spec} and Tab.~\ref{tab:thermList}). We find that the observational data implies a Lorentz factor $\Gamma \lesssim 4$ and a radius of $\approx 10^{12}$~cm for such a thermal emitter.

In Sec.~\ref{sec:originprFPA} we present a theoretical treatment using a new relativistic hydrodynamical code to simulate the interaction of the $e^+e^-$-baryon plasma with the high-density regions of the SN ejecta. We first test the code in the same low-density domain of validity describing the prompt emission phase, and then we apply it in the high-density regime of the propagation of the plasma inside the SN ejecta which we use for the theoretical interpretation of the X-ray flares. Most remarkably, the theoretical code leads to a thermal emitter at transparency with a Lorentz factor $\Gamma \lesssim 4$ and a radius of $\approx 10^{12}$~cm. The agreement between these theoretically derived values and the ones obtained from the observed thermal emission validates the model and the binary nature of the BdHN progenitors, in clear contrast with the traditional ultra-relativistic jetted models.

In Sec.~\ref{sec:conclusions} we present our conclusions. We first show how  the traditional model, describing GRBs as a single system with ultra-relativistic jetted emission extending from the prompt emission all the way to the final phases of the afterglow and of the GeV emission, is in conflict with the X-ray flare observations. We also present three main new results which illustrate new perspectives opened by our alternative approach based on the BdHNe.

A standard flat ${\Lambda}$CDM cosmological model with $\Omega_M=0.27$, $\Omega_\Lambda=0.73$, and $H_0=71$ km s$^{-1}$ Mpc$^{-1}$ is adopted throughout the paper, while Table~\ref{acronyms} summarizes the acronyms we have used.
\begin{table}
\centering
\begin{tabular}{lc}
\hline\hline
Extended wording & Acronym \\
\hline
Binary-driven hypernova & BdHN \\
Black hole                    & BH \\
Carbon-oxygen core      & CO$_{\rm core}$ \\
Circumburst medium     & CBM \\
flare-Plateau-Afterglow & FPA \\
Gamma-ray burst         & GRB \\
Gamma-ray flash          & GRF \\
Induced gravitational collapse & IGC \\
Massive neutron star     & MNS \\
Neutron star                & NS \\
New neutron star          & $\nu$NS \\
Proper gamma-ray burst & P-GRB \\
Short gamma-ray burst  & S-GRB \\
Short gamma-ray flash  & S-GRF \\
Supernova                  & SN \\
Ultrashort gamma-ray burst & U-GRB \\
White dwarf                & WD \\
X-ray flash                  & XRF \\
\hline
\end{tabular}
\caption{Alphabetic ordered list of the acronyms used in this work.}
\label{acronyms}
\end{table}

\section{Background for the observational identification of the X-ray flares}
\label{sec:obsbackground}

The discovery of GRBs by the Vela satellites \citep{1973ApJ...182L..85K} was presented at the AAAS meeting in February 1974 in San Francisco \citep{GurskyRuffini1975}. The Vela satellites were operating in gamma rays in the $150$--$750$~keV energy range and only marginally in X-rays ($3$--$12$~keV, \citealt{1979ApJ...229L..47C}). Soon after it was hypothesized from first principles that GRBs may originate from an $e^+e^-$ plasma in the gravitational collapse to a Kerr-Newman BH, implying an energy $\sim 10^{54} M_{BH}/M_\odot$~erg (\citealp{1975PhRvL..35..463D}, see also \citealp{1998bhhe.conf..167R}).

Since 1991 the BATSE detectors on the Compton Gamma-Ray Observatory \citep[CGRO, see][]{1993A&AS...97....5G} led to the classification of GRBs on the basis of their spectral hardness and of their observed $T_{90}$ duration in the
$50$--$300$~keV energy band into short/hard bursts ($T_{90}<2$~s) and long/soft bursts ($T_{90}>2$~s \citep{Mazets1981,Klebesadel1992,Dezalay1992,Koveliotou1993,Tavani1998}.
Such an emission was later called the GRB ``prompt emission". In a first attempt it was proposed that short GRBs originate from merging binary NSs \citep[see, e.g.,][]{1986ApJ...308L..47G,1986ApJ...308L..43P,Eichler:1989jb,1991ApJ...379L..17N,1992ApJ...395L..83N,1538-4357-482-1-L29} and long GRBs originate from a single source with ultra-relativistic jetted emission \citep{1993ApJ...405..273W,Paczynski:1998ey,MacFadyen:1999wc,2013ApJ...764..179B}.

\begin{figure}
\centering
\includegraphics[width=0.8\hsize,clip]{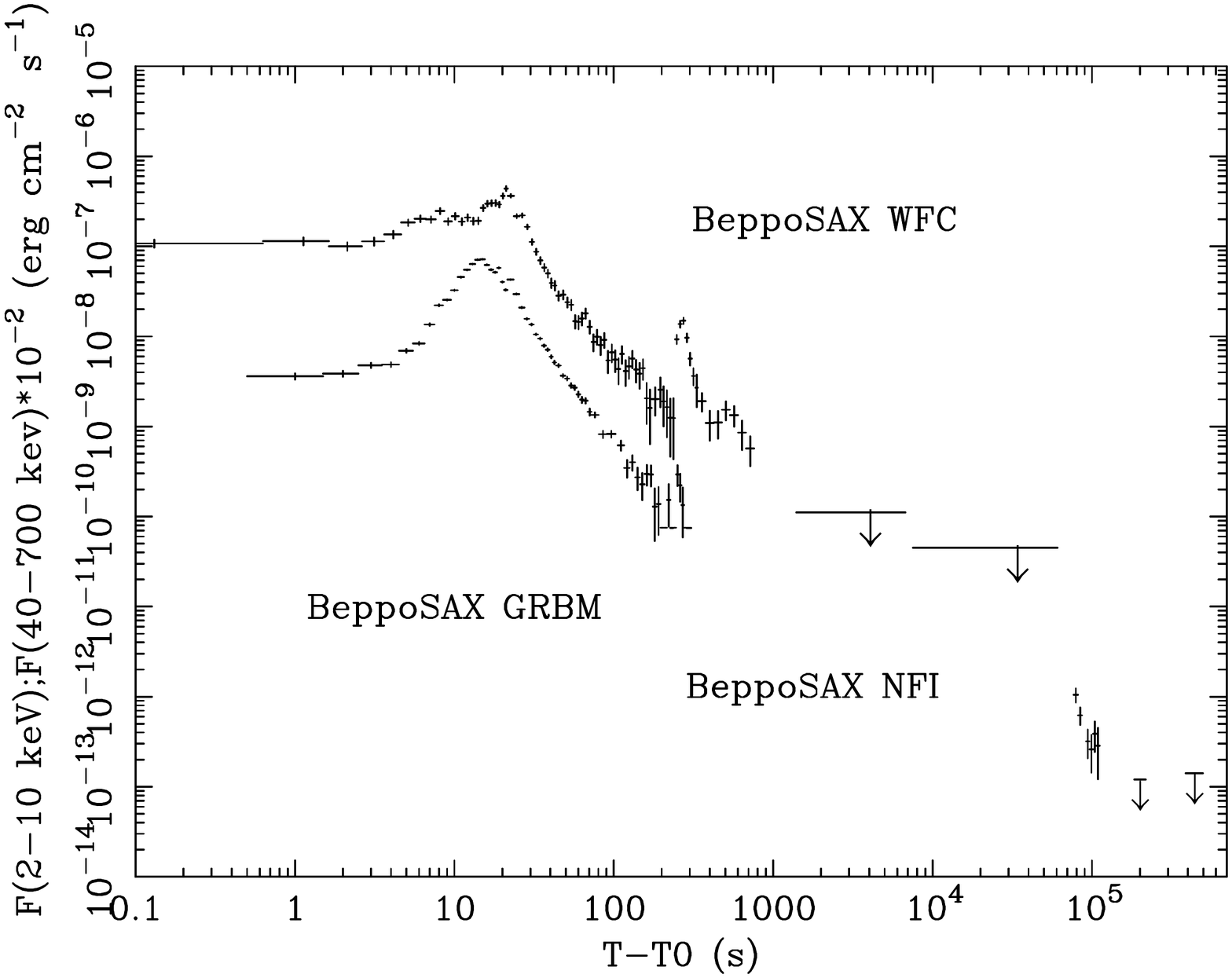}
\caption{First X-ray flare observed by BeppoSAX in GRB 011121. Reproduced from \citet{2005ApJ...623..314P}.}
\label{fig:piro}
\end{figure}

The Beppo-SAX satellite, operating since 1996, joined the expertise of the X-ray and gamma-ray communities. Its gamma-ray burst monitor (GRBM) operating in the $40$--$700$~keV energy band determined the trigger of the GRB and two wide field cameras (WFCs) operating in the $2$--$30$~keV X-ray energy band allowed the localization of the source within an arc minute resolution. This enabled a follow-up with the narrow field instruments (NFI) in the $2$--$10$~keV energy band. Beppo-SAX discovered the X-ray afterglow \citep{Costa1997}, characterized by an X-ray luminosity decreasing with a constant index of $\sim-1.3$ (see \citealp{DePasquale2006}, as well as \citealp{2016ApJ...833..159P}). This emission was detected after an ``8 hour gap," following the prompt emission identified by BATSE. The consequent determination of the accurate positions by the NFI, transmitted to the optical \citep{vanParadjis1997} and radio telescopes \citep{1997Natur.389..261F}, allowed the determination of the GRB cosmological redshifts \citep{1997Natur.387..878M}. The derived distances of $\approx5$--$10$~Gpc confirmed their cosmological origin and their unprecedented energetics $\approx 10^{50}$--$10^{54}$~erg, thus validating our hypothesis derived from first principles \citep{1975PhRvL..35..463D,1998bhhe.conf..167R}.

To Beppo-SAX goes the credit of the discovery of the temporal and spatial coincidence of GRB 980425 with SN 1998bw \citep{1998Natur.395..670G}, which suggested the connection between GRBs and SNe, soon supported by many additional events \citep[see e.g.][]{2006ARA&A..44..507W,2011IJMPD..20.1745D,2012grb..book..169H}.
Beppo-SAX also discovered the first ``X-ray flare" in GRB 011121 closely following the prompt emission \citep{2005ApJ...623..314P}, see Fig.~\ref{fig:piro}.
Our goal in this paper is to show how the X-ray flares, thanks to the observational campaign of the Swift satellite, have become the crucial test for understanding the astrophysical nature of the GRB-SN connection.

The Swift burst alert telescope (BAT), operating in the $15$--$150$~keV energy band, can detect GRB prompt emissions and accurately determine their position in the sky within 3 arcmin. Within $90$~s Swift can re-point the narrow-field X-ray telescope (XRT), operating in the $0.3$--$10$~keV energy range, and relay the burst position to the ground. This overcame the ``$8$ hour gap" in the Beppo-SAX data.

\begin{figure}
\centering
\includegraphics[angle=270,width=0.8\hsize,clip]{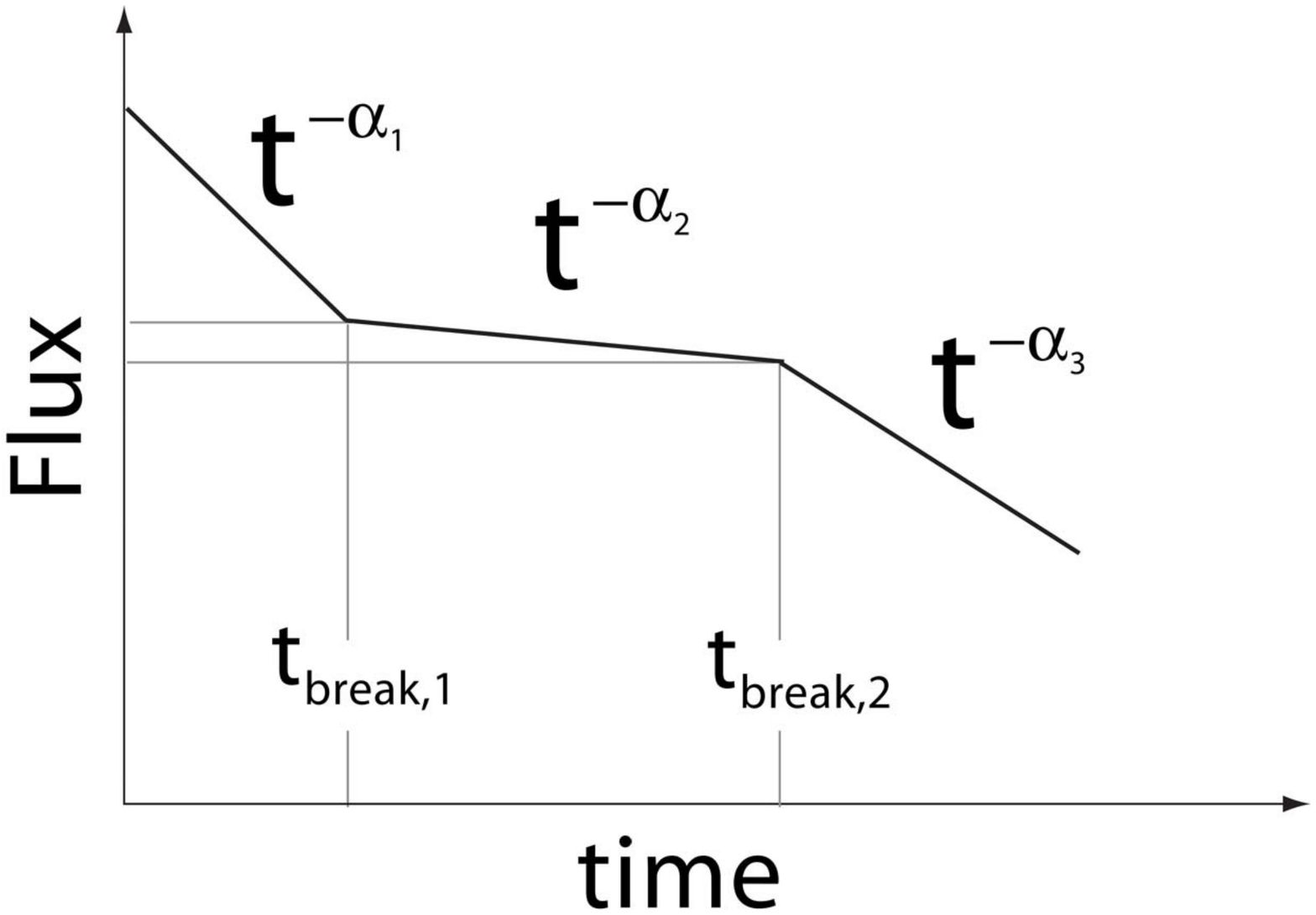}
\caption{Schematic diagram of the X-ray light-curve composed of three power-law segments with different slopes ($3\lesssim\alpha_1\lesssim 5,0.5\lesssim\alpha_2\lesssim 1.0,1\lesssim\alpha_3\lesssim 1.5$). Figure taken from \citet{Nousek2006}.}
\label{fig:lightCurveCartoon}
\end{figure}

Thanks to the Swift satellite, the number of detected GRBs increased rapidly to $480$ sources with known redshifts. By analyzing the light-curve of some long GRBs including the data in the ``$8$ hour gap" of Beppo-SAX, \citet{Nousek2006} and \citet{Zhang2006} discovered three power-law segments in the XRT flux light-curves of some long GRBs. We refer to these as the ``Nousek-Zhang power laws'' (see Fig.~\ref{fig:lightCurveCartoon}). The nature of this feature has been the subject of a long debates, still ongoing, and finally resolved in this article.

\begin{figure}
\centering
\includegraphics[width=0.8\hsize,clip]{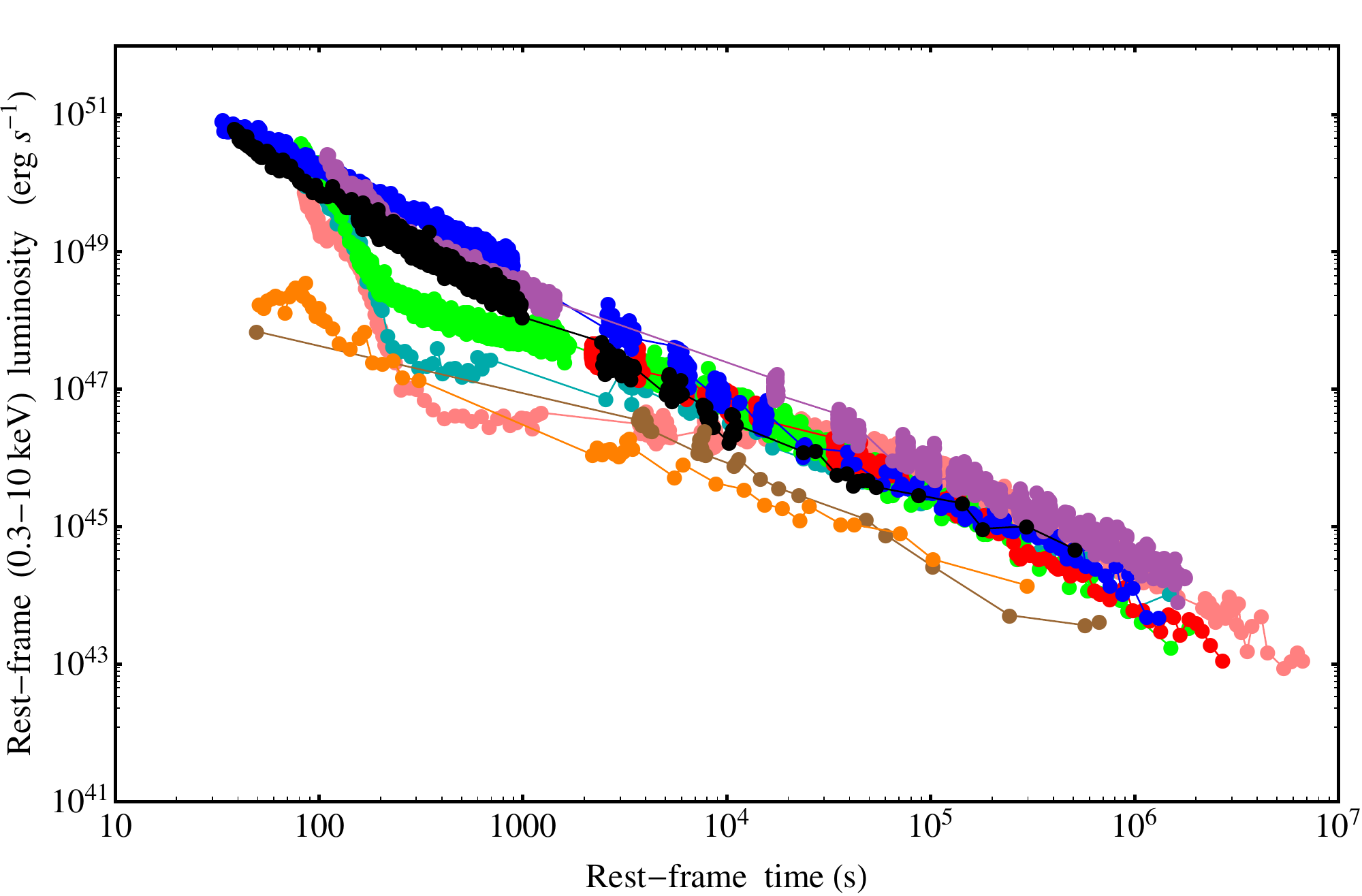}
\includegraphics[width=0.8\hsize,clip]{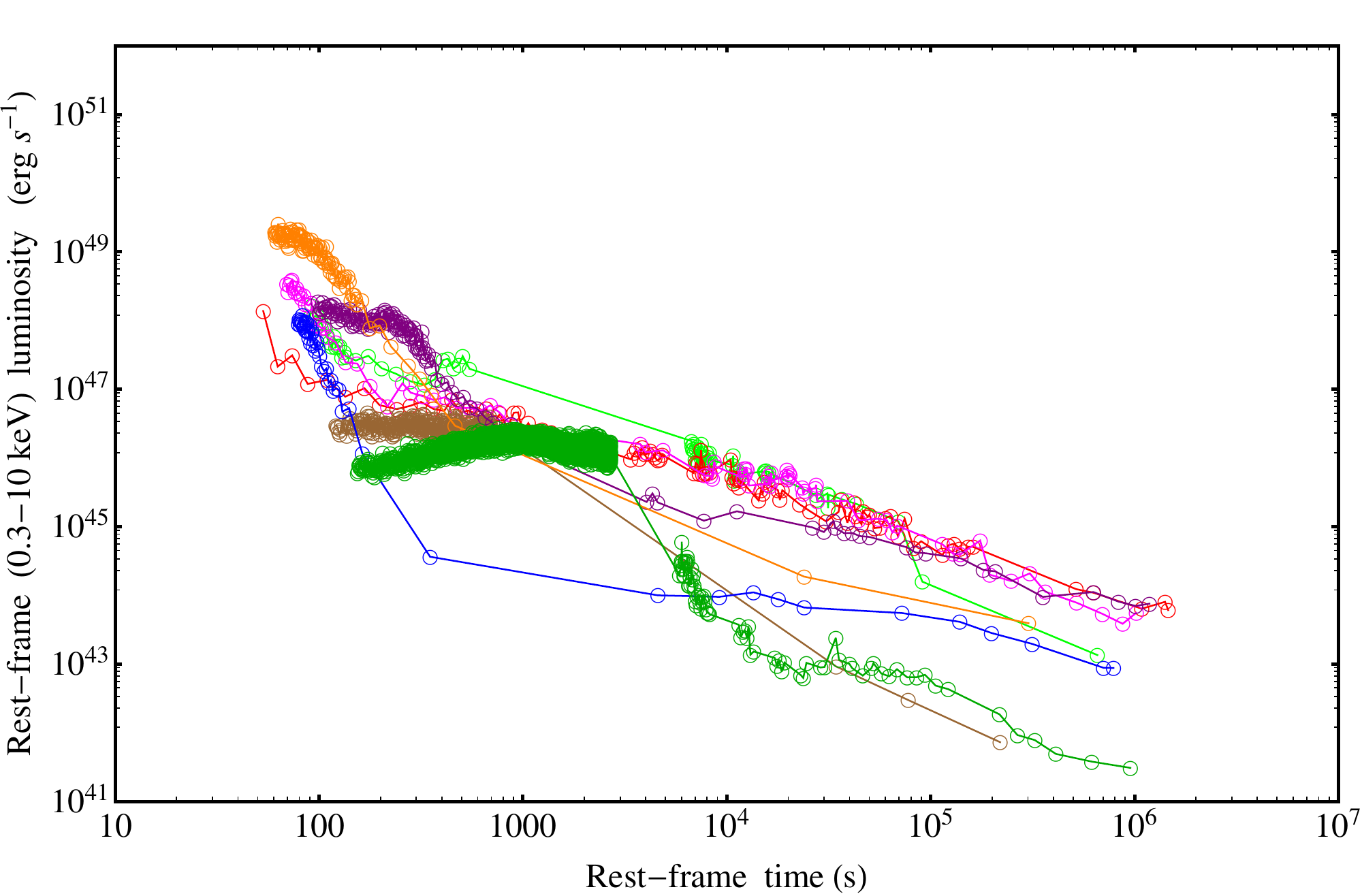}
\caption{X-ray light-curves of long GRBs observed by Swift. Top panel: BdHNe 050525 (brown), 060729 (pink), 061007 (black), 080319B (blue), 090618 (green), 091127 (red), 100816A (orange), 111228A (light blue), and 130427A (purple). Bottom panel: XRFs 050416A (red), 060218 (dark green), 070419A (orange), 081007 (magenta), 100316D (brown), 101219B (purple), and 130831A (green). XRFs have generally lower and more scattered light-curves. All these GRBs have known redshift, and the light-curves have been transformed to their cosmological rest frames.}
\label{fig:compareXRF}
\end{figure}

We have used Swift-XRT data in differentiating two distinct subclasses of long GRBs, the XRFs with $E_{iso}\lesssim10^{52}$~erg and the BdHNe with $E_{iso}\gtrsim10^{52}$~erg (see Sec.~\ref{sec:Theory}). An additional striking difference appears between the XRT luminosities of these two subclasses when measured in their cosmological rest frames: in the case of BdHNe the light-curves follow a  specific behavior which conforms to the Nousek-Zhang power-law \citep[see e.g.][]{Penacchioni2011,2013A&A...551A.133P,Pisani2013,2016ApJ...833..159P,2014A&A...565L..10R}. None of these features are present in the case of XRFs (see Fig.~\ref{fig:compareXRF}).

Finally, the Fermi satellite \citep{2009ApJ...697.1071A} launched in 2008 detects ultra-high energy photons from $20$~MeV to $300$~GeV with the large area telescope (LAT), and detects photons from $8$~keV to $30$~MeV with the gamma-ray burst monitor (GBM). For the purposes of this article addressing long GRBs, the Fermi observations have been prominent in further distinguishing XRFs and BdHNe: the Fermi-LAT GeV emission has been observed only in BdHNe and never in XRFs. 

\section{Background for the theoretical interpretation of X-ray flares and their dynamics}
\label{sec:Theory}

\subsection{The classification of GRBs}\label{sec:3.1}

\begin{figure}
\centering
\includegraphics[width=0.8\hsize,clip]{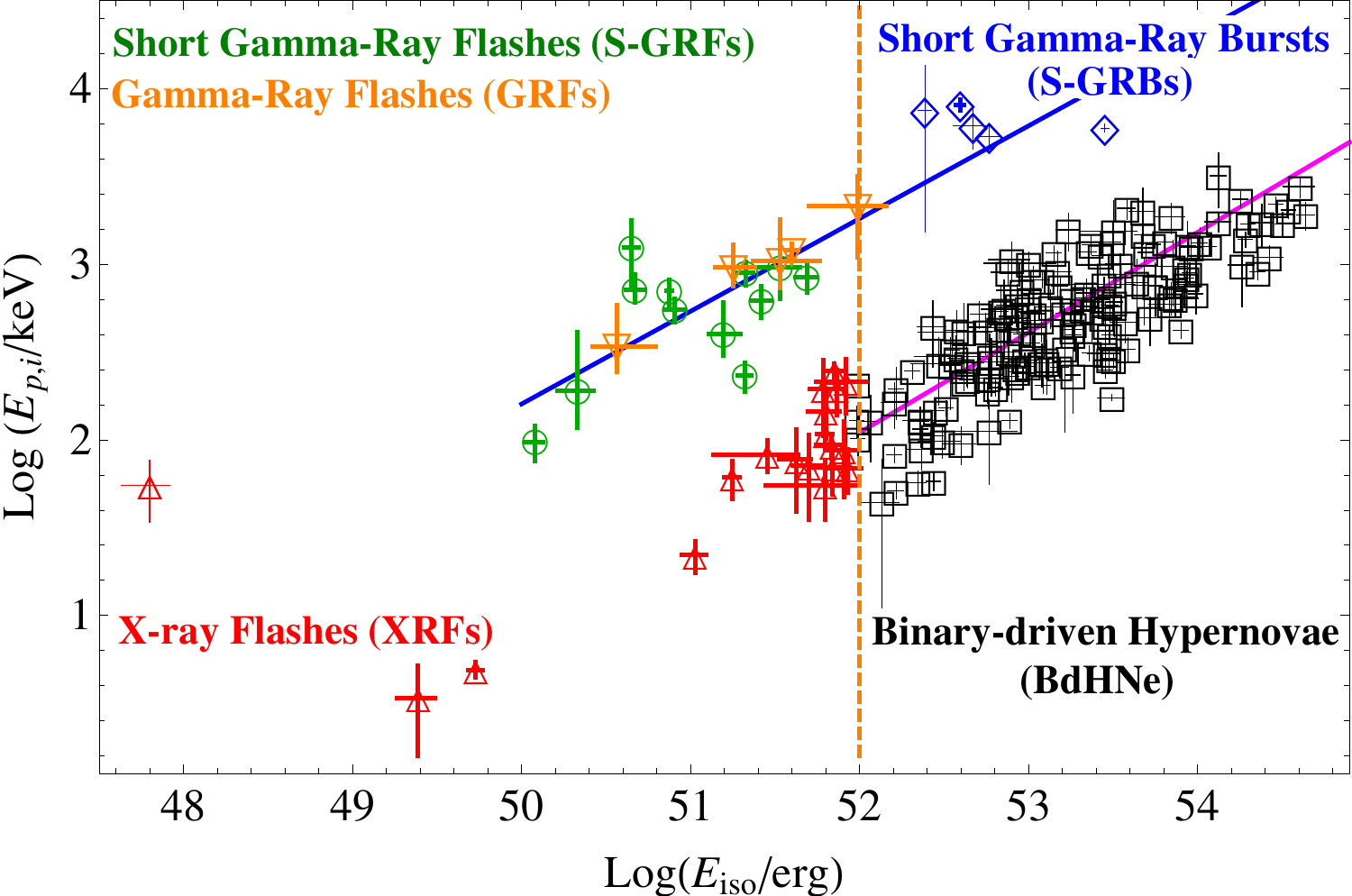}
\caption{The updated $E_{\rm p,i}$--$E_{\rm iso}$ plane for the subclasses defined in \citet{2016ApJ...832..136R}: XRFs (red triangles) cluster in the region defined by $E_{\rm p,i}\lesssim200$~keV and $E_{\rm iso}\lesssim 10^{52}$~erg. BdHNe (black squares) cluster in the region defined by $E_{\rm p,i}\gtrsim200$~keV and $E_{\rm iso}\gtrsim 10^{52}$~erg and fulfill the Amati relation \citep[solid magenta line with slope $\alpha=0.57\pm0.06$ and extra scatter $\sigma=0.25$, see e.g.][]{2013IJMPD..2230028A,Calderone2014}. S-GRFs (green circles) and the initial spike-like emission of the GRFs (orange reverse triangles) are concentrated in the region defined by $E_{\rm p,i}\lesssim2$~MeV and $E_{\rm iso}\lesssim 10^{52}$~erg, while S-GRBs (blue diamonds) in the region defined by $E_{\rm p,i}\gtrsim2$~MeV and $E_{\rm iso}\gtrsim 10^{52}$~erg. Short bursts and GRFs fulfill the MuRuWaZha relation \citep[blue solid line with slope $\alpha=0.53\pm0.07$ and extra scatter $\sigma=0.24$, see e.g.][]{2015ApJ...808..190R,Calderone2014,2012ApJ...750...88Z,2016ApJ...831..178R}. BH-SN and U-GRB subclasses \citep[see Tab.~\ref{tab:rates}][for details]{2016ApJ...832..136R} are not in the plot since their observational identification is still pending. The crucial difference between BdHNe and XRFs, and SGRBs and SGRF, is that BdHNe and S-GRBs form a BH, their energy is $\gtrsim10^{52}$~erg, and exhibit the GeV emission.}
\label{fig:EpEiso}
\end{figure}

The very extensive set of observations carried out by the above satellites in coordination with the largest optical and radio telescopes over a period of almost $40$ years has led to an impressive set of data on $480$ GRBs, all characterized by spectral, luminosity, and time variability information, and each one with a well established cosmological redshift. By classifying both the commonalities and the differences among all GRBs it has been possible to create ``equivalence relations" and divide GRBs into a number of subclasses, each one identified by a necessary and sufficient number of observables. We recall in Tab.~\ref{tab:rates} and Fig.~\ref{fig:EpEiso} the binary nature of all GRB progenitors and their classification into seven different subclasses \citep[see e.g.][]{2016ApJ...832..136R}.
In Tab.~\ref{tab:rates} we indicate the number of sources in each subclass, the nature of their progenitors and final outcomes of their evolution, their rest-frame $T_{90}$, their rest frame spectral peak energy $E_\mathrm{p,i}$ and $E_\mathrm{iso}$, as well as the isotropic energy in X-rays $E_\mathrm{iso,X}$ and in GeV emission $E_\mathrm{iso,GeV}$, and finally their local observed number density rate.
In Fig.~\ref{fig:EpEiso} we mention for these sources the $E_\mathrm{p,i}$--$E_\mathrm{iso}$ relations, including the Amati one for BdHNe and the MuRuWaZha one for the short bursts \citep[see][]{2016ApJ...832..136R,2016ApJ...831..178R},
comprising short gamma-ray flashes (S-GRFs) with $E_{iso}\lesssim10^{52}$~erg, authentic short GRBs (S-GRBs) with $E_{iso}\gtrsim10^{52}$~erg, and gamma-ray flashes (GRFs),  sources with hybrid short/long burst properties in their gamma-ray light curves, i.e., an initial spike-like harder emission followed by a prolonged softer emission observed up to $\sim100$~s, originating from NS--white dwarf binaries \citep{2009A&A...498..501C,2010A&A...521A..80C,2016ApJ...832..136R}. We have no evidence for an $E_\mathrm{p,i}$ and $E_\mathrm{iso}$ relation in the XRFs (see Fig.~\ref{fig:EpEiso}). The Amati and the MuRuWaZha relations have not yet been theoretically understood, and as such they have no predictive power.

\begin{table*}
\scriptsize
\caption{\label{tab:rates} Summary of the seven GRB subclasses (XRFs, BdHNe, BH-SN, short gamma-ray flashes (S-GRFs), authentic short GRBs (S-GRBs), ultrashort GRBs (U-GRB), and GRFs) and of their observational properties. In the first five columns we indicate the GRB subclasses and their corresponding number of sources with measured $z$, \emph{in-states} and \emph{out-states}. In the following columns we list the ranges of $T_{90}$ in the rest-frame, rest-frame spectral peak energies $E_{\rm p,i}$ and $E_{\rm iso}$ (rest-frame $1$--$10^4$~keV), the isotropic energy of the X-ray data $E_{\rm iso,X}$ (rest-frame $0.3$--$10$~keV), and the isotropic energy of the GeV emission $E_{\rm iso,GeV}$ (rest-frame $0.1$--$100$~GeV).  In  the last column we list, for each GRB subclass, the local observed number density rate $\rho_{\rm GRB}$ obtained in \citet{2016ApJ...832..136R}. For details see \citet{2014A&A...565L..10R,2015ApJ...808..190R,2015ApJ...798...10R,2015PhRvL.115w1102F,2016ApJ...831..178R,2016ApJ...832..136R,2016ApJ...833..107B}.}
\begin{ruledtabular}
\begin{tabular}{llccccccccc}
  & Subclass  &  number  & \emph{In-state}  & \emph{Out-state} &  $T_{90}$  &  $E_{\rm p,i}$ &  $E_{\rm iso}$ & $E_{\rm iso,X}$ &  $E_{\rm iso,Gev}$ & $\rho_{\rm GRB}$ \\
& & & (\emph{Progenitor}) & (\emph{Final outcome}) &  (s) & (MeV) & (erg) & (erg) & (erg) & (Gpc$^{-3}$yr$^{-1}$) \\
\hline
I    & XRFs & $82$ & CO$_{\rm core}$-NS  & $\nu$NS-NS &  $\sim2$--$10^3$  & $\lesssim0.2$  &  $\sim 10^{48}$--$10^{52}$ &  $\sim 10^{48}$--$10^{51}$ &  $-$ & $100^{+45}_{-34}$ \\
II   & BdHNe  & $345$ & CO$_{\rm core}$-NS  & $\nu$NS-BH &  $\sim2$--$10^2$  &  $\sim0.2$--$2$ &  $\sim 10^{52}$--$10^{54}$ &  $\sim 10^{51}$--$10^{52}$ &  $\lesssim 10^{53}$ & $0.77^{+0.09}_{-0.08}$ \\
III  & BH-SN & $-$ & CO$_{\rm core}$-BH  & $\nu$NS-BH &  $\sim2$--$10^2$  &  $\gtrsim2$ &  $>10^{54}$ & $\sim 10^{51}$--$10^{52}$ &  $\gtrsim 10^{53}$ & $\lesssim 0.77^{+0.09}_{-0.08}$ \\
IV   & S-GRFs & $33$ & NS-NS & MNS  &  $\lesssim2$  &  $\lesssim2$ &  $\sim 10^{49}$--$10^{52}$ &  $\sim 10^{49}$--$10^{51}$ &  $-$ &  $3.6^{+1.4}_{-1.0}$ \\
V    & S-GRBs  & $7$ & NS-NS & BH  &  $\lesssim2$  &  $\gtrsim2$ &  $\sim 10^{52}$--$10^{53}$ &  $\lesssim 10^{51}$ & $\sim 10^{52}$--$10^{53}$ & $\left(1.9^{+1.8}_{-1.1}\right)\times10^{-3}$ \\
VI   & U-GRBs & $-$ & $\nu$NS-BH & BH &  $\ll2$&  $\gtrsim2$ &  $>10^{52}$ &  $-$ & $-$ & $\gtrsim 0.77^{+0.09}_{-0.08}$ \\
VII  & GRFs  & $13$ & NS-WD & MNS &  $\sim2$--$10^2$  &  $\sim0.2$--$2$ &  $\sim 10^{51}$--$10^{52}$ &  $\sim 10^{49}$--$10^{50}$ & $-$ & $1.02^{+0.71}_{-0.46}$ \\
\end{tabular}
\end{ruledtabular}
\end{table*}

\subsection{The role of the time parametrization in GRBs}\label{sec:3.2}

Precise general relativistic rules in the space-time parameterization of GBRs are needed \citep{Ruffini2001a}. Indeed, there are four time variables entering this discussion which have to be properly distinguished one from another: 1) the comoving time $t_{com}$, which is the time used to compute the evolution of the thermodynamical quantities (density, pressure, temperature); 2) the laboratory time $t=\Gamma t_{com}$, where as usual the Lorentz gamma factor is $\Gamma=(1-\beta^2)^{-1/2}$ and $\beta=v/c$ is the expansion velocity of the source; 3) the arrival time $t_a$ at which each photon emitted by the source reaches an observer in the cosmological rest frame of the source, given by \citep[see also][]{2001A&A...368..377B,2002ApJ...581L..19R,Bianco2005a}:
\begin{equation}
t_a=t-\frac{r(t)}{c}\cos\vartheta\ ,
\end{equation}
where $r(t)$ is the radius of the expanding source in the laboratory frame and $\vartheta$ is the displacement angle of the normal to the emission surface from the line of sight; and 4) the arrival time at the detector on the Earth $t_a^d=t_a(1+z)$ corrected for cosmological effects, where $z$ is the source redshift, needed in order to compare GRBs at different redshifts $z$. As emphasized in \citet{Ruffini2001a}, ``the bookkeeping of these four different times and the corresponding space variables must be done carefully in order to keep the correct causal relation in the time sequence of the events involved". The chain of relations between these four times is given by \citep[see e.g.][and see also Secs.~\ref{sec:thermalflare} and \ref{sec:thermalflare2} for the dynamics of the flares]{2001A&A...368..377B,Ruffini2001a,2002ApJ...581L..19R,Bianco2005a}:
\begin{equation}
t_a^d = (1+z) t_a = (1+z)\left(t-\frac{r(t)}{c}\cos\vartheta\right)=(1+z)\left(\Gamma t_{com}-\frac{r(\Gamma t_{com})}{c}\cos\vartheta\right).
\label{tadef}
\end{equation}

The proper use of these four time variables is mandatory in modeling GRB sources, especially when we are dealing with a model not based on a single component but on multiple components, each characterized by a different world-line and a different Lorentz gamma factor, as it is the case for BdHNe (see Secs.~\ref{sec:sample} and \ref{sec:XRTlum}).

\subsection{The role of the GRBs cosmological rest-frame}\label{sec:3.3}

In addition to all the above, in order to compare the luminosities of different GRBs at different redshifts we need to express the observational data in the cosmological rest frames of each source (where the arrival time is $t_a$), and correspondingly apply the K-correction to luminosities and spectra (see Sec.~\ref{sec:sample}). This formalism is at the very foundation of the treatment presented in this paper and has been systematically neglected in the great majority of current GRB models.

\subsection{The Episode 1 :The hypercritical accretion process}
\label{sec:accretion}

In order to describe the dynamics of BdHNe a number of different episodes involving different physical conditions have to be described.
Episode 1 is dominated by the IGC paradigm: the hypercritical accretion of a SN ejecta on the companion binary NS \citep[see, e.g.,][]{2014ApJ...793L..36F,2015ApJ...812..100B,2015PhRvL.115w1102F,2016ApJ...833..107B}. Weak interactions and neutrinos \citep[see e.g.][]{1934NCim...11....1F}, which play a fundamental role in SN through the URCA process \citep{1940PhRv...58.1117G,1941PhRv...59..539G}, are also needed in the case of hypercritical accretion processes onto a NS in a SN fallback \citep{1971ApJ...163..221C,1972SvA....16..209Z,1973PhRvL..31.1362R}.
They are especially relevant in the case of BdHNe where the accretion rate onto the NS companion from the CO$_\mathrm{core}$ can reach up to $\dot{M} = 0.1~M_\odot$~s$^{-1}$ \citep{Rueda2012,2014ApJ...793L..36F,2015ApJ...812..100B,2016ApJ...833..107B}. Due to weak interactions, $e^+e^-$ pairs annihilate to $\nu \bar\nu$ pairs with a cross-section $\sigma\sim G_F \langle E_e \rangle^2$ \citep{1985ApJ...296..197M,1989ApJ...339..354I}. In the thermal system of $e^+e^-$ pairs at large temperature $k T > m_e c^2$ and density $n_e~\sim T^3$, the neutrino emissivity of the $e^+e^-$ annihilation is $\epsilon_{e^+e^-} \sim n_e^2 \langle \sigma v_e \rangle \langle E_e \rangle \sim 10^{25} (kT/\mathrm{MeV})^9$~erg~s$^{-1}$~cm$^{-3}$, leading to neutrino luminosities $L_\nu \sim R_\mathrm{NS}^3\epsilon_{e^+e^-} \sim 10^{52}$~erg~s$^{-1}$, which dominate over other microscopic processes for cooling \citep{2016ApJ...833..107B}. Thus the $e^+e^-$ pair annihilation to $\nu \bar\nu$ is the main process for cooling, allowing the process of hypercritical accretion to convert gravitational energy to thermal energy, to build up high temperature and consequently to form an $e^+e^-$ plasma. Only at the end of this Episode 1, as the critical mass of the companion NS is reached, a BH is formed with the additional $e^+e^-$ pairs linked to the BH electrodynamical process \citep{1975PhRvL..35..463D,2009PhRvD..79l4002C}.

\subsection{The $e^+e^-$ pairs colliding with the SN ejecta}
\label{sec:dynamicse+e-}

Episode 2. This Episode is dominated by the new phenomena of the impact of the $e^+e^-$ pairs generated in the GRB with the SN ejecta. We describe this process within the fireshell model. Two main differences exist between the fireshell and the fireball model. In the fireshell model the $e^+e^-$ plasma is initially in thermal equilibrium and undergoes an ultra-relativistic expansion keeping this condition of thermal equilibrium all the way to reaching transparency  (\citealp{1998bhhe.conf..167R}, see also \citealp{2007PhRvL..99l5003A,2010PhR...487....1R} and references therein) while, in the fireball model \citet{1978MNRAS.183..359C}, the $e^+e^-$ pairs undergo an initial annihilation process producing the photons driving the fireball. An additional basic difference is that the evolution of the $e^+e^-$ plasma is not imposed by a given asymptotic solution but integrated following the relativistic fluidoynamics equations. The plasma, with energy $E_{e^{+}e^{-}}$, goes first through an initial acceleration phase \citep{RSWX2}. After colliding with the baryons (of total mass $M_B$), characterized by the baryon load parameter $B=M_Bc^2/E_{e^+e^-}$, the optically thick plasma keeps accelerating until reaching transparency and emitting a proper gamma-ray burst \citep[P-GRB, see][]{RSWX}. The accelerated baryons then interact with the circumburst medium (CBM) clouds \citep{Ruffini2001b}; the equation of motion of the plasma has been integrated leading to results which differ from the ones of the \citet{1976PhFl...19.1130B} self-similar solution \citep[see][]{Bianco2004,Bianco2005a,2005ApJ...633L..13B,2006ApJ...644L.105B}.
By using Eq.~(\ref{tadef}) defining the ``equitemporal surfaces" \citep[see][]{2001A&A...368..377B,Bianco2004,Bianco2005a,2005ApJ...633L..13B,2006ApJ...644L.105B} it has been possible to infer the structure of the gamma-ray spikes in the prompt emission, which for the most part has been applied to the case of BdHNe \citep[see, e.g.,][]{2002ApJ...581L..19R,2005ApJ...634L..29B,Patricelli,2012A&A...543A..10I,Penacchioni2011,2013A&A...551A.133P,2016ApJ...831..178R}. For typical baryon loads $10^{-4} \lesssim B \lesssim 10^{-2}$ leading to Lorentz gamma factors  $\Gamma \approx 10^2$--$10^3$ at transparency for the $e^+e^-$-baryon plasma, characteristic distances from the BH of $\approx10^{15}$--$10^{17}$~cm have been derived \citep[see, e.g.,][and references therein]{2016ApJ...832..136R}. Those procedures are further generalised in this paper to compute the propagation of the $e^+e^-$ through the SN ejecta (see Sec.~\ref{sec:originprFPA}), after having computed their density profiles (see Fig.~\ref{fig:model2}) and the corresponding baryon load (see Fig.~\ref{fig:model1}).  The equations have been integrated all the way up to reaching the condition of transparency (see Figs.~\ref{fig:model5} and \ref{fig:model4}).

\subsection{Episode 3: the ongoing research on the gamma-ray flares, afterglow and GeV emission}\label{sec:3.6}

We have exemplified the necessary steps in the analysis of each episode which include: to determine the physical nature of each episode, the corresponding world-line with the specific time-dependent Lorentz gamma factor and so determining, using Eq.~(\ref{tadef}), the arrival time at the detector which has to agree, for consistency, with the one obtained from the observations. This program is applied in this article specifically for the analysis of early X-ray flares (see Sec.~\ref{sec:thermalflare} and \ref{sec:thermalflare2}). We will follow the same procedures for: 1) the more complex analysis of gamma-ray flares, 2) the analysis of the afterglow consistent with the constraints on the X-ray flares observations, and 3) the properties of the GeV emission, common to BdHNe and S-GRBs \citep{2015ApJ...798...10R,2016ApJ...831..178R}. Having established the essential observational and theoretical background in Secs.~\ref{sec:obsbackground} and \ref{sec:Theory}, we proceed to the data analysis of the early X-ray flares (see Secs.~\ref{sec:sample}--\ref{sec:originprFPA}).

\section{The early flares and sample selection}
\label{sec:sample}

With the increase in the number of observed GRBs, an attempt was made to analyze the X-ray flares and other processes considered to be similar in the observer reference frame, independent of the nature of the GRB type and of the value of their cosmological redshift or the absence of such a value. This goal of this  attempt was to identify their ``standard" properties, following a statistical analysis methodology often applied in classical astronomy (see \citealp{2007ApJ...671.1903C,2007ApJ...671.1921F,2010MNRAS.406.2149M} as well as the review articles \citealp{Piran1999,2004RvMP...76.1143P,Meszaros2002,Meszaros2006,2014ARA&A..52...43B,2015PhR...561....1K}). We now summarize our alternative approach, having already given in the introduction and in Sec.~\ref{sec:obsbackground} and \ref{sec:Theory} the background for the observational identification and the theoretical interpretation of the X-ray flares.

As a first step, we only consider GRBs with an observed cosmological redshift. Having ourselves proposed the classification of all GRBs into seven different subclasses (see Sec.~\ref{sec:Theory}), we have given preliminary attention to verifying whether X-ray flares actually occur preferentially in some of these subclasses and  if so, to identify the physical reasons determining such a correlation. We have analyzed all X-ray flares and  found, a posteriori, that X-ray flares only occur in BdHNe. No X-ray flare has been identified in any other GRB subclass, either long or short. A claim of their existence in short bursts \citep{2005Natur.438..994B,2005ApJ...635L.129F,2006Sci...311.1127D} has been superseded: 
GRB 050724 with $T_{90} \sim 100$~s is not a short GRB, but actually a GRF, expected to originate in the merging of a neutron star and a white dwarf (see Fig.~\ref{fig:EpEiso}), the X-ray data for this source from XRT  is sufficient to assert that there is no evidence of an X-ray flare as defined in this section. GRB 050709 is indeed a short burst. It has been classified as S-GRF \citep{2017ApJ...844...83A}, and has been observed by HETE with very sparse X-ray data \citep{2005GCN..3570....1B},  and no presence of an X-ray flare can be inferred; the Swift satellite pointed at this source too late, $38.5$ hours after the HETE trigger \citep{2005GCN..3577....1M}.

As a second step, since all GRBs have a different redshift $z$, in order to compare them we need a description of each one of them in its own cosmological rest frame. The luminosities have to be estimated after doing the necessary K-corrections and the time coordinate in the observer frame has to be corrected by the cosmological redshift $t_a^d = (1+z) t_a$.
This also affects the determination of the $T_{90}$ of each source (see e.g. Fig.~\ref{fig:t90Histogram} in Sec.~\ref{sec:conclusions} where the traditional approach by \citet{Koveliotou1993,2013ApJ...764..179B} has been superseded by ours).

As a third step, we recall an equally important distinction from the traditional fireball approach with a single ultra relativistic jetted emission.  Our GRB analysis  envisages the existence of different episodes within each GRB, each one characterized by a different physical process and needing the definition of its own world-line and corresponding gamma factors, essential for estimating the time parametrization in the rest-frame of the observer (see Sec.~\ref{sec:obsbackground}). 

These three steps are applied in the present article, which specifically addresses the study of the early X-ray flares and their fundamental role in establishing the physical and astrophysical nature of BdHNe and in distinguishing our binary model from the traditional one.

Before proceeding let us recall the basic point of the K-correction. All the observed GRBs have a different redshift. In order to compare them it is necessary to refer each one of them to its cosmological rest frame. This step has often been neglected in the current literature \citep{2007ApJ...671.1903C,2007ApJ...671.1921F,2010MNRAS.406.2149M}. Similarly, for the flux observed by the above satellites in Sec.~\ref{sec:obsbackground}, each instrument is characterized by its fixed energy window $[\epsilon_{obs,1};\epsilon_{obs,2}]$. The observed flux $f_{obs}$, defined as the energy per unit area and time in a fixed instrumental energy window $[\epsilon_{obs,1};\epsilon_{obs,2}]$, is expressed in terms of the observed photon number spectrum $n_{obs}$ (i.e., the number of observed photons per unit energy, area and time) as
\begin{equation}
    f_{obs,[\epsilon_{obs,1};\epsilon_{obs,2}]} = \int_{\epsilon_{obs,1}}^{\epsilon_{obs,2}} \epsilon~n_{obs}(\epsilon) d \epsilon\, .
\label{eq:fluxIntegration}
\end{equation}
It then follows that the luminosity $L$ of the source (i.e., the total emitted energy per unit time in a given bandwidth), expressed by definition in the source cosmological rest frame, is related to $f_{obs}$ through the luminosity distance $D_L(z)$:
\begin{equation}
    L_{[\epsilon_{obs,1}(1+z);\epsilon_{obs,2}(1+z)]} = 4 \pi D_L^2(z) f_{obs,[\epsilon_{obs,1};\epsilon_{obs,2}]}\, .
\label{eq:CorrectedLuminosity}
\end{equation}
The above Eq.(\ref{eq:CorrectedLuminosity}) gives the luminosities in different cosmological rest frame energy bands, depending on the source redshift. To express the luminosity $L$ in a fixed cosmological rest frame energy band, e.g., $[E_1;E_2]$, common to all sources, we can rewrite Eq.(\ref{eq:CorrectedLuminosity}) as:
\begin{equation}
   L_{[E_1;E_2]} = 4 \pi D_L^2 f_{obs,\left[\frac{E_1}{1+z};\frac{E_2}{1+z}\right]} = 4 \pi D_L^2 k[\epsilon_{obs,1};\epsilon_{obs,2};E_1;E_2;z] f_{obs,[\epsilon_{obs,1};\epsilon_{obs,2}]}\, ,
\label{eq:kCorrectedLuminosity}
\end{equation}
where we have defined the K-correction factor:
\begin{equation}
   k[\epsilon_{obs,1};\epsilon_{obs,2};E_1;E_2;z] = \frac{f_{obs,\left[\frac{E_1}{1+z};\frac{E_2}{1+z}\right]}}{f_{obs,[\epsilon_{obs,1};\epsilon_{obs,2}]}} = \frac{\int_{E_1/(1+z)}^{E_2/(1+z)} \epsilon~n_{obs}(\epsilon) d \epsilon}{\int_{\epsilon_{obs,1}}^{\epsilon_{obs,2}} \epsilon~n_{obs}(\epsilon) d \epsilon}\, .
\label{eq:kCorrectionFactor}
\end{equation}

If the energy range $[\frac{E_1}{1+z};\frac{E_2}{1+z}]$ is not fully inside the instrumental energy band $[\epsilon_{obs,1};\epsilon_{obs,2}]$, It may well happen that we need to extrapolate $n_{obs}$ within the integration boundaries $[\frac{E_1}{1+z};\frac{E_2}{1+z}]$.

Finally we express each luminosity in a rest frame energy band which coincides with the energy window of each specific instrument.

\begin{figure}
\centering
\includegraphics[width=0.7\hsize,clip]{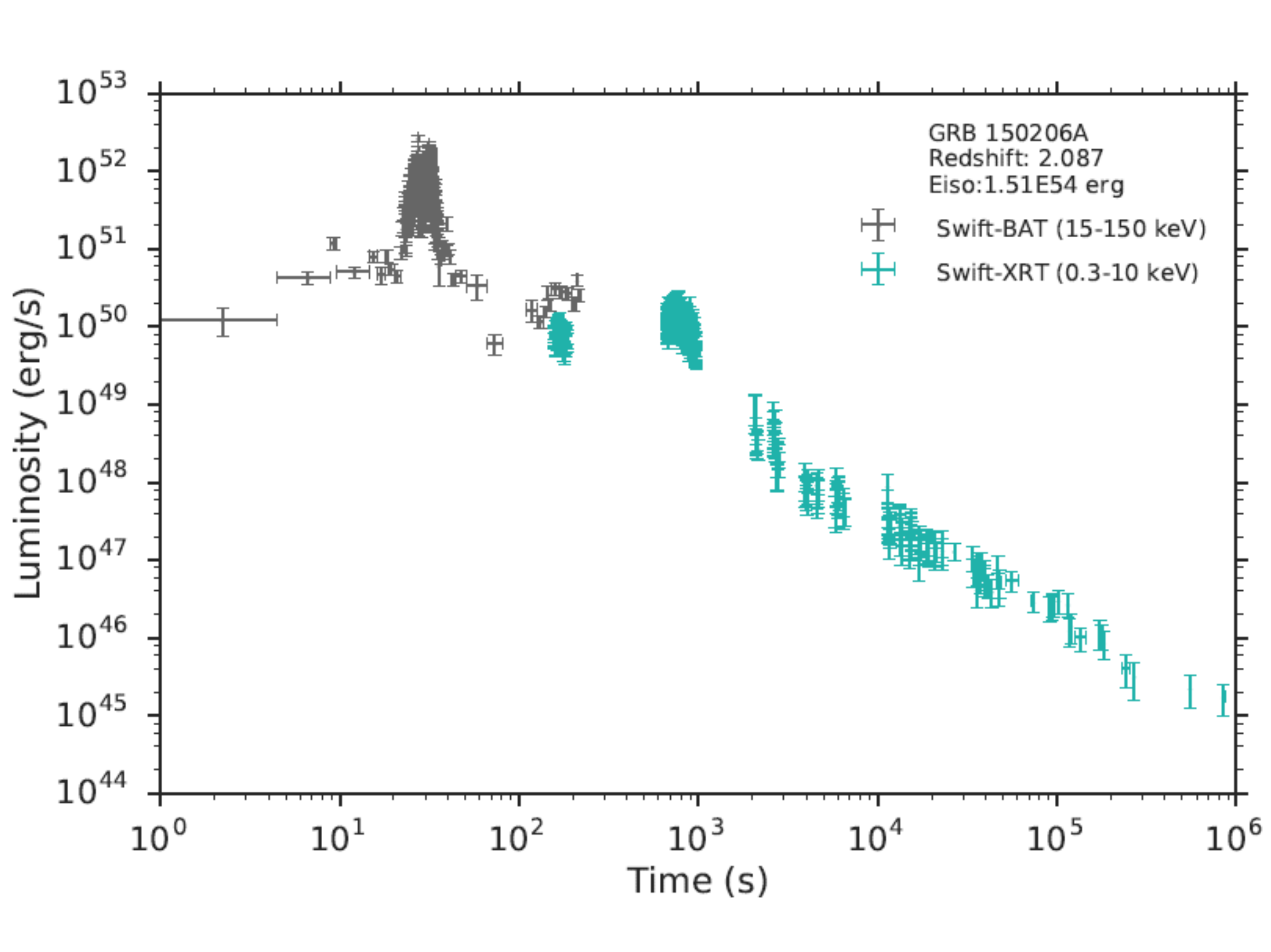}
\caption{GRB 150206A is an example of a GRB with incomplete data, which therefore must be excluded. It only has $30$~s Swift-\emph{XRT} observations in the early $300$~s. The flare determination is not possible under these conditions.}
\label{fig:dataflare}
\end{figure}

\begin{figure}
\centering
\includegraphics[width=0.7\hsize,clip]{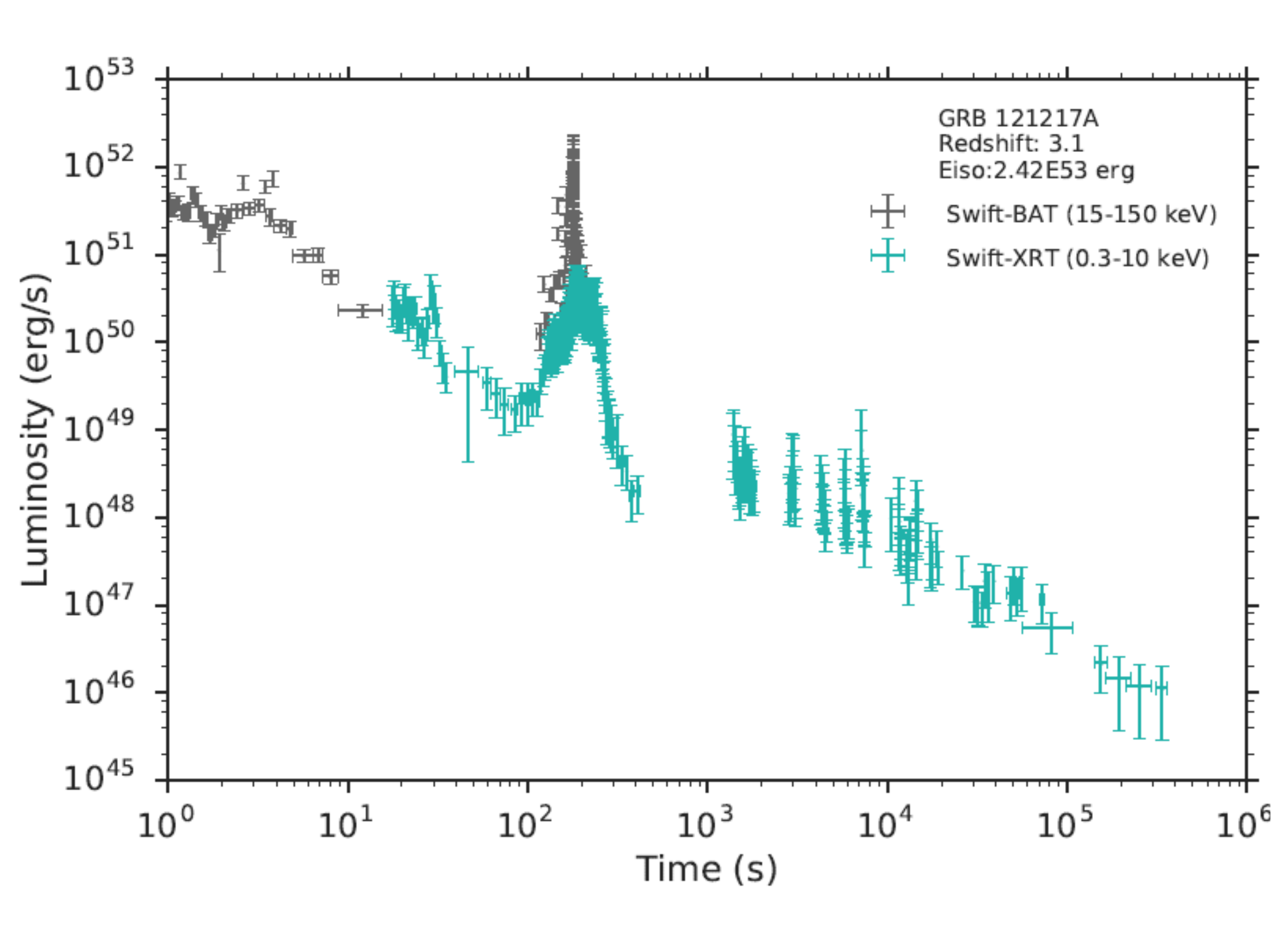}
\caption{GRB 121217A clearly shows a gamma-ray flare observed by Swift-BAT which coincides with a soft X-ray component observed by Swift-XRT. From the spectral analysis, it has a soft power-law photon index, and most of the energy deposited in high energy gamma-rays. This is an indication that the soft X-ray component is likely the low energy part of a gamma-ray flare. For these reasons, we neglect it in our sample.}
\label{fig:spikeflare}
\end{figure}

\begin{figure}
\centering
\includegraphics[width=0.7\hsize,clip]{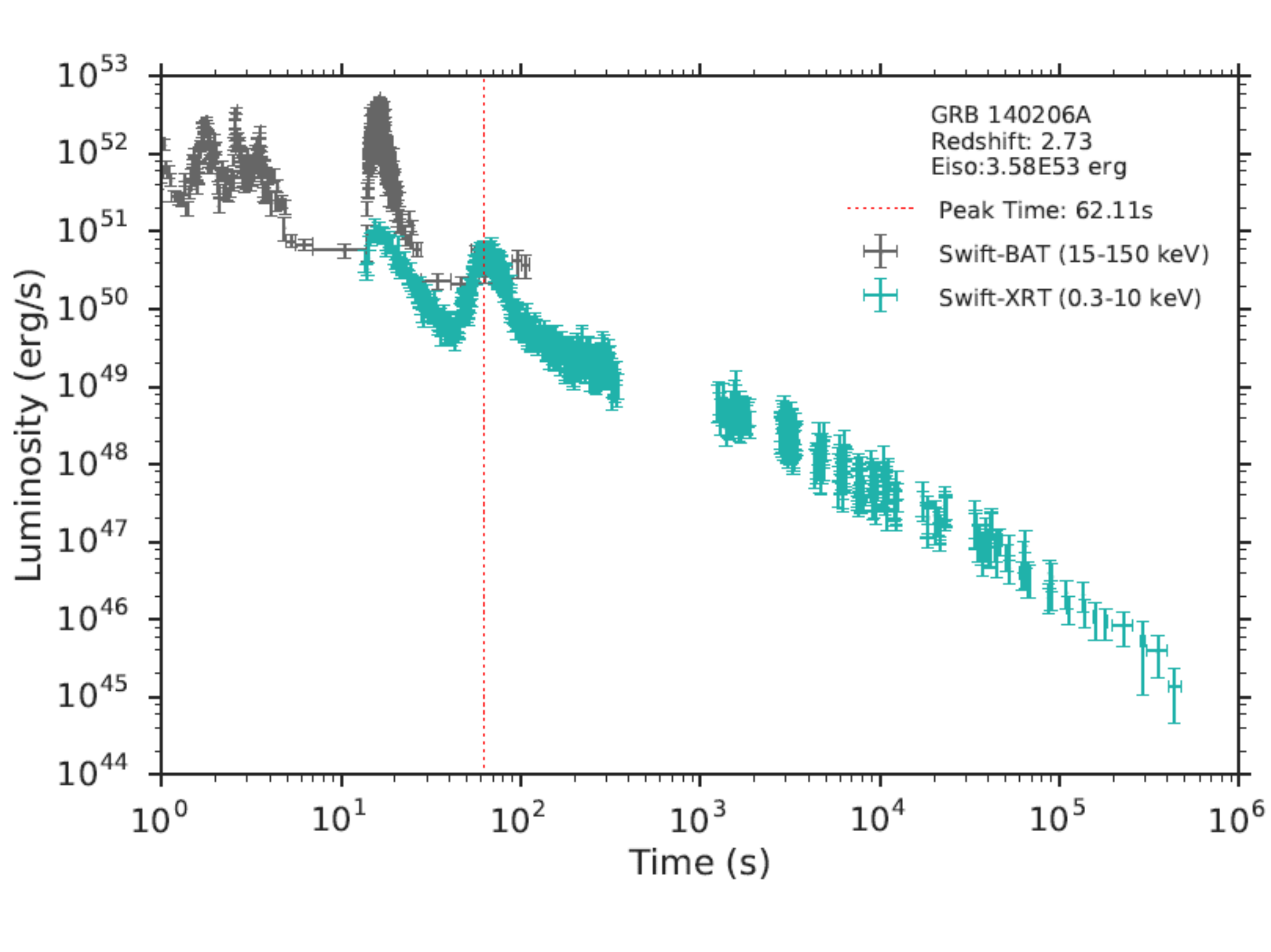}
\caption{GRB 140206A has two flares. A gamma-ray flare coincides with the first flare while it is dim in the second one. The spectral analysis, using both Swift-XRT and Swift-BAT data, indicates a power-law index $-0.88 \pm 0.03$ for the first flare. While the second flare requires an additional blackbody component; its power-law index is $-1.73 \pm 0.06$ and its blackbody temperature is $0.54 \pm 0.07$~keV. Clearly, the energy of the first flare is contributed mainly by gamma-ray photons---it is a gamma-ray flare, and the second flare is an X-ray flare that we consider in this article.}
\label{fig:twoflare}
\end{figure}

\begin{figure}
\centering
\includegraphics[width=0.8\textwidth]{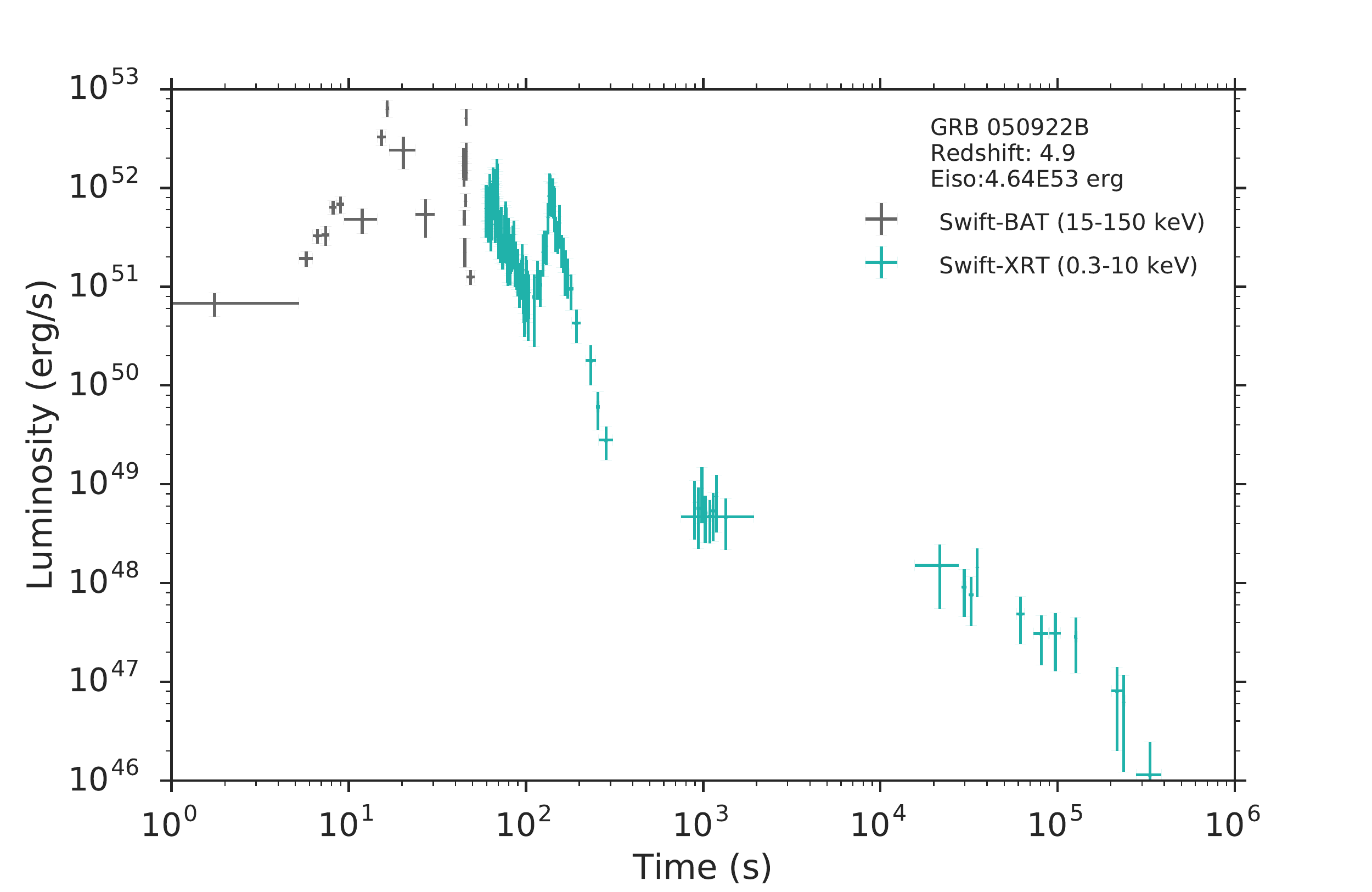}
\caption{The Swift-BAT data of GRB 050922B has poor resolution, it cannot provide valid information after $50$~s; the energy observed in its energy band $15$--$150$ keV during this $50$~s duration is $1.19 \times 10^{53}$~erg. The energy observed by Swift-XRT is higher; the energy of the flares ($60-200$~s) in the Swift-XRT band $0.3$--$10$ keV is $3.90 \times 10^{53}$~erg. These results imply that the Swift-BAT observations may not cover the entire prompt emission phase; the isotropic energy computed from the Swift-BAT data is not reliable, and consequently the Swift-XRT observed partial prompt emission which brings complexity to the X-ray light-curve makes the identification of the authentic X-ray flare more difficult.}
\label{fig:GRB050922B}
\end{figure}

We turn now to the selection procedure for the early X-ray flares. We take the soft X-ray flux light curves of each source with known redshift from the Swift-\emph{XRT} repository \citep{2007A&A...469..379E,2009MNRAS.397.1177E}. We then apply the above K-correction to obtain the corresponding luminosity light curves in the rest frame $0.3$--$10$ keV energy band. Starting from $421$ Swift-XRT light curves, we found in $50$ sources X-ray flare structures in the early $200$~s. Remarkably, all of them are in BdHNe. We further filter our sample by applying the following criteria:
\begin{enumerate}
\item
We exclude GRBs with flares having low ($<20$) signal to noise ratio (SNR), or with an incomplete data coverage of the early X-ray light curve --- 14 GRBs are excluded (see e.g., Fig.~\ref{fig:dataflare}).

\item
We consider only X-ray flares and do not address here the gamma-ray flares which will be studied in a forthcoming article --- 8 GRBs having only gamma-ray flares are temporarily excluded (see e.g., Fig.~\ref{fig:spikeflare}). In Fig.~\ref{fig:twoflare} we show an illustrative example of the possible co-existence of a X-ray flare and a gamma-ray flare, and a way to distinguish them. 

\item
We also neglect here the late X-ray flare, including the ultra-long GRB, which will be discussed in a forthcoming paper --- 6 GRBs are consequently excluded.  

\item 
We neglect the GRBs for which the soft X-ray energy observed by Swift-XRT ($0.3 -- 10$~keV) before the plateau phase is higher than the gamma-ray energy observed by Swift-BAT ($15 -- 150$~keV) during the entire valid Swift-BAT observation. This Swift-BAT anomaly points to an incomplete coverage of the prompt emission -- 6 GRBs are excluded (see e.g., Fig.~\ref{fig:GRB050922B}).

\end{enumerate}

Finally, we have found $16$ BdHNe satisfying all the criteria to be included in our sample. Among them, $7$ BdHNe show a single flare. The other $9$ BdHNe contain two flares: generally we exclude the first one, which appears to be a component from the gamma-ray spike or gamma-ray flare, and therefore select  the second one for analysis (see, e.g., Fig.~\ref{fig:twoflare}).

These $16$ selected BdHNe cover a wide range of redshifts. The closest one is GRB 070318 with redshift $z=0.84$, and the farthest one is GRB 090516A with redshift $z=4.11$. Their isotropic energy is also distributed over a large range: 5 GRBs have energies of the order of $10^{52}$~erg, 9 GRBs of the order of $10^{53}$~erg, and 2 GRBs have extremely high isotropic energy $E_{iso} > 10^{54}$~erg. Therefore, this sample is well-constructed although the total number is limited.

\section{The XRT luminosity light curves of the 16 BdHN sample}
\label{sec:XRTlum}

\begin{figure*}
\centering
\includegraphics[width=0.7\hsize,clip]{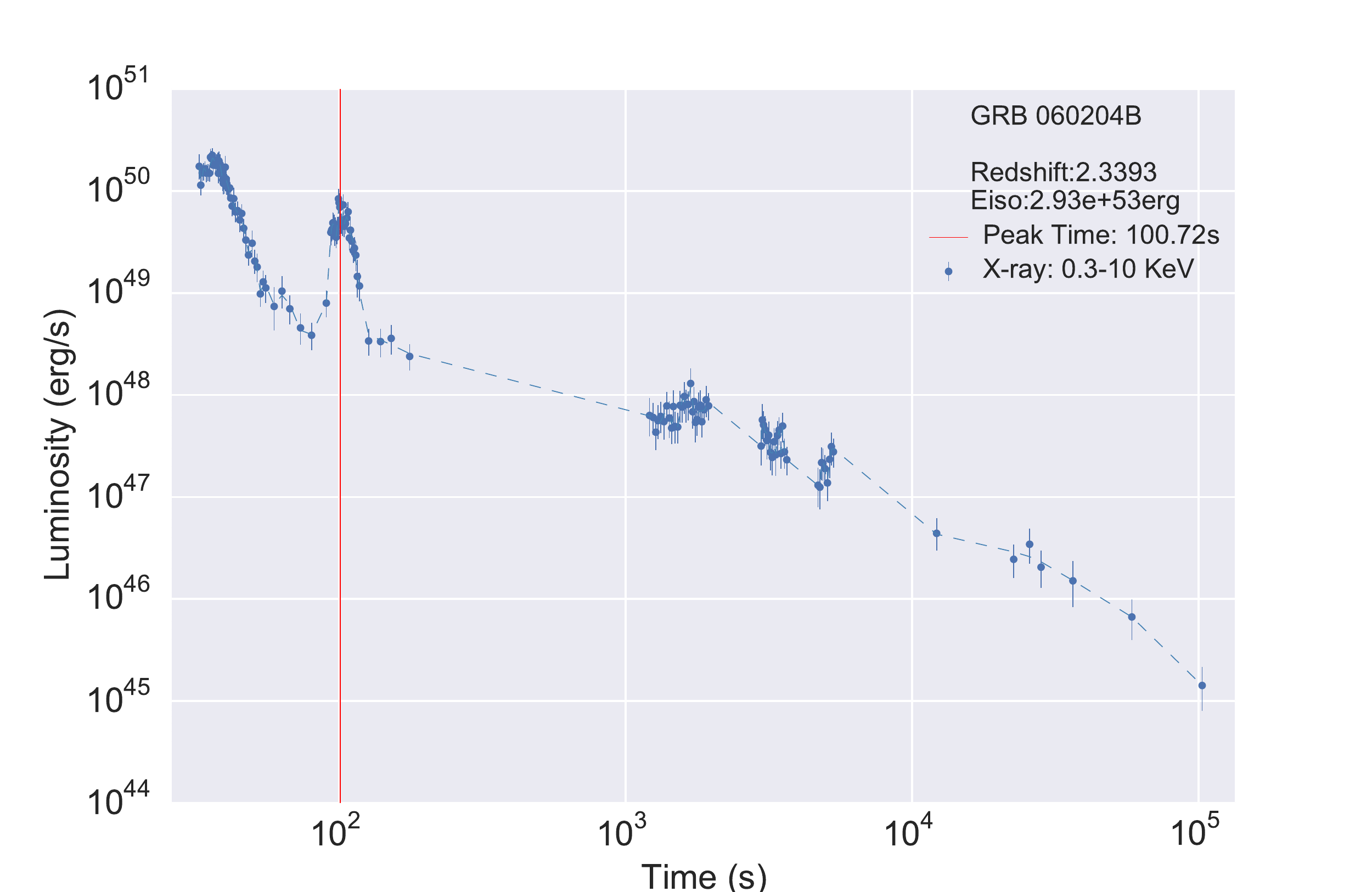}
\caption{\textbf{060204B}: this GRB was triggered by Swift-BAT \citep{2006GCN..4655....1F}; Swift-XRT began observing $28.29$ sec after the BAT trigger. There is no observation from the Fermi satellite. X-shooter found its redshift at $2.3393$ based on the host galaxy \citep{2016ApJ...817....7P}. The isotropic energy of this GRB reaches $2.93 \times 10^{53}$~erg computed from Swift-BAT data.}
\label{Samplein}
\includegraphics[width=0.7\hsize,clip]{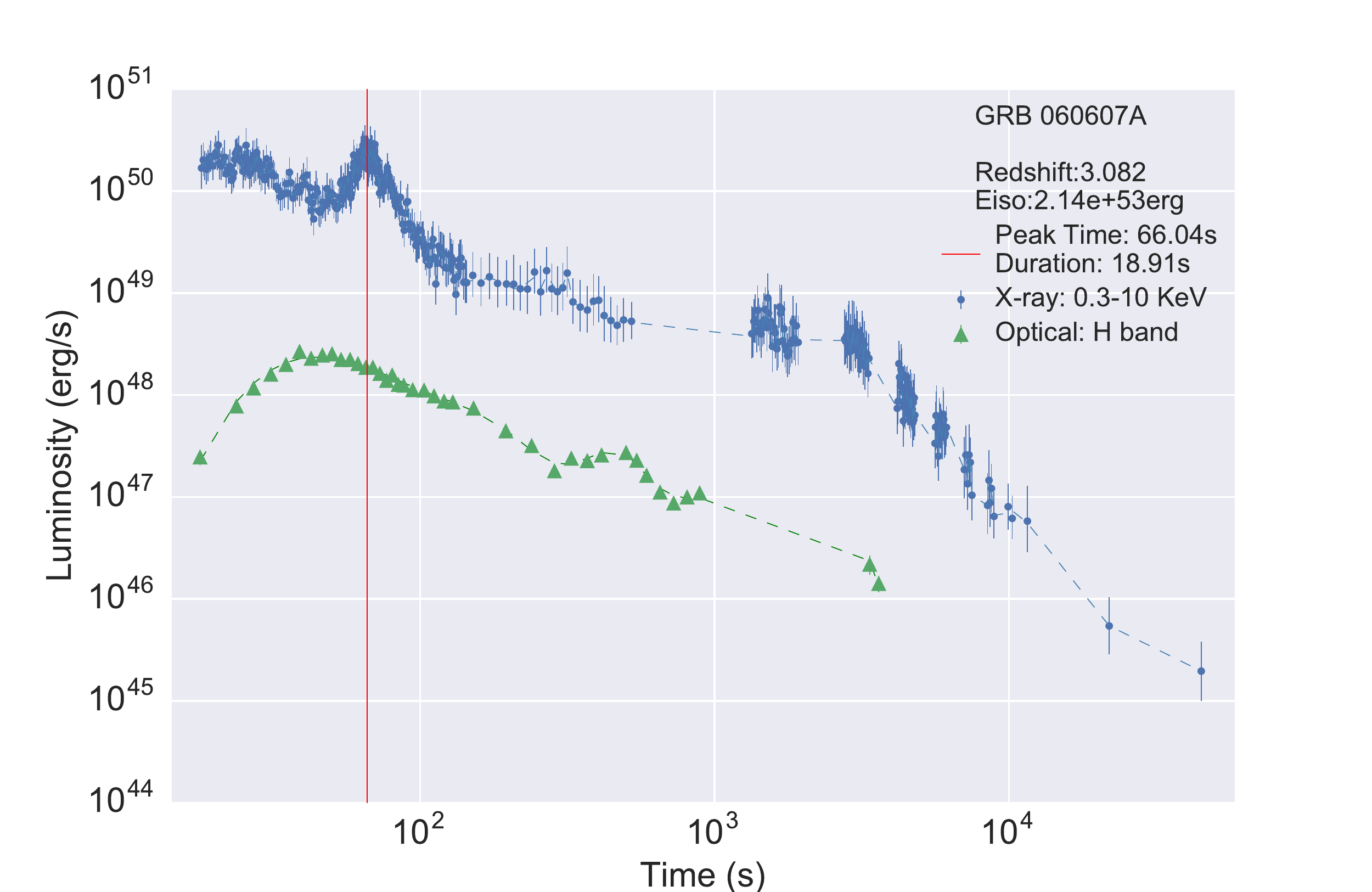}
\caption{\textbf{060607A}: this source has been detected by the {Swift} satellite \citep{2006GCN..5233....1Z}. It has a bright optical counterpart \citep{2006GCN..5233....1Z}. It is located at a redshift $z=3.082$ \citep{2006GCN..5237....1L}. The prompt light curve presents a doubled-peaked emission that lasts around $10$ s, plus a second emission at $\sim 25$~s of $2.5$~s duration. The isotropic energy is $E_{iso}=2.14 \times 10^{53}$~erg. Optical data is from \citet{2009ApJ...693.1417N}}
\end{figure*}

We now turn to the light curves of each one of the 16 GRBs composing our sample (see Figs.~\ref{Samplein}--\ref{Sample}). Blue curves represent the X-rays observed by {Swift}-XRT, and green curves are the corresponding optical observations when available. All the values are in the rest-frame and the X-ray luminosities have been K-corrected. The red vertical lines indicate the peak time of the X-ray flares. The rest-frame luminosity light curves of some GRBs show different flare structures compared to the observed count flux light-curves. An obvious example is GRB 090516A, as follows by comparing Fig.~\ref{fig:090516A} in this paper with Fig.~1 in \citet{2015ApJ...803...10T}. The details of the FPA, as well as their correlations or absence of correlation with $E_{iso}$, are given in the next section.

\begin{figure*}
\centering
\includegraphics[width=0.7\hsize,clip]{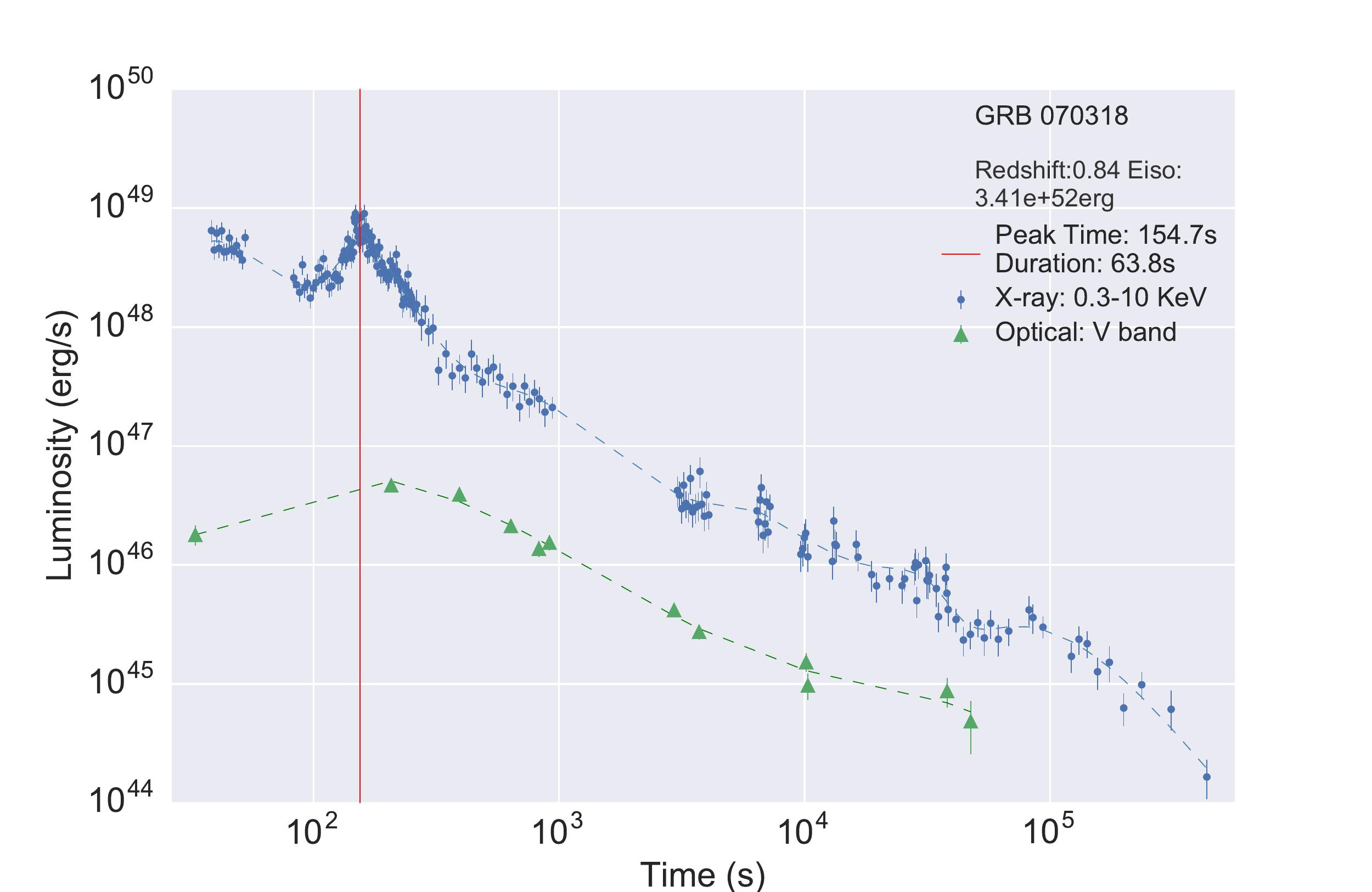}
\caption{\textbf{070318}: this source has been detected by the {Swift} satellite \citep{2007GCN..6210....1C}. It has a spectroscopic redshift of $z=0.836$ \citep{2007GCN..6216....1J}. The prompt light curve shows a peak with a typical fast-rise exponential-decay (FRED) behavior lasting about $55$ s. XRT began observing the field $35$ s after the BAT trigger. The isotropic energy is $E_{iso}= 3.64 \times 10^{52}$~erg. From the optical observation at $\sim$ 20 days, no source or host galaxy is detected at the position of the optical afterglow, indicating that the decay rate of the afterglow must have steepened after some hours \citep{2007GCN..6296....1C}. Its optical data is from \citet{2008AIPC.1000..421C}}
\includegraphics[width=0.7\hsize,clip]{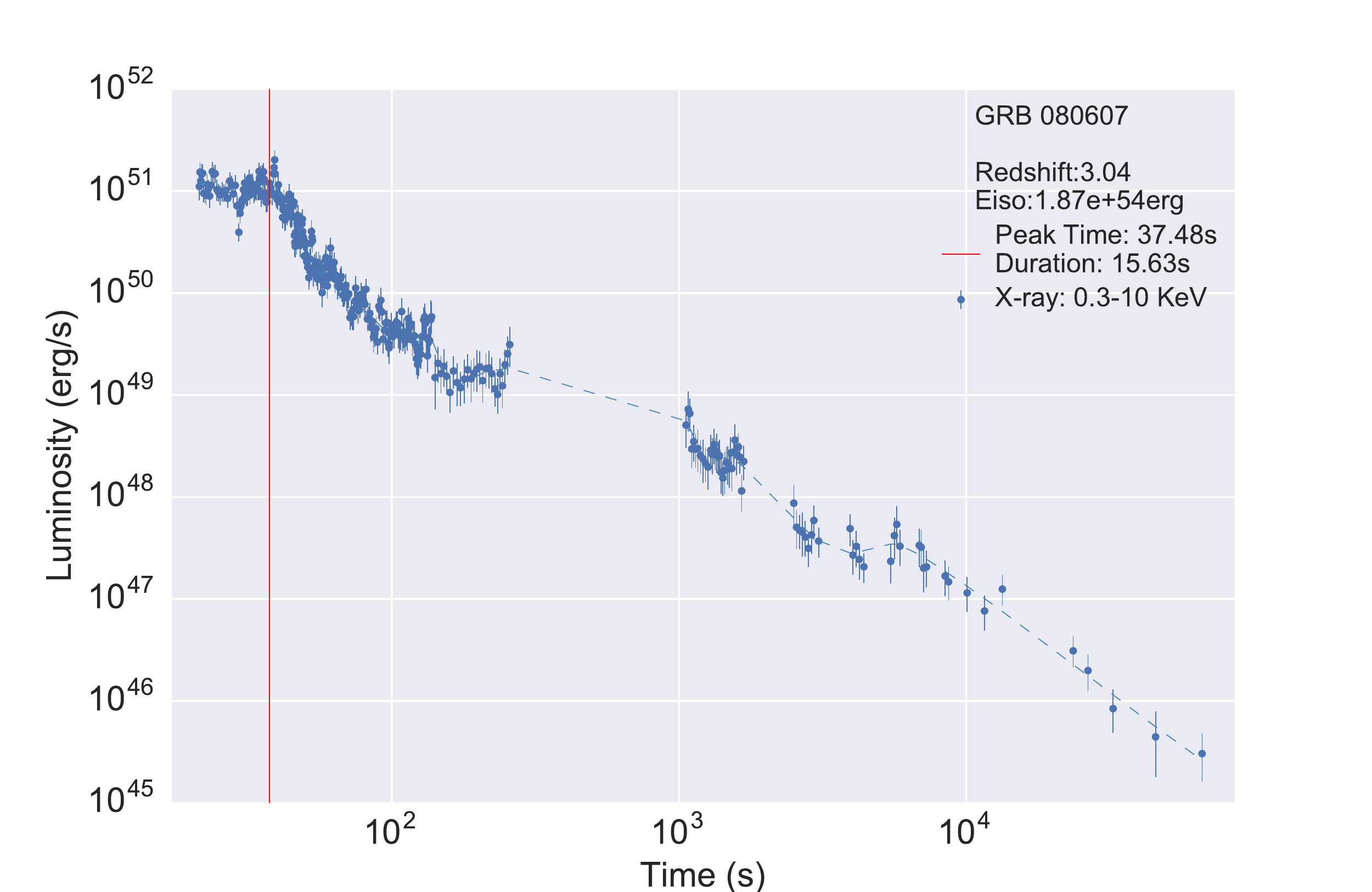}
\caption{\textbf{080607}: this source has been observed by AGILE \citep{2008GCN..7866....1M}, Konus-Wind \citep{2008GCN..7862....1G} and \textit{Swift} \citep{2008GCN..7847....1M}. UVOT detected only a faint afterglow, since the source is located at a redshift $z=3.04$. The isotropic energy is $E_{iso}=1.87 \times 10^{54}$~erg. The BAT prompt light curve shows a very pronounced peak that lasts $\sim10$ s, followed by several shallow peaks until $25$~s. The {Swift} localization is at about $113^\mathrm{o}$ off-axis with respect to the AGILE pointing, so well out of the field of view of the AGILE gamma-ray imaging detector (GRID), which does not show any detection. The Konus-Wind light curve in the $50$--$200$~keV range shows a multiple peak emission lasting $15$~s.}
\end{figure*}

\begin{figure*}
\centering
\includegraphics[width=0.7\hsize,clip]{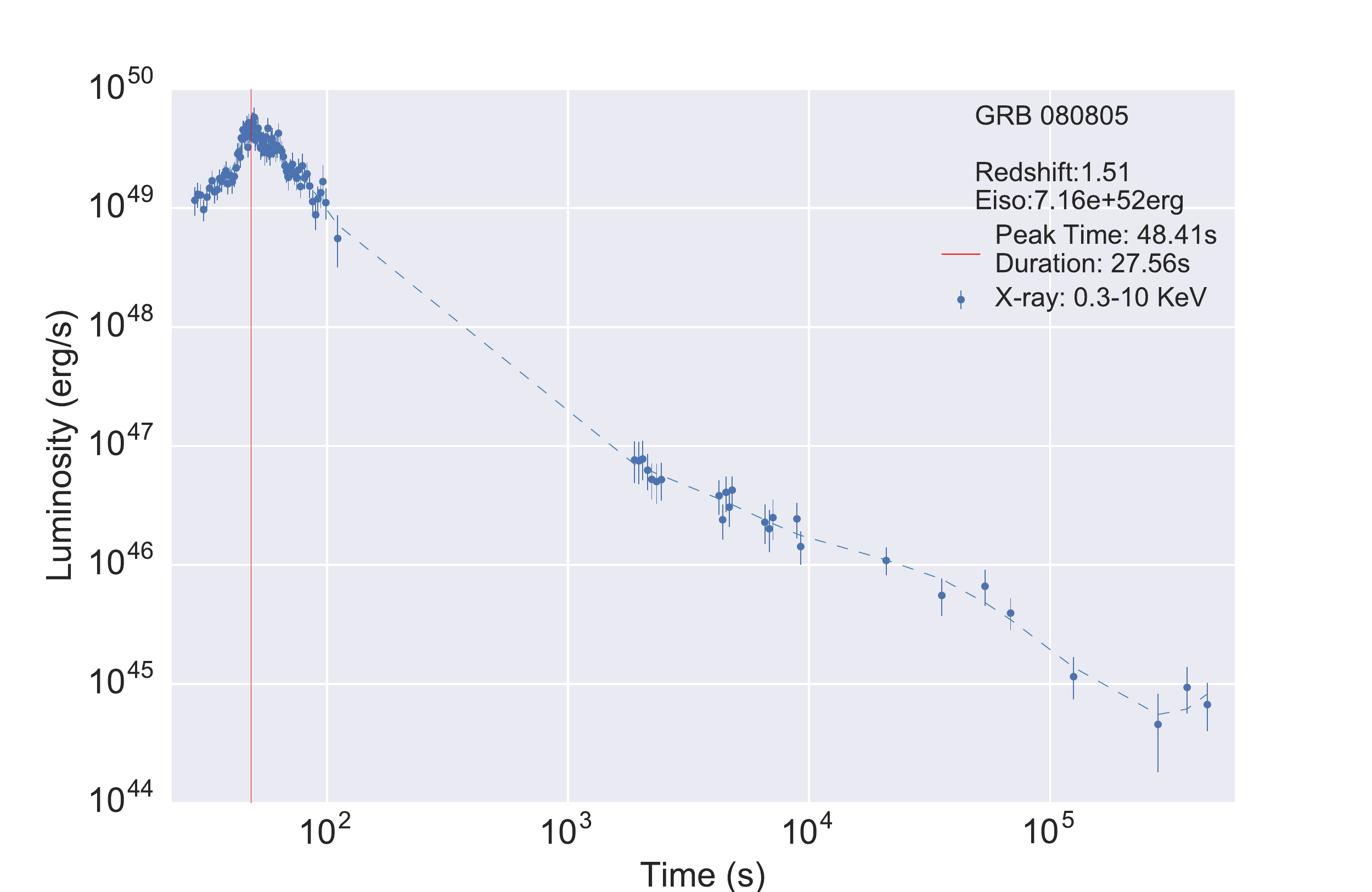}
\caption{\textbf{080805}: this source was detected by \textit{Swift} \citep{2008GCN..8059....1P}. The prompt light curve shows a peak with a FRED behavior lasting about $32$~s. The redshift is $z=1.51$, as reported by VLT \citep{2008GCN..8077....1J}, and the isotropic energy is $E_{iso}=7.16 \times 10^{52}$~erg.}
\includegraphics[width=0.7\hsize,clip]{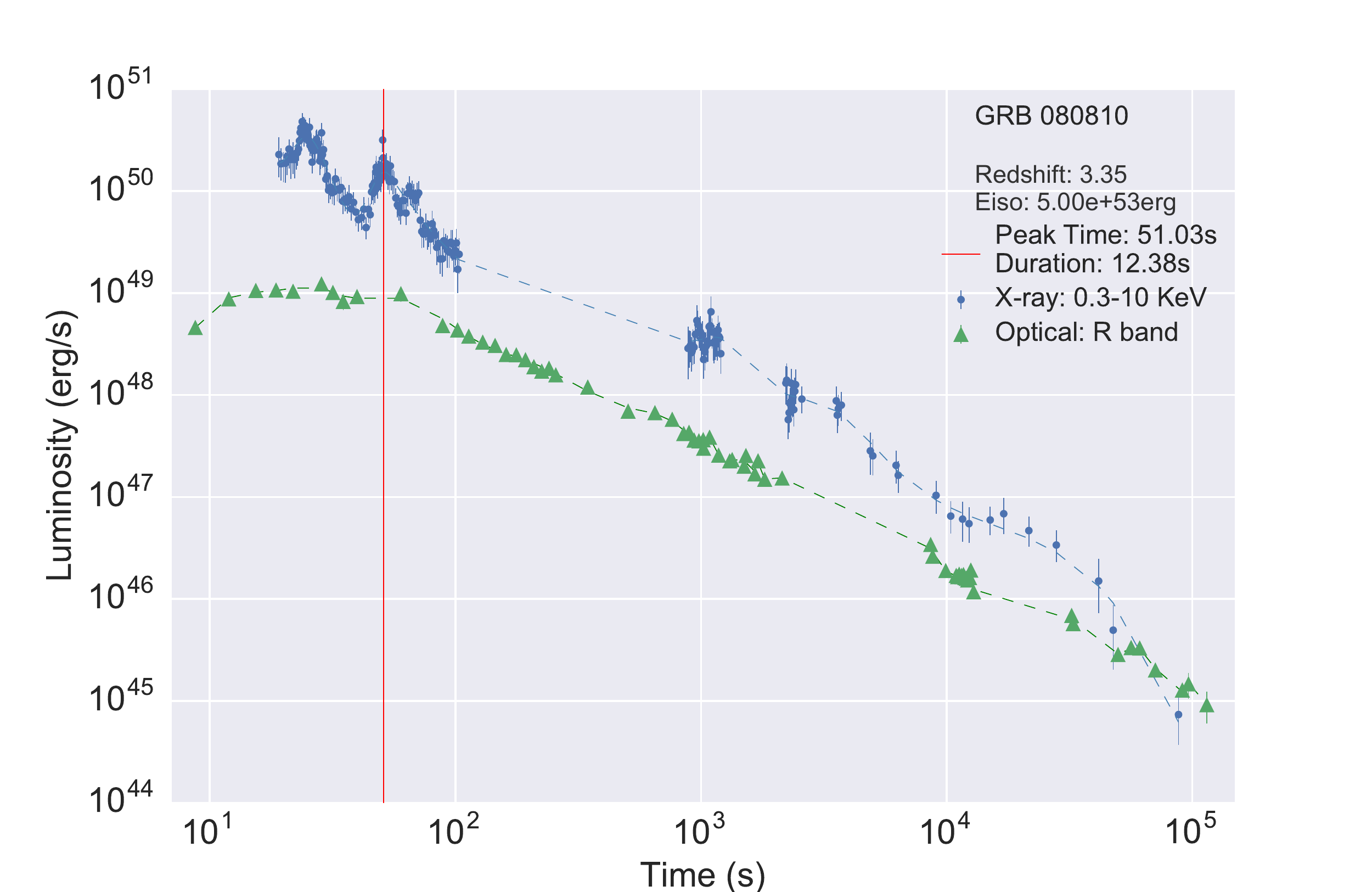}
\caption{\textbf{080810}: this source was detected by \textit{Swift} \citep{2008GCN..7862....1G}. The BAT light curve shows a multiple-peaked structure lasting about $23$~s. XRT began observing the field $76$~s after the BAT trigger. The source is located at a redshift of $z=3.35$ and has an isotropic energy $E_{iso}=3.55 \times 10^{53}$~erg. Optical date is taken from \citet{2009MNRAS.400..134P}.}
\end{figure*}

\begin{figure*}
\centering
\includegraphics[width=0.7\hsize,clip]{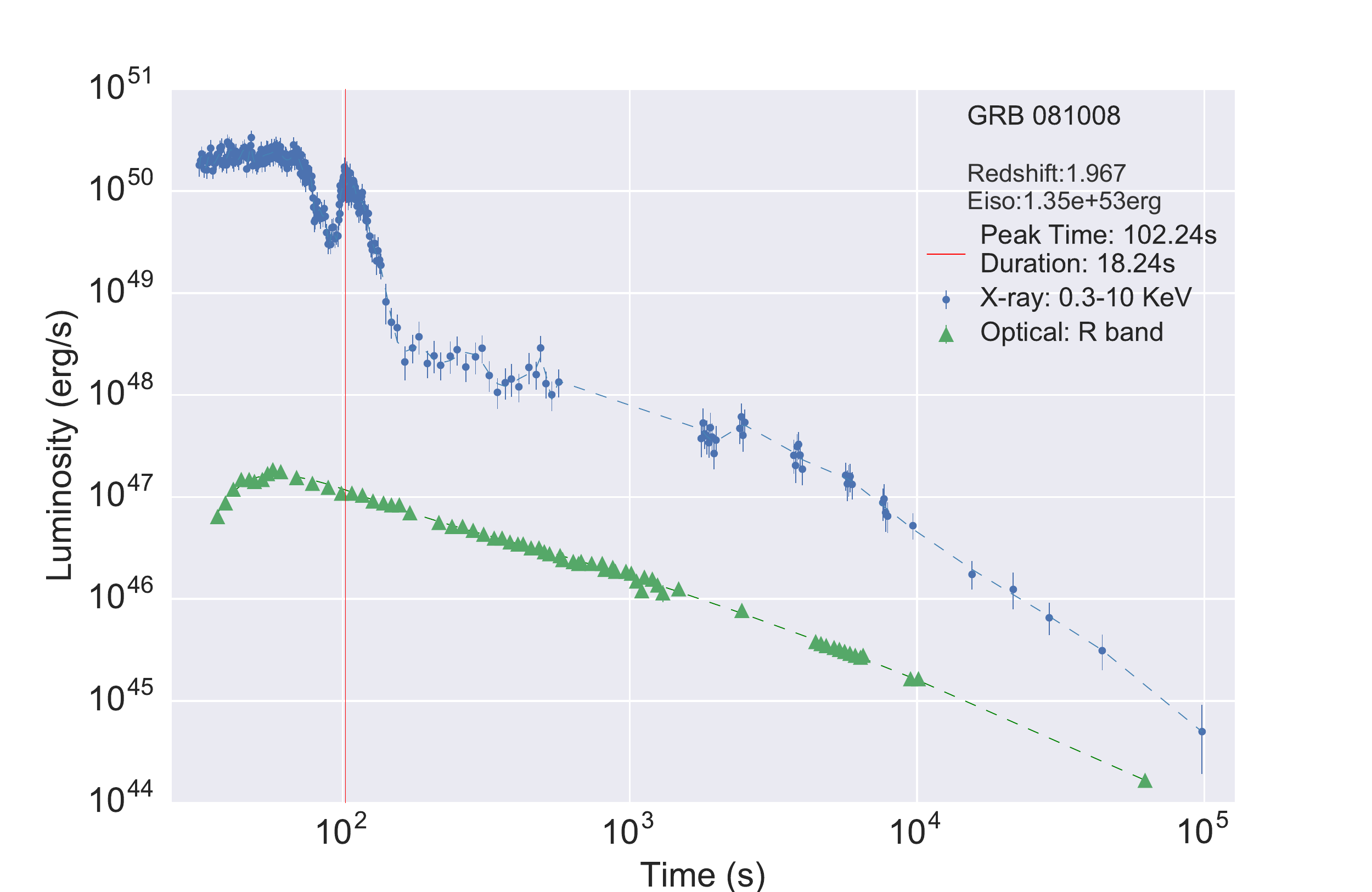}
\caption{\textbf{081008}: this source was detected by \textit{Swift} \citep{2008GCN..8344....1R}. The prompt emission lasts about $60$~s and shows two peaks separated by $13$~s. It is located at $z=1.967$, as reported by VLT \citep{2008GCN..8350....1D}, and has an isotropic energy $E_{iso}=1.07 \times 10^{53}$~erg. Optical data is from \citet{2010ApJ...711..870Y}.}
\includegraphics[width=0.7\hsize,clip]{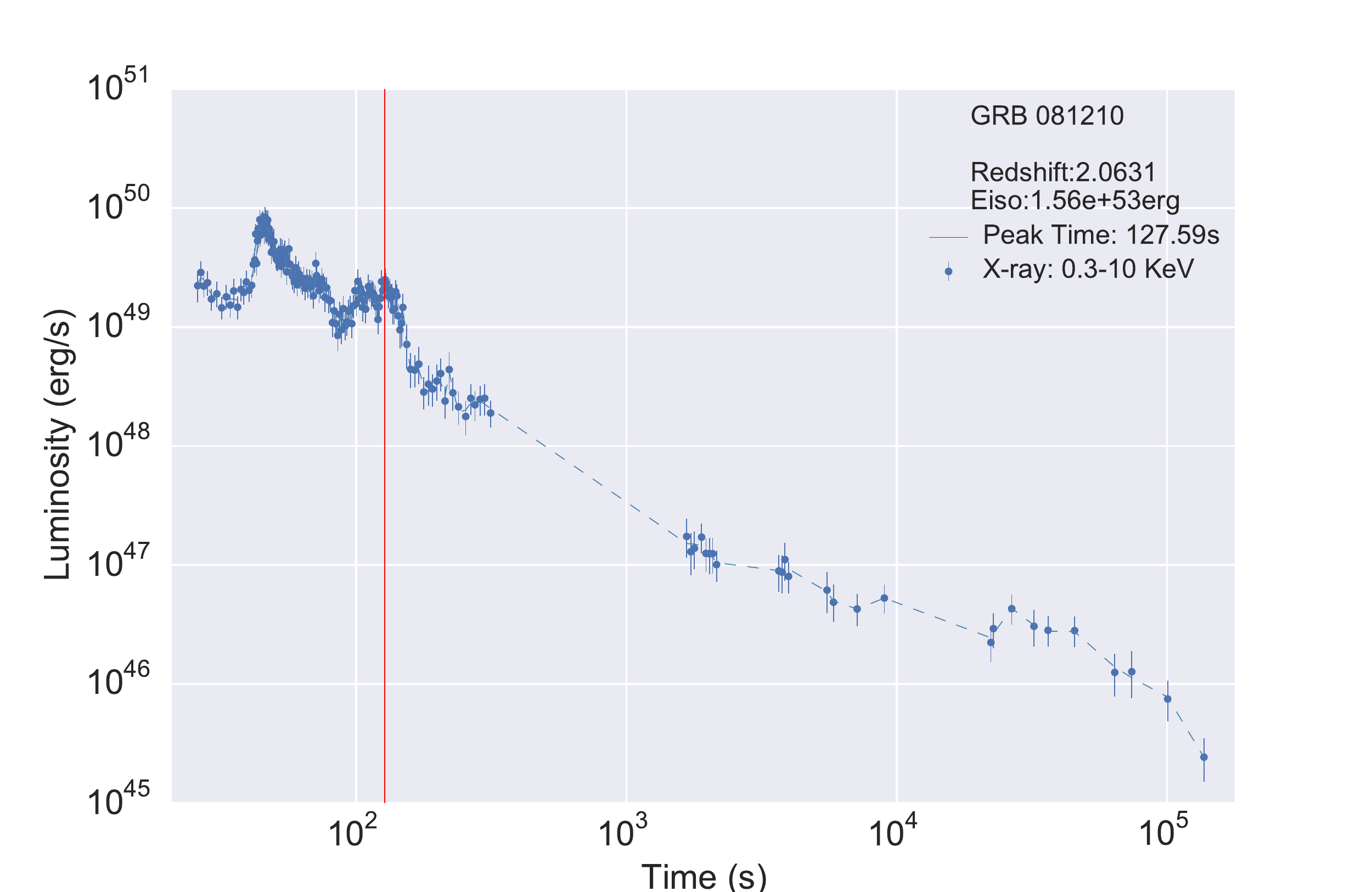}
\caption{\textbf{081210}: this GRB was detected by Swift-BAT \citep{2008GCN..8648....1K}, Swift-XRT began observing at $23.49$~s after the BAT trigger. The BAT light curve begins with two spikes with a total duration of about $10$~s, and an additional spike at $45.75$~s. There is no observation from the Fermi satellite. X-shooter found its redshift to be $2.0631$ \citep{2016ApJ...817....7P}. The isotropic energy of this GRB is $1.56 \times 10^53$~erg.}
\end{figure*}

\begin{figure*}
\centering
\includegraphics[width=0.7\hsize,clip]{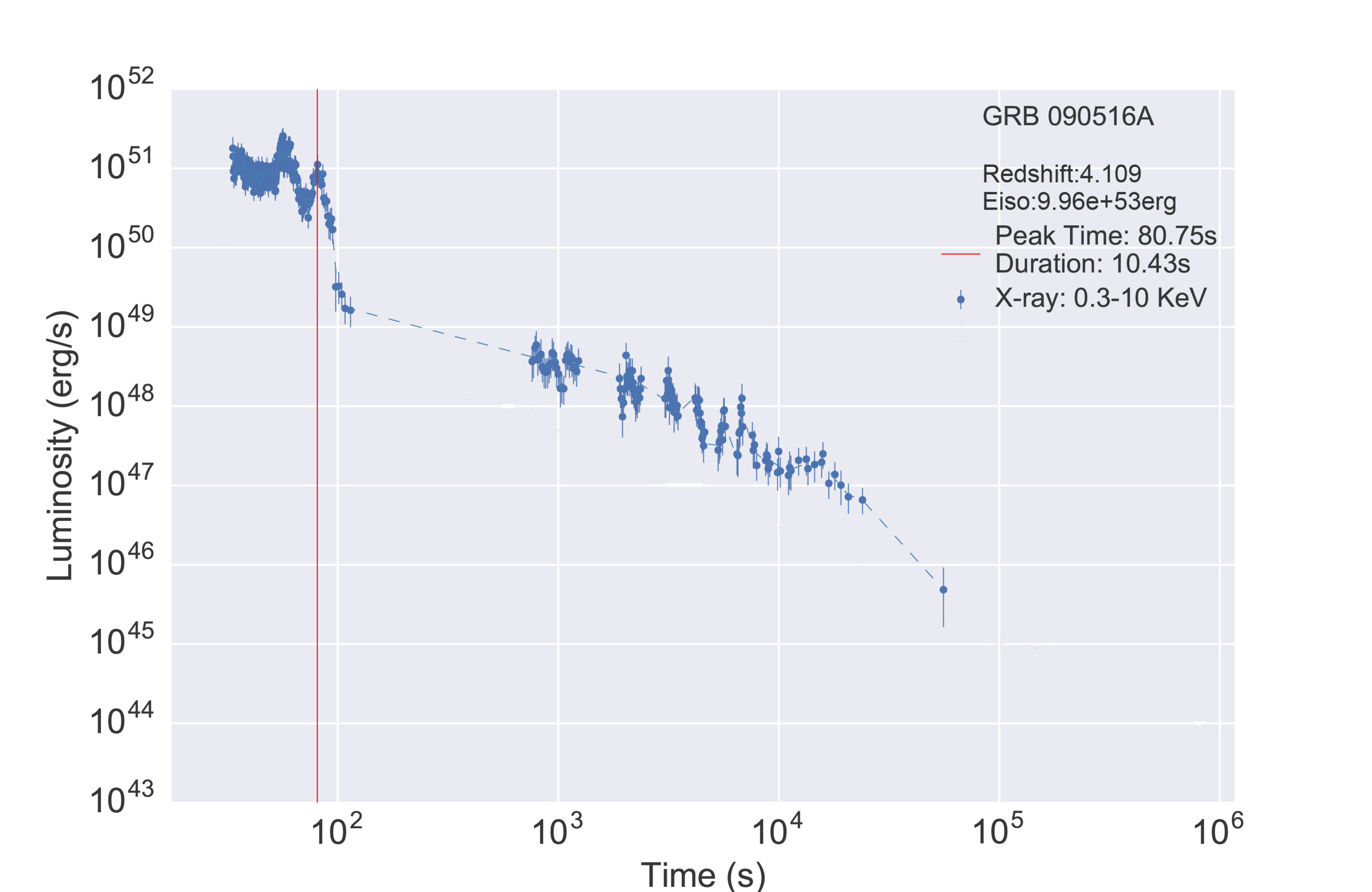}
\caption{\textbf{090516A}: this source was detected by {Swift} \citep{2009GCN..9374....1R}, Konus Wind and {Fermi}/GBM \citep{2009GCN..9415....1M}. The BAT prompt light curve is composed of two episodes, the first starting $2$~s before the trigger and lasting up to $10$~s after the trigger, while the second episode starts at $17$ s and lasts approximately $2$~s. The GBM light curve consists of about five overlapping pulses from $T_\mathrm{F,0}-10$~s to $T_\mathrm{F,0}+21$~s (where $T_\mathrm{F,0}$ is the trigger time of the \textit{Fermi}/GBM). Konus-Wind observed this GRB in the waiting mode. VLT identified the redshift of the afterglow as $z=4.109$ \citep{2012A&A...548A..11D}, in agreement with the photometric redshift obtained with GROND \citep{2009GCN..9382....1R}. Fermi-LAT was inside of the field of view, following the standard Fermi-LAT likelihood analysis in \href{https://fermi.gsfc.nasa.gov/ssc/data/analysis/scitools/likelihood\_tutorial.html}{https://fermi.gsfc.nasa.gov/ssc/data/analysis/scitools/likelihood\_tutorial.html}, the upper limit of observed count flux is $4.76 \times 10^{-6}$~photons~cm$^{-2}$~s$^{-1}$, no GeV photon was found for this high redshift and low observed fluence GRB. The isotropic energy is $E_{iso}=6.5\times 10^{53}$~erg.}
\includegraphics[width=0.7\hsize,clip]{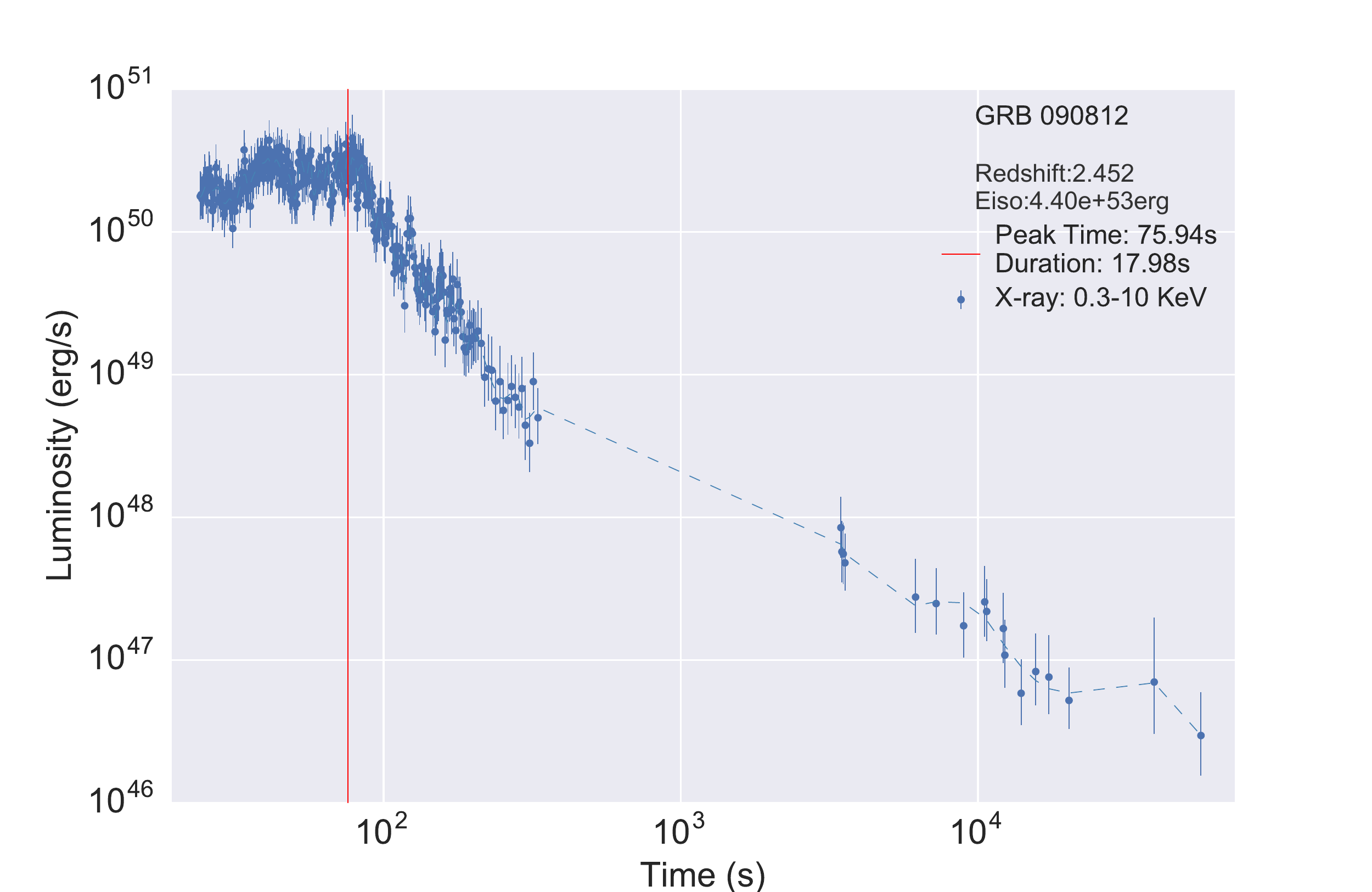}
\caption{\textbf{090812}: this source was detected by \textit{Swift} \citep{2009GCN..9768....1S}. It has a redshift $z=2.452$ as confirmed by VLT \citep{2012A&A...548A..11D} and an isotropic energy $E_{iso}=4.75\times 10^{53}$~erg. The BAT light curve shows three successive bumps lasting $\sim 20$~s in total. XRT began observing the field $22$~s after the BAT trigger \citep{2009GCN..9768....1S}. The BAT light curve shows a simple power-law behavior.}
\label{fig:090516A}
\end{figure*}

\begin{figure*}
\centering
\includegraphics[width=0.7\hsize,clip]{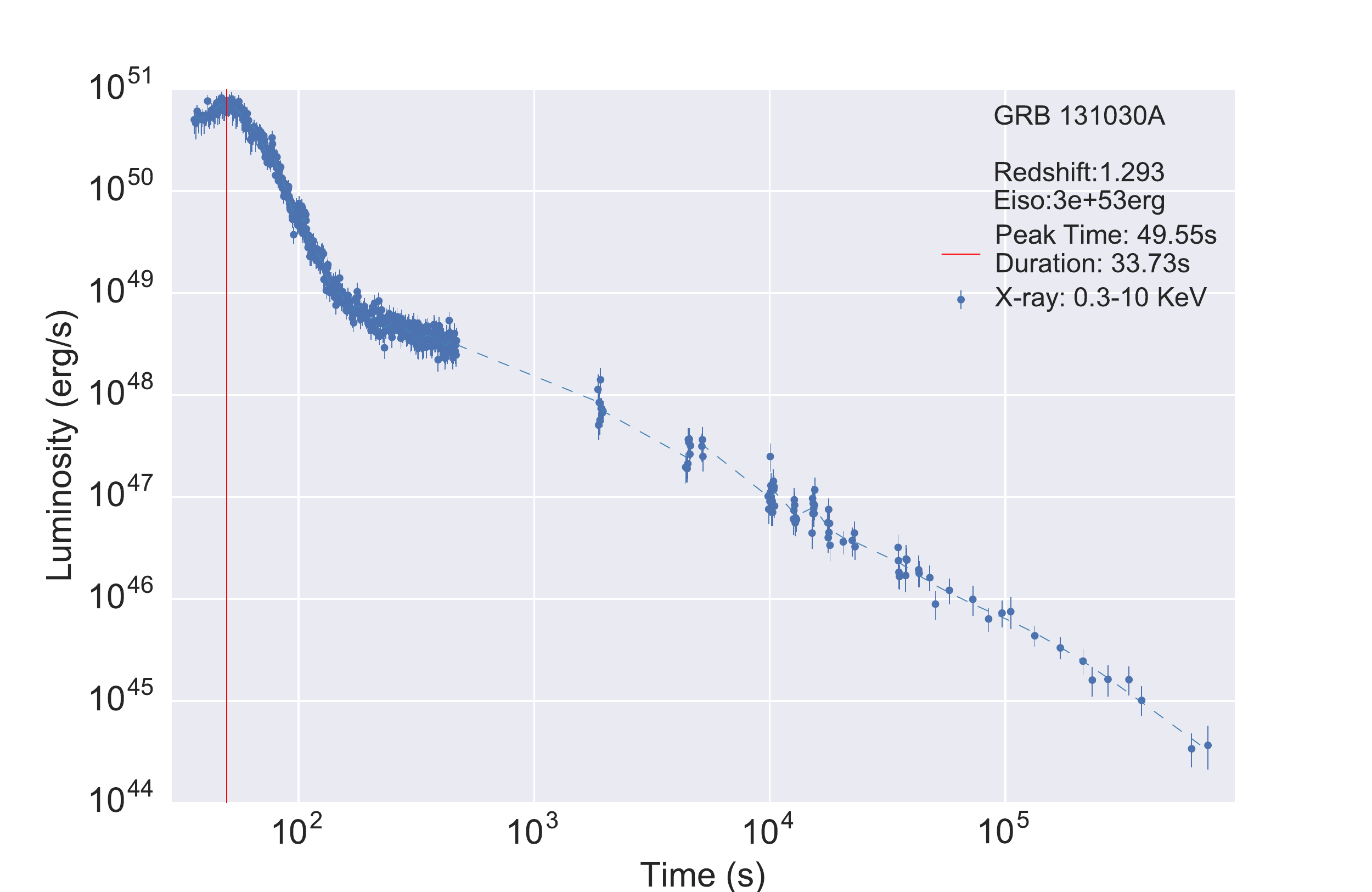}
\caption{\textbf{131030A}: this source was observed by \textit{Swift} \citep{2013GCN..15402...1T} and Konus-Wind \citep{2013GCN..15413...1G}. The BAT light curve shows two overlapping peaks starting, with respect to the {Swift}-BAT trigger $T_\mathrm{B,0}$, at $\sim T_\mathrm{B,0}-3.5$~s and peaking at $\sim T_\mathrm{B,0}+4.4$~s \citep{2013GCN..15456...1B}. The duration is $18$~s in the $15$--$350$~keV band. The Konus-Wind light curve shows a multi-peaked pulse from $\sim T_\mathrm{KW,0}-1.3$ s till $\sim T_\mathrm{KW,0}+11$ s (where $T_\mathrm{KW,0}$ is the Konus-Wind trigger time). The redshift of this source is $z=1.293$, as determined by NOT \citep{2013GCN..15407...1X}. The isotropic energy is $E_{iso}=3\times 10^{53}$~erg.}
\includegraphics[width=0.7\hsize,clip]{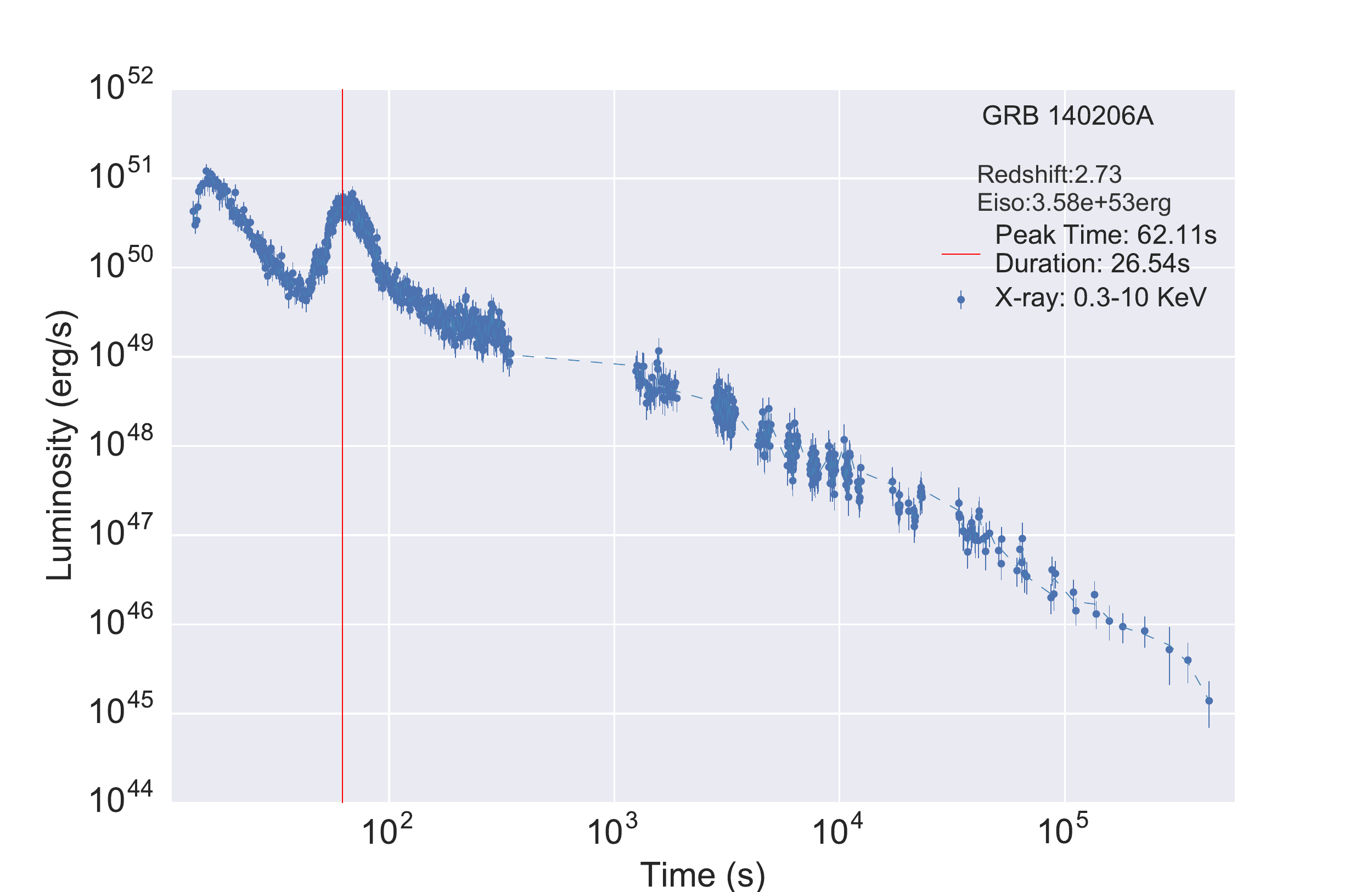}
\caption{\textbf{140206A}: this source was detected by all the instruments onboard {Swift} \citep{2014GCN..15784...1L} and by Fermi/GBM \citep{2014GCN..15796...1V}. The GBM light curve shows a single pulse with a duration of $\sim 7$~s ($50$--$300$~keV). The source was outside of the field of view, $123^\mathrm{o}$ from the LAT bore-sight at the time of the trigger. The BAT light curve shows a multi-peaked structure with roughly three main pulses \citep{2014GCN..15805...1S}. The source duration in the $15$--$350$~keV band is $25$~s. The redshift, as observed by NOT \citep{2014GCN..15800...1M} is $z=2.73$, and the isotropic energy is $E_{iso}=4.3\times 10^{53}$~erg.}
\end{figure*}

\begin{figure*}
\centering
\includegraphics[width=0.7\hsize,clip]{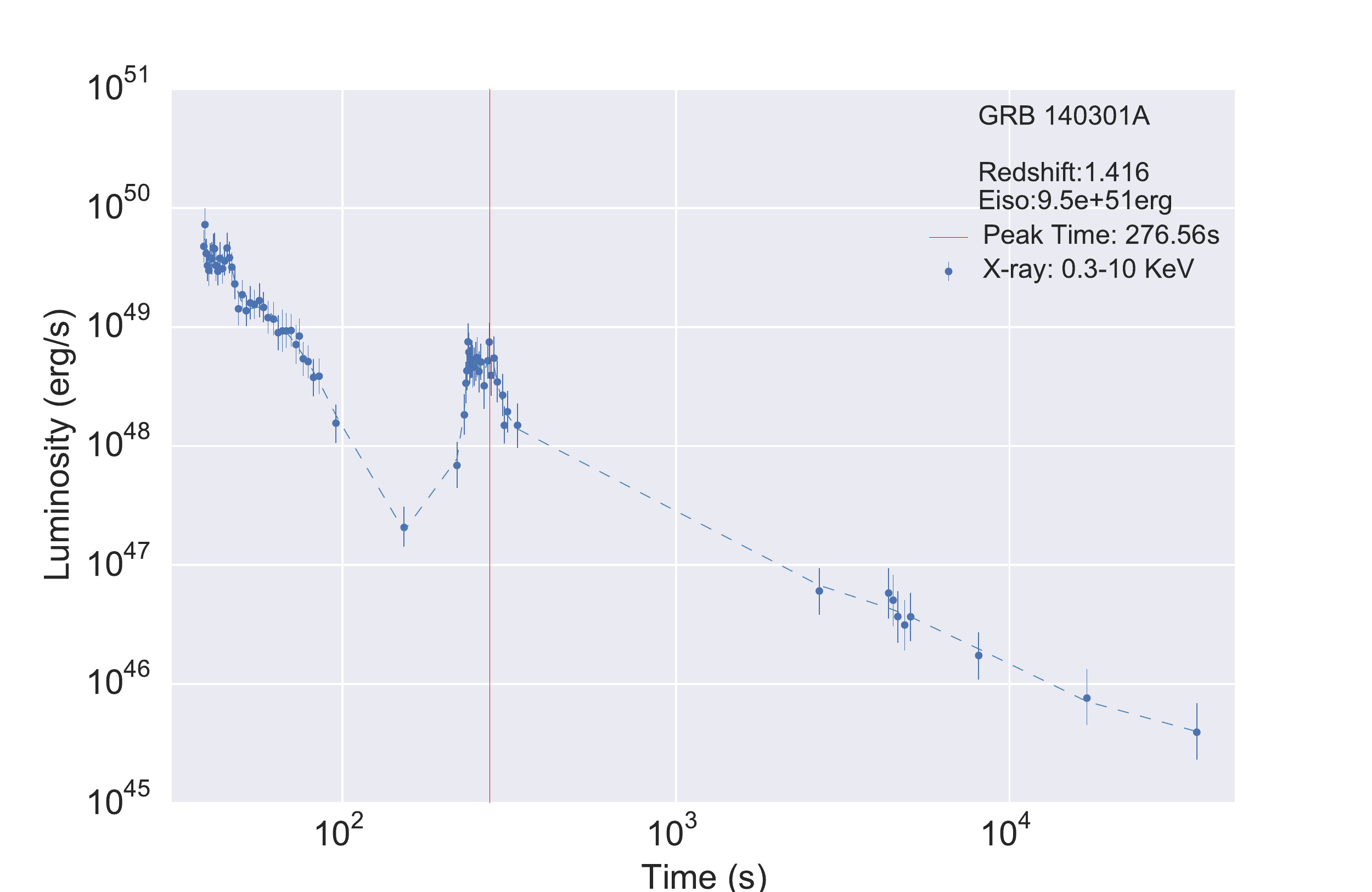}
\caption{\textbf{140301A}: this GRB was triggered by Swift-BAT \citep{2014GCNR..460....1P}; the BAT light-curve has a single spike with a duration of about $4$~s. XRT started to observe $35.63$~s after the BAT trigger. There is no observation from the Fermi satellite. From the X-shooter spectrum analysis, redshift was revealed at $1.416$ \citep{2014GCN..15900...1K}. The isotropic energy of this GRB is $9.5 \times 10^{51}$~erg.}
\includegraphics[width=0.7\hsize,clip]{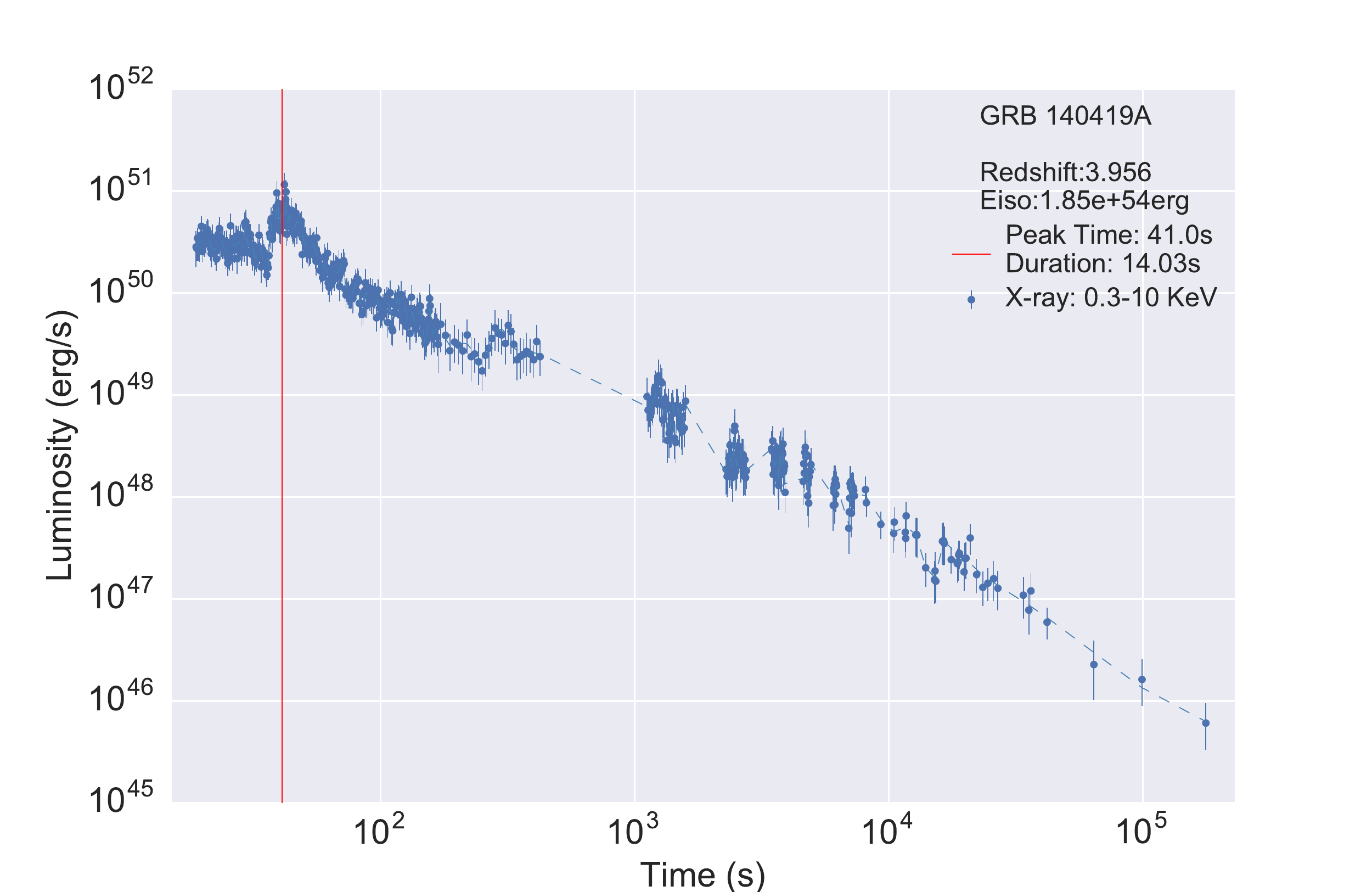}
\caption{\textbf{140419A}: this source was detected by Konus Wind \citep{2014GCN..16134...1G} and {Swift} \citep{2014GCN..16118...1M}. The Konus Wind light curve shows a broad pulse from $\sim T_\mathrm{KW,0}-2$~s to $\sim T_\mathrm{KW,0}+8$~s, followed by softer pulses around $\sim T_\mathrm{KW,0}+10$~s. The total duration of the burst is $\sim 16$ s. The BAT light curve shows two slightly overlapping clusters of peaks, starting at $\sim T_\mathrm{B,0}-2$~s, peaking at $\sim T_\mathrm{B,0}+2$~s and $\sim T_\mathrm{B,0}+10$~s, and ending at $\sim T_\mathrm{B,0}+44$~s \citep{2014GCN..16127...1B}. The total duration (in the $15$--$350$~keV) is $19$ s. The redshift of this source, as determined by Gemini, is $z=3.956$ \citep{2014GCN..16125...1T} and its isotropic energy is $E_{iso}=1.85\times 10^{54}$~erg.}
\end{figure*}

\begin{figure*}
\centering
\includegraphics[width=0.7\hsize,clip]{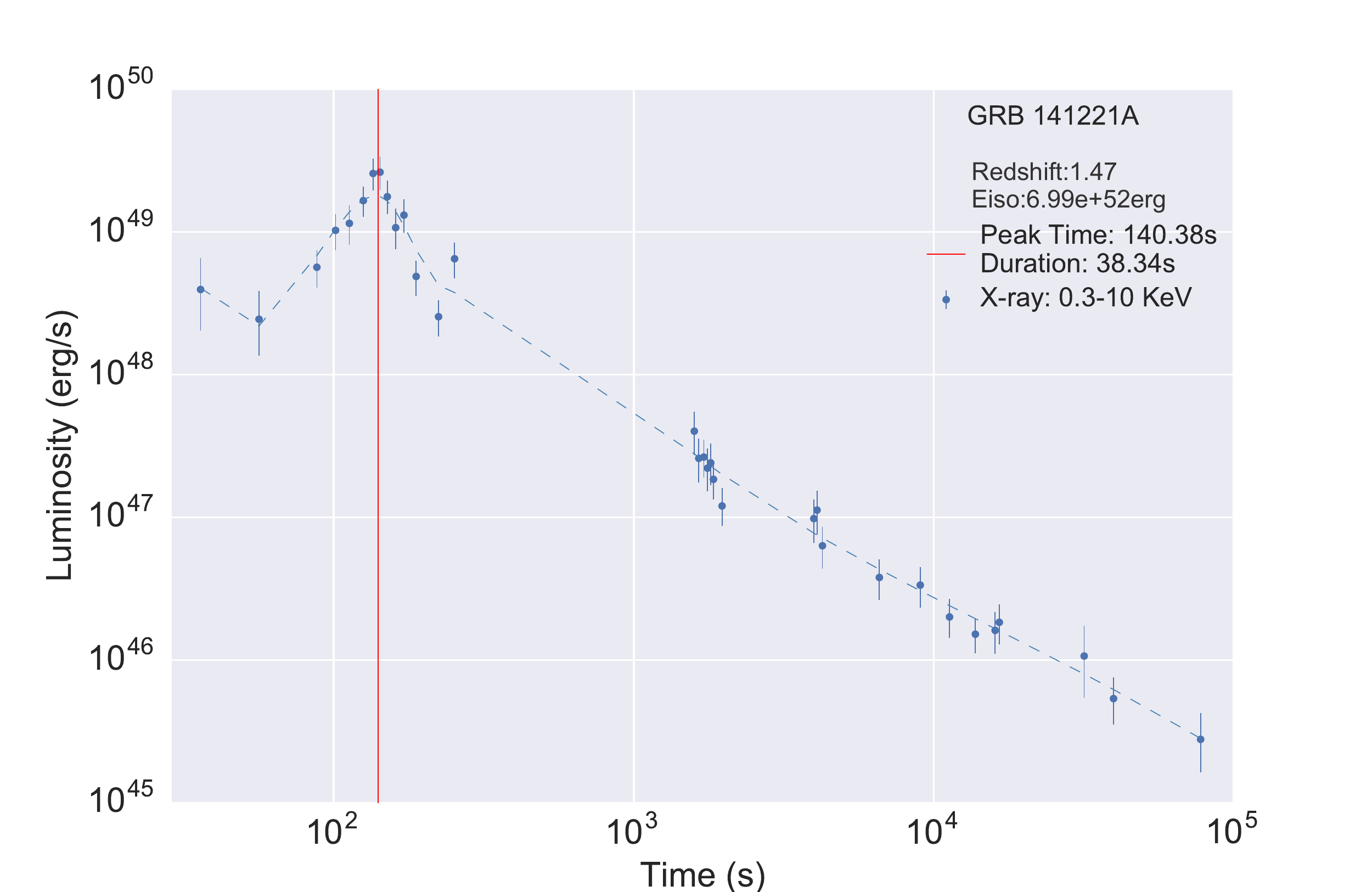}
\caption{\textbf{141221A}: this source is located at a spectroscopic redshift $z=1.47$, as determined by Keck \citep{2014GCN..17228...1P}. Its isotropic energy is $E_{iso}=1.91\times 10^{52}$~erg. The emission was detected by all the instruments onboard Swift \citep{2014GCN..17206...1S} and by Fermi/GBM \citep{2014GCN..17216...1Y}. The GBM light curve consists of two pulses with a duration of about $10$ s ($50$--$300$~keV). The source was $76^\mathrm{o}$ from the LAT boresight at the time of the trigger, out of the field of view. The BAT light curve showed a double-peaked structure with a duration of about $8$~s. XRT began observing the field $32$ s after the BAT trigger.}
\includegraphics[width=0.7\hsize,clip]{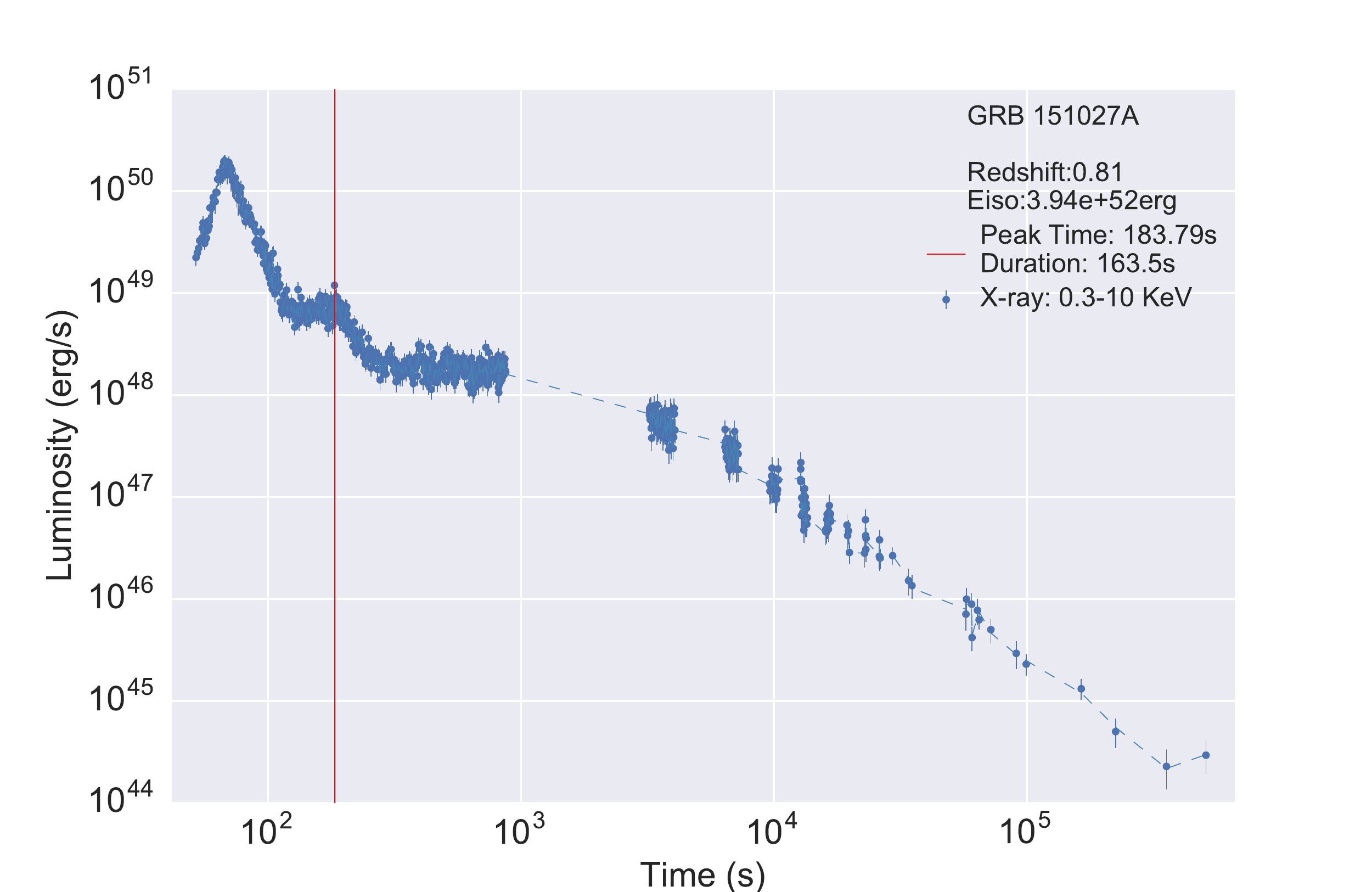}
\caption{\textbf{151027A}: this source was detected by MAXI \citep{2015GCN..18525...1M}, Konus-Wind \citep{2015GCN..18516...1G}, {Swift} \citep{2015GCN..18478...1M} and Fermi/GBM \citep{2015GCN..18492...1T}. It is located at a redshift $z=0.81$, as determined by Keck/HIRES \citep{2015GCN..18487...1P}, and the isotropic energy is $E_{iso}=3.94\times 10^{52}$~erg. The LAT boresight of the source was $10^\mathrm{o}$ at the time of the trigger, there are no clear associated high energy photons; an upper limit of observed count flux is computed as $9.24 \times 10^{-6}$~photons~cm$^{-2}$~s$^{-1}$ following the standard Fermi-LAT likelihood analysis. The BAT light curve showed a complex peaked structure lasting at least $83$ seconds. XRT began observing the field $48$~s after the BAT trigger. The GBM light curve consists of three pulses with a duration of about $68$~s in the $50$--$300$~keV band. The Konus-Wind light curve consists of at least three pulses with a total duration of $\sim 66$~s. The MAXI detection is not significant, but the flux is consistent with the interpolation from the \textit{Swift}/XRT light curve.}
\label{Sample}
\end{figure*}

We then conclude that in our sample, there are {Swift} data for all the GRBs, {Konus-Wind} observed GRB 080607, 080810, 090516A, 131030A, 140419A, 141221A and 151027A while {Fermi} detected GRB 090516A, 140206, 141221A, 151027A. The energy coverage of the available satellites is limited, as mentioned in Sec.~\ref{sec:obsbackground}: {Fermi} detects the widest photon energy band, from $8$~keV to $300$~GeV, \emph{Konus-Wind} observes from $20$~keV to $15$~MeV, {Swift}-BAT has a narrow coverage from $15$~keV to $150$~keV. No GeV photons were observed, though GRB 090516A and 151027 were in the Fermi-LAT field of view. This contrasts with the observations of S-GRBs for which, in all of the sources so far identified and within the Fermi-LAT field of view, GeV photons were always observed \citep{2016ApJ...831..178R,2016ApJ...832..136R} and can always freely reach a distant observer. These observational facts suggest that NS-NS (or NS-BH) mergers leading to the formation of a BH leave the surrounding environment poorly contaminated by the material ejected in the merging process ($\lesssim10^{-2}$--$10^{-3}~M_{\odot}$) and therefore the GeV emission, originating from accretion on the BH formed in the merger process \citep{2016ApJ...831..178R} can be observed. On the other hand, BdHNe originate in CO$_\mathrm{core}$-NS binaries in which the material ejected from the CO$_\mathrm{core}$ explosion ($\approx M_{\odot}$) greatly pollutes the environment where the GeV emission has to propagate to reach the observer (see Sec.~\ref{sec:Theory}). This together with the asymmetries of the SN ejecta \citep[see Sec.~\ref{sec:Theory} and][]{2016ApJ...833..107B} lead to the possibility that the GeV emission in BdHNe can be ``obscured'' by the material of the SN ejecta, explaining the absence of GeV photons in the above cases of GRB 090516A and 151027.

We derive the isotropic energy $E_{iso}$ by assuming the prompt emission to be isotropic and by integrating the prompt photons in the rest-frame energy range from $1$~keV to $10$~MeV \citep{2001AJ....121.2879B}. None of the satellites is able to cover the entire energy band of $E_{iso}$, so we need to fit the spectrum and find the best-fit function, then extrapolate the integration of energy by using this function. This method is relatively safe for GRBs observed by Fermi and Konus-Wind, but 6 GRBs in our sample have been observed only by Swift, so we uniformly fit and extrapolate these 6 GRBs by power-laws and cutoff power-laws; then we take the average value as $E_{iso}$. In general, our priority in computing $E_{iso}$ is Fermi, Konus-Wind, then Swift. In order to take into account the expansion of the universe, all of our computations consider K-correction. The formula of K-correction for $E_{iso}$ varies depending on the best-fit function. The energy in the X-ray afterglow is computed in the cosmological rest-frame energy band from $0.3$~keV to $10$~keV. We smoothly fit the luminosity light-curve using an algorithm named locally weighted regression \citep{10.2307/2289282} which provides a sequence of power-law functions. The corresponding energy in a fixed time interval is obtained by summing up all of the integrals of the power-laws within it. This method is applied to estimate the energy of the flare $E_f$, as well as the energy of the FPA phase up to $10^9$~s, $E_{FPA}$. An interesting alternative procedure was used in \citet{2014ApJ...788...30S} to fit the light-curve and determine the flaring structure with a Bayesian Information method. On this specific aspect the two treatments are equally valid and give compatible results.

Tab.~\ref{tab:grbList} contains the relevant energy and time information of the $16$ BdHNe of the sample: the cosmological redshift $z$, $E_{iso}$, the flare peak time $t_p$, the corresponding peak luminosity $L_p$, the flare duration $\Delta t$, and the energy of the flare $E_f$. To determine $t_p$ we apply a locally weighted regression, which results in a smoothed light-curve composed of power-law functions: the flare peak is localized where the power-law index is zero. Therefore $t_p$ is defined as the time interval between the flare peak and the trigger time of \textit{Swift}-BAT \footnote{In reality, the GRB occurs earlier than the trigger time, since there is a short period when the flux intensity is lower than the satellite trigger threshold \citep{2003AIPC..662..491F}}. Correspondingly, we find the peak luminosity $L_p$ at $t_p$ and its duration $\Delta t$ which is defined as the time interval between a start time and an end time at which the luminosity is half of $L_p$. We have made public the entire details including the codes online\footnote{\href{https://github.com/YWangScience/AstroNeuron}{https://github.com/YWangScience/AstroNeuron}}.
\begin{table}
\centering
\begin{tabular}{c@{\hskip 0.08in}c@{\hskip 0.08in}c@{\hskip 0.08in}cccc@{\hskip 0.08in}c@{\hskip 0.08in}c}
\hline\hline
GRB & z & $T_{90}$ (s) & $E_{iso}$ (erg) & $t_p$ (s) & $L_p$ (erg/s) & $\Delta t$ (s) & $E_{f}$ (erg) & $\alpha_f$\\
\hline
060204B & 2.3393 & 40.12 & $2.93(\pm0.60)\times10^{53}$  & $100.72\pm6.31  $ & $7.35(\pm2.05)\times10^{49}$  & $17.34\pm6.83 $ & $8.56(\pm0.82)\times 10^{50}$ & 2.73 \\
060607A & 3.082  & 24.49 & $2.14(\pm1.19)\times10^{53}$  & $66.04\pm4.98  $ & $2.28(\pm0.48)\times10^{50}$  & $18.91\pm3.84 $ & $3.33(\pm0.32)\times 10^{51}$ & 1.72 \\
070318  & 0.84   & 28.80 & $3.41(\pm2.14)\times10^{52}$  & $154.7\pm12.80 $ & $6.28(\pm1.30)\times10^{48}$  & $63.80\pm19.82$ & $3.17(\pm0.37)\times 10^{50}$ & 1.84 \\
080607  & 3.04   & 21.04 & $1.87(\pm0.11)\times10^{54}$  & $37.48\pm3.60  $ & $1.14(\pm0.27)\times10^{51}$  & $15.63\pm4.32 $ & $1.54(\pm0.24)\times 10^{52}$ & 2.08 \\
080805  & 1.51   & 31.08 & $7.16(\pm1.90)\times10^{52}$  & $48.41\pm5.46  $ & $4.66(\pm0.59)\times10^{49}$  & $27.56\pm9.33 $ & $9.68(\pm1.24)\times 10^{50}$ & 1.25 \\
080810  & 3.35   & 18.25 & $5.00(\pm0.44)\times10^{53}$  & $51.03\pm6.49  $ & $1.85(\pm0.53)\times10^{50}$  & $12.38\pm4.00 $ & $1.80(\pm0.17)\times 10^{51}$ & 2.37 \\
081008  & 1.967  & 62.52 & $1.35(\pm0.66)\times10^{53}$  & $102.24\pm5.66 $ & $1.36(\pm0.33)\times10^{50}$  & $18.24\pm3.63 $ & $1.93(\pm0.16)\times 10^{51}$ & 2.46 \\
081210  & 2.0631 & 47.66 & $1.56(\pm0.54)\times10^{53}$  & $127.59\pm13.68 $ & $2.23(\pm0.21)\times10^{49}$  & $49.05\pm6.49$ & $8.86(\pm0.54)\times 10^{50}$ & 2.28 \\
090516A & 4.109  & 68.51 & $9.96(\pm1.67)\times10^{53}$  & $80.75\pm2.20  $ & $9.10(\pm2.26)\times10^{50}$  & $10.43\pm2.44 $ & $7.74(\pm0.63)\times 10^{51}$ & 3.66 \\
090812  & 2.452  & 18.77 & $4.40(\pm0.65)\times10^{53}$  & $77.43\pm16.6  $ & $3.13(\pm1.38)\times10^{50}$  & $17.98\pm4.51 $ & $5.18(\pm0.61)\times 10^{51}$ & 2.20 \\
131030A & 1.293  & 12.21 & $3.00(\pm0.20)\times10^{53}$  & $49.55\pm7.88  $ & $6.63(\pm1.12)\times10^{50}$  & $33.73\pm6.55 $ & $3.15(\pm0.57)\times 10^{52}$ & 2.22 \\
140206A & 2.73   & 7.24 & $3.58(\pm0.79)\times10^{53}$  & $62.11\pm12.26 $ & $4.62(\pm0.99)\times10^{50}$  & $26.54\pm4.31 $ & $1.04(\pm0.59)\times 10^{51}$ & 1.73 \\
140301A  & 1.416 & 12.83 & $9.50(\pm1.75)\times10^{51}$  & $276.56\pm15.50  $ & $5.14(\pm1.84)\times10^{48}$  & $64.52\pm10.94 $ & $3.08(\pm0.22)\times 10^{50}$ & 2.30\\
140419A & 3.956  & 16.14 & $1.85(\pm0.77)\times10^{54}$  & $41.00\pm4.68  $ & $6.23(\pm1.45)\times10^{50}$  & $14.03\pm5.74 $ & $7.22(\pm0.88)\times 10^{51}$ & 2.32\\
141221A & 1.47   & 9.64 & $6.99(\pm1.98)\times10^{52}$  & $140.38\pm5.64 $ & $2.60(\pm0.64)\times10^{49}$  & $38.34\pm9.26 $ & $7.70(\pm0.78)\times 10^{50}$& 1.79\\
151027A & 0.81   & 68.51 & $3.94(\pm1.33)\times10^{52}$  & $183.79\pm16.43$ & $7.10(\pm1.75)\times10^{48}$  & $163.5\pm30.39$ & $4.39(\pm2.91)\times 10^{51}$& 2.26\\
\hline
\end{tabular}
\caption{GRB sample properties of the prompt and flare phases. This table contains: the redshift $z$, the $T_{90}$ in the rest frame, the isotropic energy $E_{iso}$, the flare peak time $t_p$ in the rest frame, the flare peak luminosity $L_p$, the flare duration of which the starting and ending time correspond to half of the peak luminosity $\Delta t$, the flare energy $E_{f}$  within the time interval $\Delta t$, and $\alpha_f$ power-law index from the fitting of the flare's spectrum.} 
\label{tab:grbList}
\end{table}

\section{Statistical Correlation}
\label{sec:correlation}

\begin{figure}
\centering
\includegraphics[width=0.8\hsize,clip]{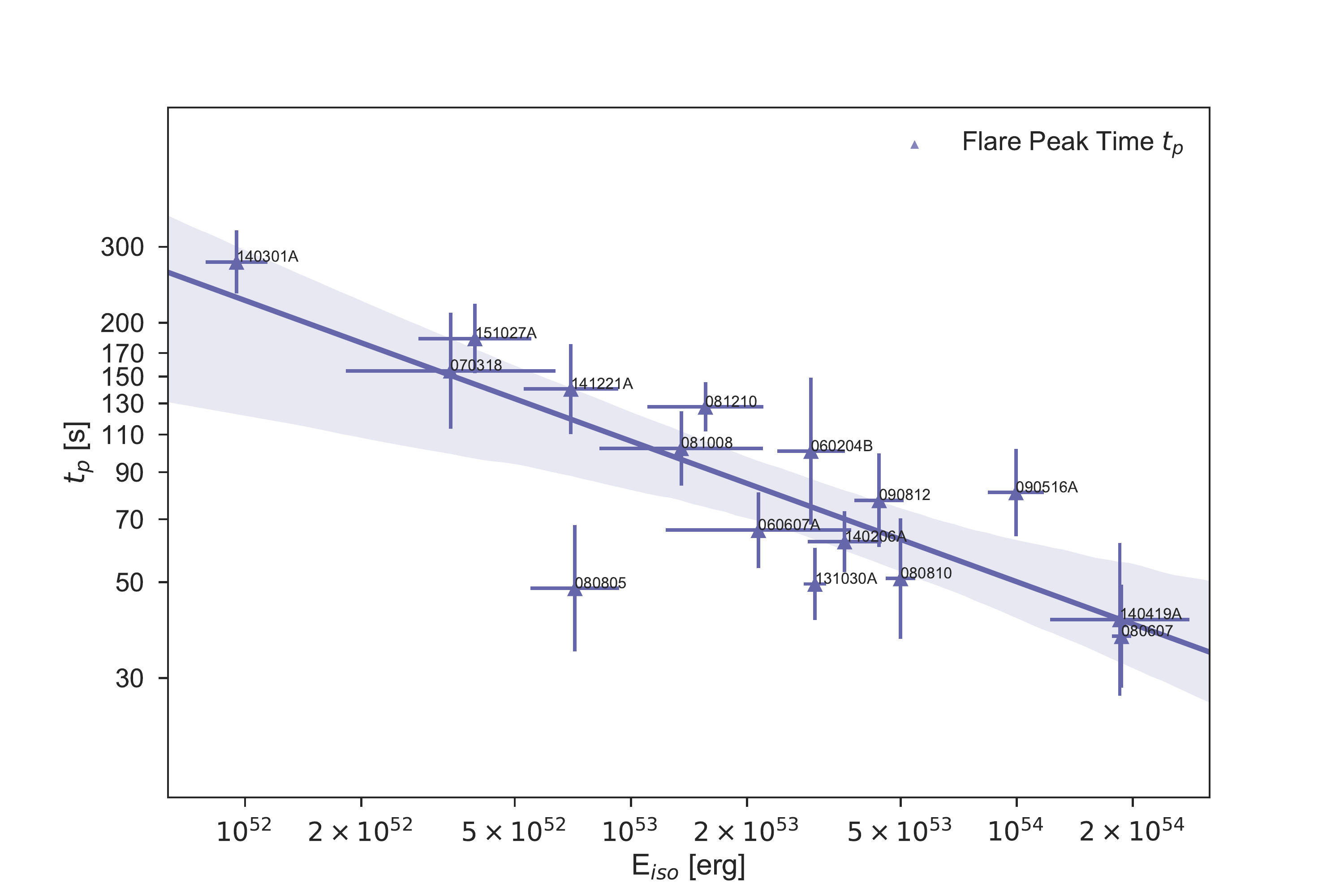}
\caption{Relation between $E_{iso}$ and $t_p$ fit by a power-law. The shaded area indicates the $95\%$ confidence level.}
\label{fig:EisoTflare}
\end{figure}

\begin{figure}
\centering
\includegraphics[width=0.8\hsize,clip]{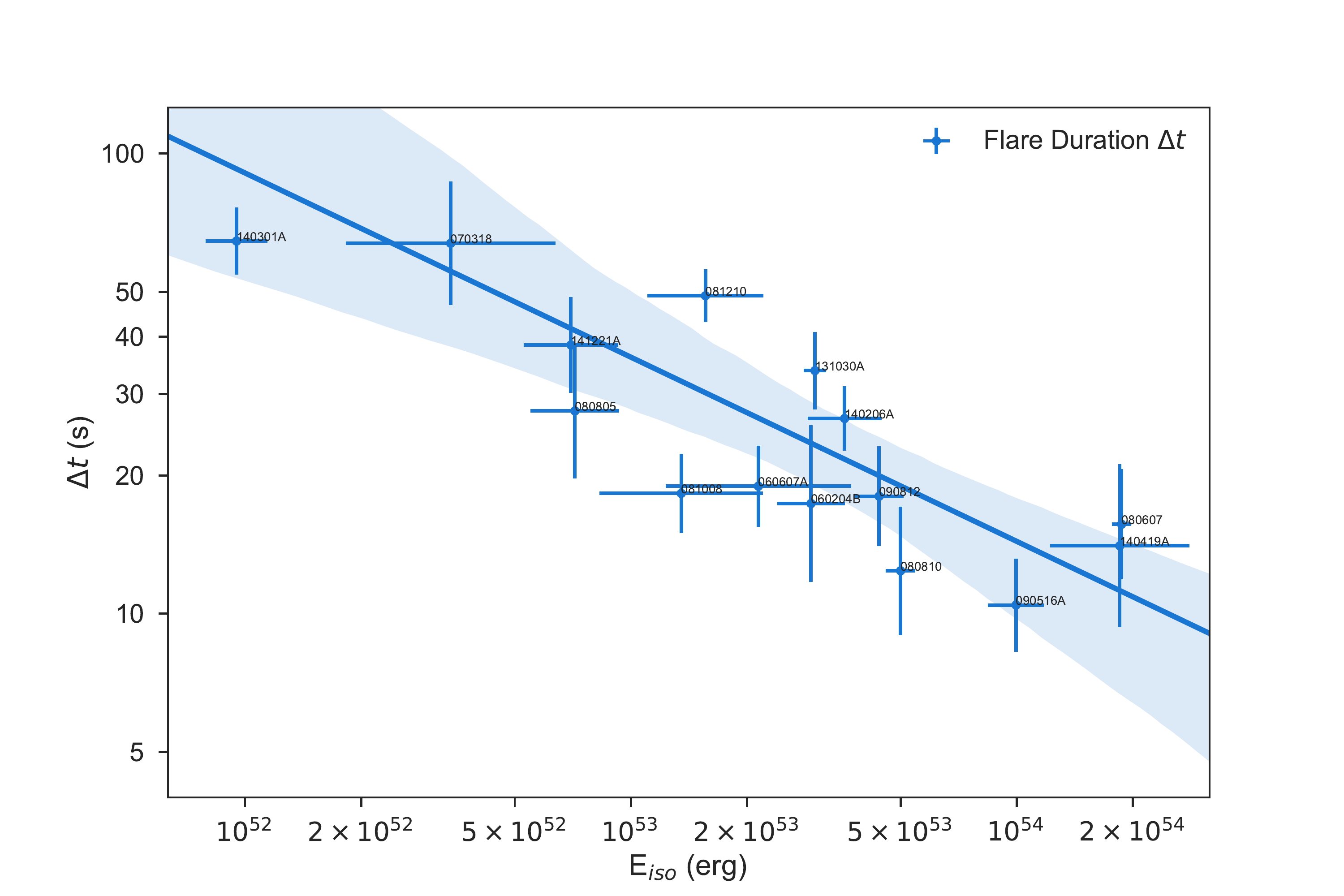}
\caption{Relation between $E_{iso}$ and $\Delta t$ fit by a power-law. The shaded area indicates the $95\%$ confidence level.}
\label{fig:EisoDeltaT}
\end{figure}

\begin{figure}
\centering
\includegraphics[width=0.8\hsize,clip]{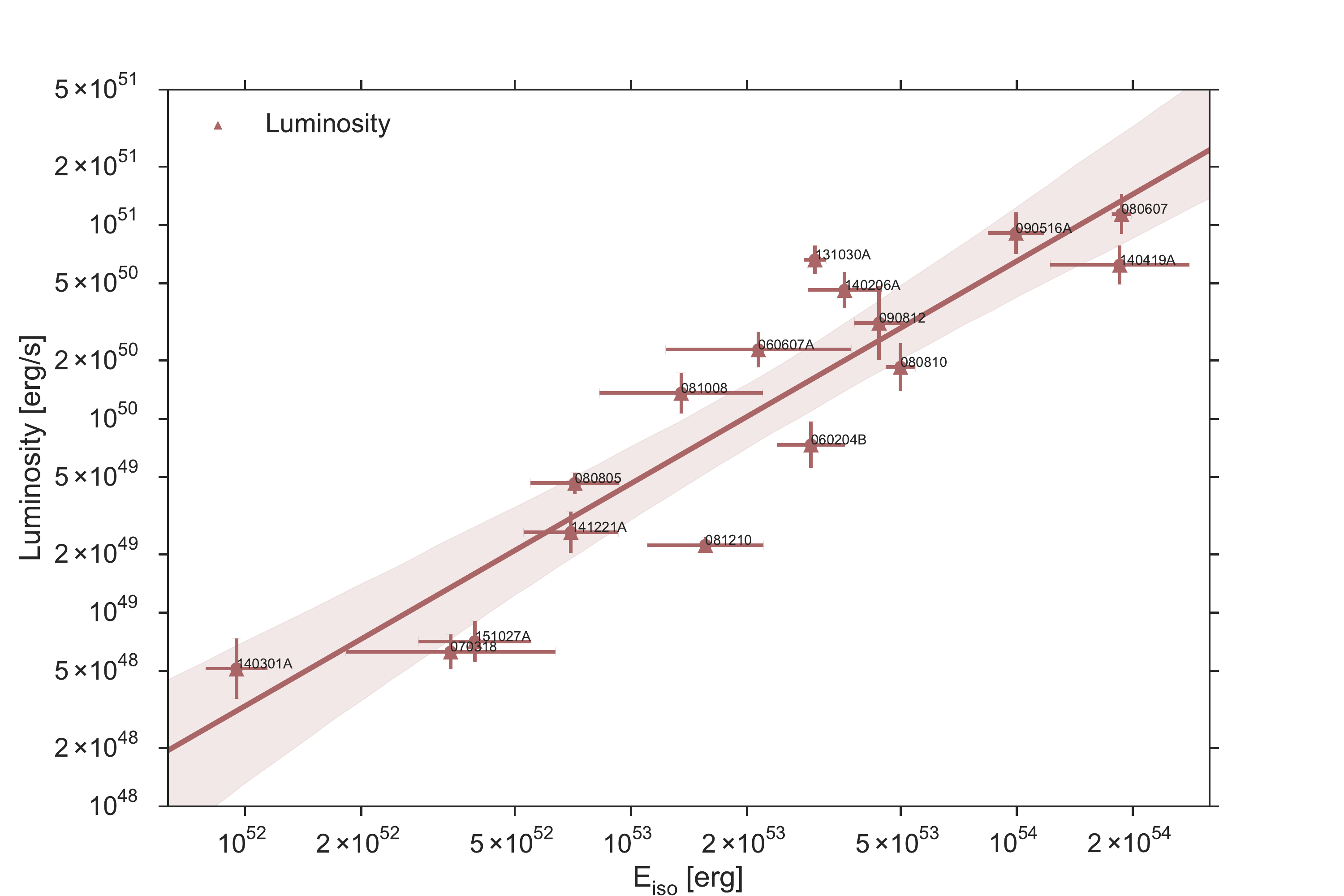}
\caption{Relation between $E_{iso}$ and $L_p$ fit by a power-law. The shaded area indicates the $95\%$ confidence level.}
\label{fig:EisoLuminosity}
\end{figure}

\begin{figure}
\centering
\includegraphics[width=0.8\hsize,clip]{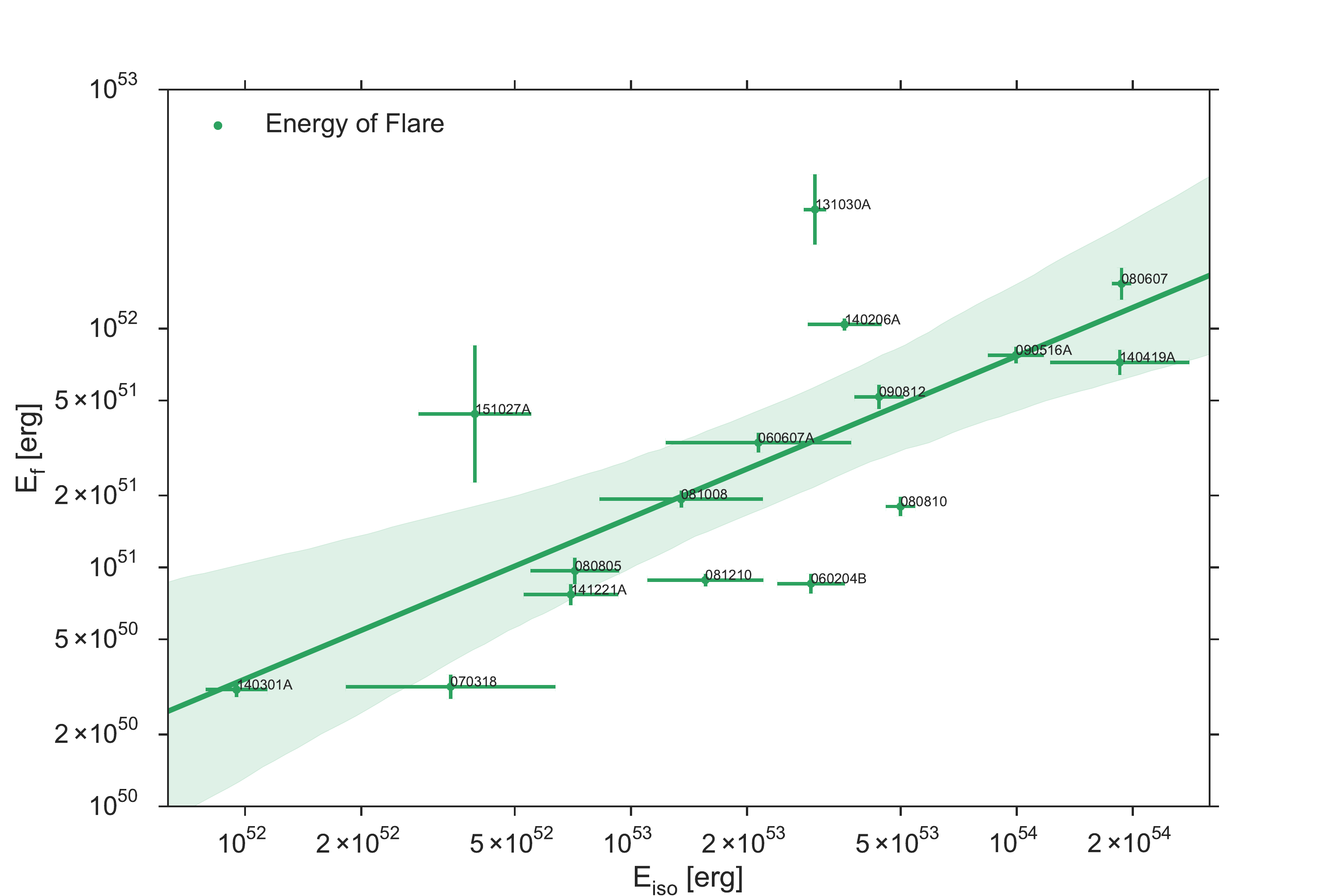}
\caption{Relation between $E_{iso}$ and $E_{f}$ fit by a power-law. The shaded area indicates the $95\%$ confidence level.}
\label{fig:EisoEflare}
\end{figure}

We then establish correlations between the above quantities characterizing each luminosity light curve of the sample with the $E_{iso}$ of the corresponding BdHN. We have relied heavily on the Markov chain Monte Carlo (MCMC) method and iterated $10^5$ times for having the best fit of the power-law and their correlation coefficient. The main results are summarized in Figs.~\ref{fig:EisoTflare}--\ref{fig:EisoEflare}. All the codes are publicly available online\footnote{\href{https://github.com/YWangScience/MCCC}{https://github.com/YWangScience/MCCC}}.
We conclude that the peak time and the duration of the flare, as well as the peak luminosity and the total energy of flare, are highly correlated with $E_{iso}$, with correlation coefficients larger than $0.6$ (or smaller than $-0.6$). The average values and the 1-$\sigma$ uncertainties are shown in Tab.~\ref{tab:correlation}.

\begin{table}
\centering
\begin{tabular}{lrr}
\hline\hline
Correlation & Power-Law Index & Coefficient \\
\hline
$E_{iso}-t_p$       & $-0.290(\pm0.010)$  & $-0.764(\pm 0.123)$  \\
$E_{iso}-\Delta t$  & $-0.461(\pm0.042)$  & $-0.760(\pm 0.138)$  \\
$E_{iso}-L_p$       & $1.186(\pm0.037)$   & $0.883(\pm 0.070)$  \\
$E_{iso}-E_f$       & $0.631(\pm0.117)$   & $0.699(\pm 0.145)$  \\
\hline
\end{tabular}
\caption{Power-law correlations among the quantities in Tab.~\ref{tab:grbList}. The values and uncertainties (at $1$--$\sigma$ confidence level) of the power-law index and of the correlation coefficient are obtained from $10^5$ MCMC iterations. All relations are highly correlated.}
\label{tab:correlation}
\end{table}

\section{The partition of electron-positron plasma energy between the prompt emission and the FPA} \label{sec:partition}

\begin{figure}
\centering
\includegraphics[width=0.8\hsize,clip]{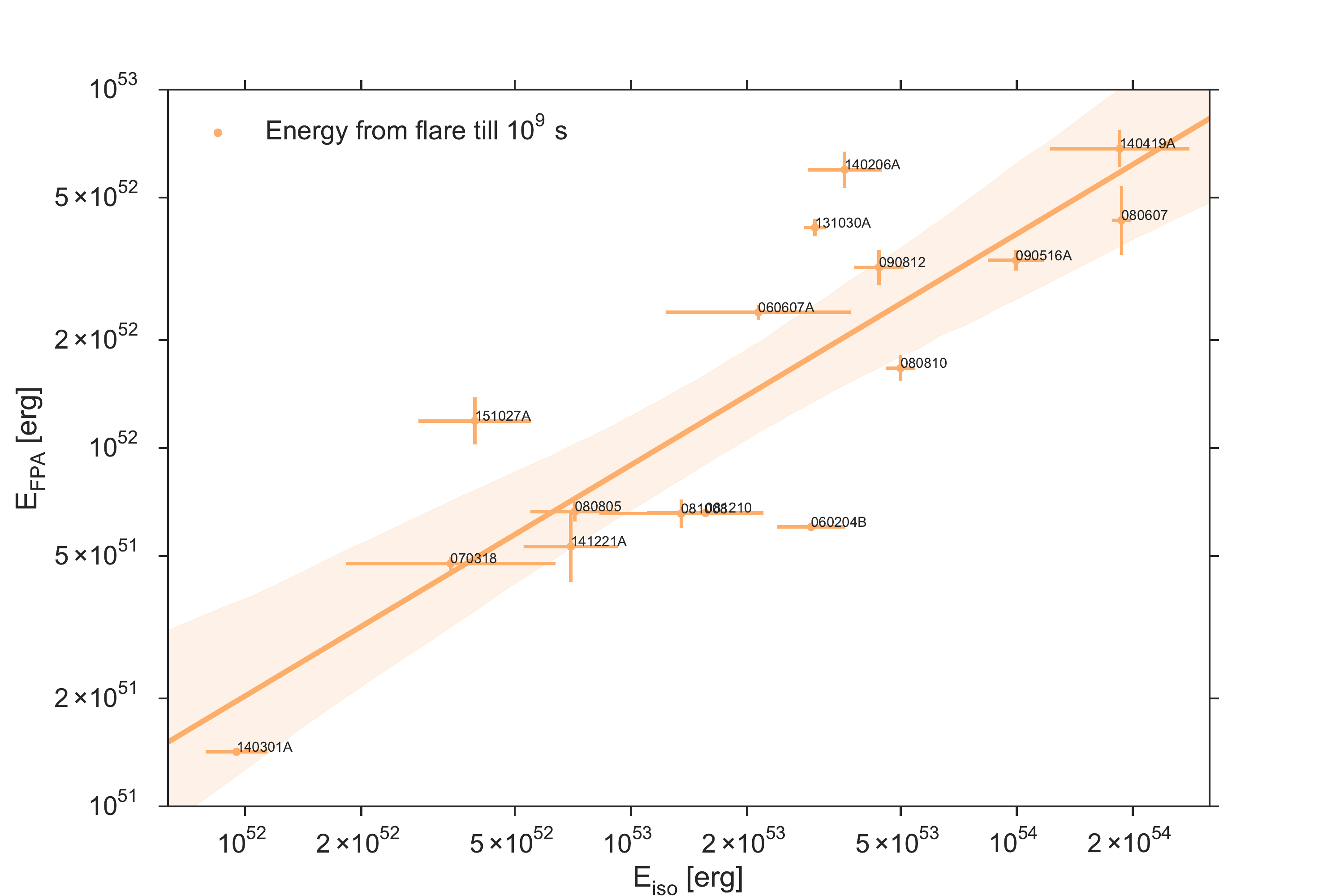}
\caption{Relation between $E_{iso}$ and $E_{FPA}$ fit by a power-law. The shaded area indicates the $95\%$ confidence level.}
\label{fig:EisoEflareEnd}
\end{figure}

\begin{figure}
\centering
\includegraphics[width=0.7\hsize,clip]{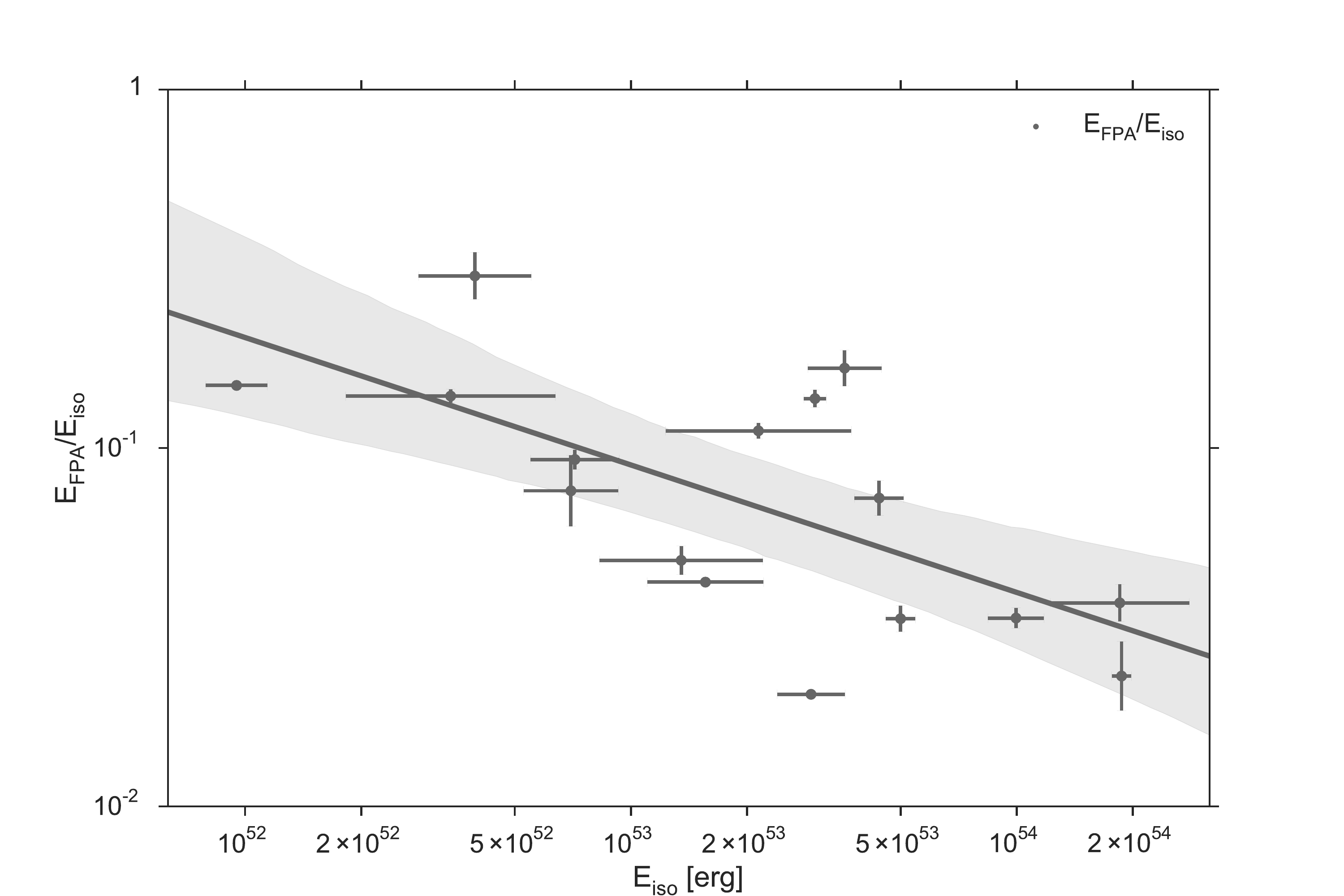}
\caption{Relation between the percentage of $E_{e^+e^-}$ going to the SN ejecta and accounting for the energy in FPA, i.e., $E_{FPA}/E_{iso}\times 100$\%, and $E_{iso}$ fit by a power-law. The shaded area indicates the $95\%$ confidence level.}
\label{fig:ratioEiso}
\end{figure}

\begin{figure}
\centering
\includegraphics[width=0.8\hsize,clip]{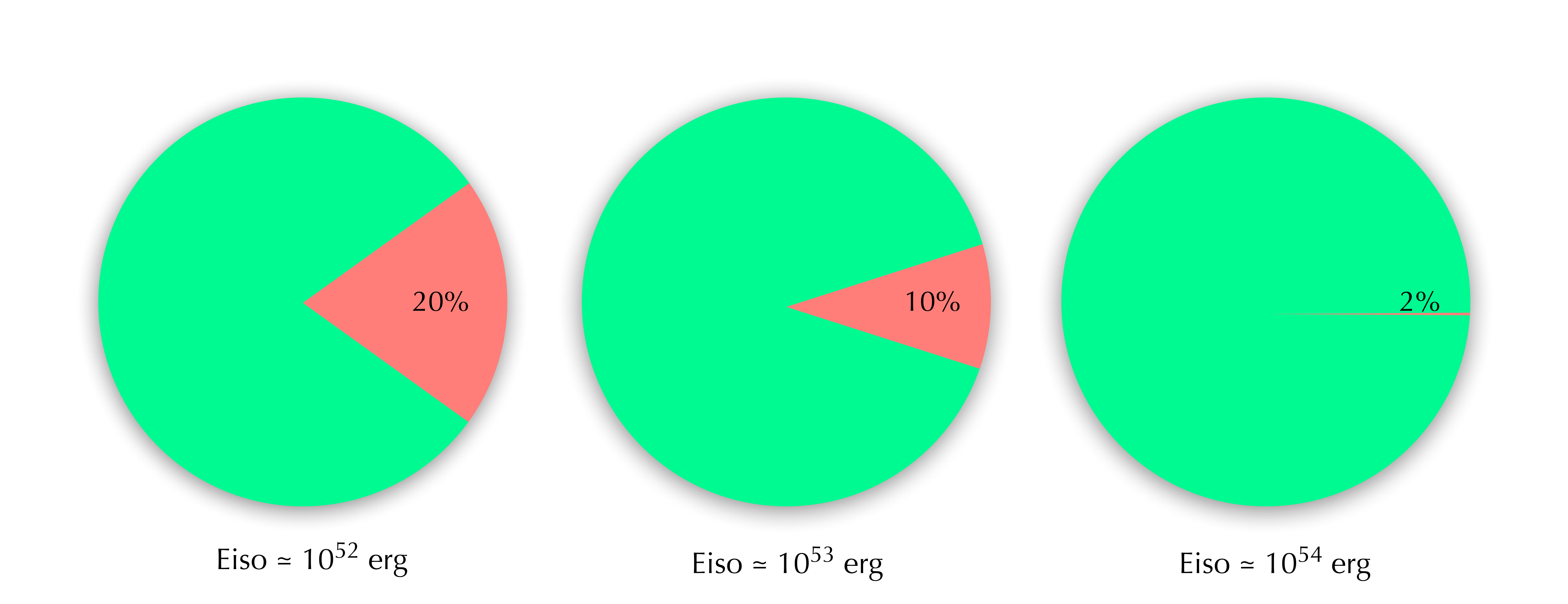}
\caption{Distribution of the GRB total energy $E_{e^+e^-}=E_{iso}$ into prompt and FPA energies. The percentage of $E_{e^+e^-}$ going to the SN ejecta accounting for the energy in FPA phase appears in red, i.e., $E_{FPA}/E_{iso}\times 100$\%. The green part is therefore the percentage of $E_{e^+e^-}$ used in the prompt emission, i.e., $E_\mathrm{prompt}/E_{iso}\times 100$\%. It can be seen that the lower the GRB energy $E_{e^+e^-}=E_{iso}$, the higher the FPA energy percentage, and consequently the lower the prompt energy percentage.}
\label{fig:percentage}
\end{figure}

The energy of the prompt emission is proportional to $E_{iso}$ if and only if spherical symmetry is assumed: this clearly follows from the prompt emission time integrated luminosity. We are now confronted with a new situation: the total energy of the FPA emission up to $10^9$~s ($E_{FPA}$) is also proportional to $E_{iso}$, following the correlation given in Tabs.~\ref{tab:grbList2} and \ref{tab:correlation2}, and Fig.~\ref{fig:EisoEflareEnd}. What is clear is that there are two very different components where the energy of the dyadosphere $E_{e^+e^-}$ is utilized: the energy $E_\mathrm{prompt}$ of the prompt emission and the energy $E_{FPA}$ of the FPA, i.e., $E_{e^+e^-}=E_{iso} = E_\mathrm{prompt} + E_{FPA}$. Fig.~\ref{fig:ratioEiso} and Fig.~\ref{fig:percentage} show the distribution of $E_{e^+e^-}=E_{iso}$ among these two components. 
\begin{table}
\centering
\begin{tabular}{cccc}
\hline\hline
GRB & z & $E_{iso}$ (erg) & $E_{FPA}$ (erg) \\
\hline
060204B & 2.3393& $2.93(\pm0.60)\times10^{53}$  & $6.02(\pm0.20)\times 10^{51}$\\
060607A & 3.082 & $2.14(\pm1.19)\times10^{53}$  & $2.39(\pm0.12)\times 10^{52}$\\
070318  & 0.84  & $3.41(\pm2.14)\times10^{52}$  & $4.76(\pm0.21)\times 10^{51}$\\
080607  & 3.04  & $1.87(\pm0.11)\times10^{54}$  & $4.32(\pm0.96)\times 10^{52}$\\
080805  & 1.51  & $7.16(\pm1.90)\times10^{52}$  & $6.65(\pm0.42)\times 10^{51}$\\
080810  & 3.35  & $5.00(\pm0.44)\times10^{53}$  & $1.67(\pm0.14)\times 10^{52}$\\
081008  & 1.967 & $1.35(\pm0.66)\times10^{53}$  & $6.56(\pm0.60)\times 10^{51}$\\
081210  & 2.0631& $1.56(\pm0.54)\times10^{53}$  & $6.59(\pm0.60)\times 10^{51}$\\
090516A & 4.109 & $9.96(\pm1.67)\times10^{53}$  & $3.34(\pm0.22)\times 10^{52}$\\
090812  & 2.452 & $4.40(\pm0.65)\times10^{53}$  & $3.19(\pm0.36)\times 10^{52}$\\
131030A & 1.293 & $3.00(\pm0.20)\times10^{53}$  & $4.12(\pm0.23)\times 10^{52}$\\
140206A & 2.73  & $3.58(\pm0.79)\times10^{53}$  & $5.98(\pm0.69)\times 10^{52}$\\
140301A & 1.416 & $9.50(\pm1.75)\times10^{51}$  & $1.42(\pm0.14)\times 10^{50}$\\
140419A & 3.956 & $1.85(\pm0.77)\times10^{54}$  & $6.84(\pm0.82)\times 10^{52}$\\
141221A & 1.47  & $6.99(\pm1.98)\times10^{52}$  & $5.31(\pm1.21)\times 10^{51}$\\
151027A & 0.81  & $3.94(\pm1.33)\times10^{52}$  & $1.19(\pm0.18)\times 10^{52}$\\
\hline
\end{tabular}
\caption{GRB sample properties of the prompt and FPA phases. This table lists: $z$, $E_{iso}$, and the FPA energy $E_{FPA}$ from the flare till $10^9$~s.}.
\label{tab:grbList2}
\end{table}

\begin{table}
\centering
\begin{tabular}{lrr}
\hline\hline
Correlation & Power-Law Index & Coefficient \\
\hline
$E_{iso}$--$E_{FPA}$& $0.613(\pm0.041)$   & $0.791(\pm 0.103)$  \\
$E_{iso}$--$E_{FPA}/E_{iso}$& $-0.005(\pm0.002)$   & $0.572(\pm 0.178)$  \\
\hline
\end{tabular}
\caption{Power-law correlations among the quantities in Tab.~\ref{tab:grbList2}. The statistical considerations of Tab.~\ref{tab:correlation} are valid here as well.}
\label{tab:correlation2}
\end{table}

As a consequence of the above, in view of the presence of the companion SN remnant ejecta \citep[see][for more details]{2016ApJ...833..107B},  we assume here that the spherical symmetry of the prompt emission is broken. Part of the energy due to the impact of the $e^+e^-$ plasma on the SN is captured by the SN ejecta, and gives origin to the FPA emission as originally proposed by \citet{2015mgm..conf..242R}.
We shall return to the study of the impact between the plasma and the SN ejecta in Sec.~\ref{sec:originprFPA} after 
studying the motion of the matter composing the FPA in the next few sections.

It can also be seen that the relative partition between $E_\mathrm{prompt}$ and $E_{FPA}$ strongly depends on the value of $E_{e^+e^-}$: the lower the GRB energy, the higher the FPA energy percentage and consequently the lower is the prompt energy percentage (see Fig.~\ref{fig:percentage}). 

In \citet{2016ApJ...833..107B} we indicate that both the value of $E_{e^+e^-}$ and the relative ratio of the above two components can in principle be explained in terms of the geometry of the binary nature of the system: the smaller the distance is between the CO$_\mathrm{core}$ and the companion NS, the shorter the binary period of the system, and the larger the value of $E_{e^+e^-}$ .

\section{On the flare thermal emission, its temperature and dynamics}
\label{sec:thermalflare}

We discuss now the profound difference between the prompt emission, which we recall is emitted at distances of the order of $10^{16}$~cm away from the newly-born BH with $\Gamma\approx 10^2$--$10^3$, and the FPA phase. We focus on a further fundamental set of data, which originates from a thermal emission associated with the flares\footnote{The late afterglow phases have been already discussed in \citet{Pisani2013,2016ApJ...833..159P}.}. Only in some cases is this emission so clear and prominent that it allows the estimation of the flare expansion speed, and the determination of its mildly relativistic Lorentz factor $\Gamma \lesssim 4$, which creates a drastic separatrix, both in the energy and in the gamma factor between the astrophysical nature of the prompt emission and of the flares.

\begin{table}
\centering
\begin{tabular}{lccc}
\hline\hline
GRB & Radius (cm) & $kT_{\text{obs}}$ (keV) & Significance\\
\hline
\textbf{060204B} & $\mathbf{1.80(\pm1.11)\times 10^{11}}$  & $\mathbf{0.60(\pm 0.15)}$  & $\mathbf{0.986}$  \\
\textbf{060607A} & $\mathbf{1.67(\pm1.01)\times 10^{11}}$  & $\mathbf{0.92(\pm 0.24)}$  & $\mathbf{0.991}$  \\
070318   & $unconstrained$  & $1.79(\pm 1.14)$  & $0.651$  \\
\textbf{080607}  & $\mathbf{1.52(\pm0.72)\times 10^{12}}$  & $\mathbf{0.49(\pm 0.10)}$  & $\mathbf{0.998}$  \\
080805   & $1.12(\pm1.34)\times 10^{11}$  & $1.31(\pm 0.59)$  & $0.809$  \\
\textbf{080810}  & $\mathbf{2.34(\pm4.84)\times 10^{11}}$  & $\mathbf{0.61(\pm 0.57)}$  & $\mathbf{0.999}$  \\
\textbf{081008}  & $\mathbf{1.84(\pm0.68)\times 10^{12}}$  & $\mathbf{0.32(\pm 0.03)}$  & $\mathbf{0.999}$  \\
081210  & $unconstrained$  & $0.80(\pm 0.51)$  & $0.295$  \\
090516A  & $unconstrained$  & $1.30(\pm 1.30)$  & $0.663$  \\
090812   & $1.66(\pm1.84)\times 10^{12}$  & $0.24(\pm 0.12)$  & $0.503$  \\
\textbf{131030A} & $\mathbf{3.67(\pm1.02)\times 10^{12}}$  & $\mathbf{0.55(\pm 0.06)}$  & $\mathbf{0.999}$  \\
\textbf{140206A} & $\mathbf{9.02(\pm2.84)\times 10^{11}}$  & $\mathbf{0.54(\pm 0.07)}$  & $\mathbf{0.999}$  \\
140301A  & $unconstrained$  & $unconstrained$  & $0.00$  \\
140419A  & $1.85(\pm1.17)\times 10^{12}$  & $0.23(\pm 0.05)$  & $0.88$  \\
141221A  & $1.34(\pm2.82)\times 10^{12}$  & $0.24(\pm 0.24)$  & $0.141$  \\
151027A  & $1.18(\pm0.67)\times 10^{12}$  & $0.29(\pm 0.06)$  & $0.941$  \\
\hline
\end{tabular}
\caption{Radii and temperatures of the thermal components detected within the flare duration $\Delta t$. The observed temperatures $k T_\mathrm{obs}$ are inferred from fitting with a power-law plus blackbody spectral model. The significance of a blackbody is computed by the maximum likelihood ratio for comparing nested models and its addition improves a fit when the significance is $>0.95$. The radii are calculated assuming mildly-relativistic motion ($\beta = 0.8$) and isotropic radiation. The GRBs listed in boldface have prominent black bodies, with radii of the order of $\sim10^{11}$--$10^{12}$~cm. Uncertainties are given at $1$--$\sigma$ confidence level.}
\label{tab:grbTemperature}
\end{table}

Following the standard data reduction procedure of {Swift}-XRT \citep{2006A&A...456..917R,2007A&A...469..379E,2009MNRAS.397.1177E}, X-ray data within the duration of flare are retrieved from the United Kingdom Swift Science Data Centre (UKSSDC) \footnote{\href{http://www.swift.ac.uk}{http://www.swift.ac.uk}} and analyzed by Heasoft\footnote{\href{http://heasarc.gsfc.nasa.gov/lheasoft/}{http://heasarc.gsfc.nasa.gov/lheasoft/}}. Tab.~\ref{tab:grbTemperature} shows the fit of the spectrum within the duration $\Delta t$ of the flare  for each BdHN of the sample. As a first approximation, in computing the radius we have assumed a constant expansion velocity of $0.8c$ indicated for some BdHNe, such as GRB 090618 \citep{2014A&A...565L..10R} and GRB 130427A \citep{2015ApJ...798...10R}. Out of $16$ sources, $7$ BdHNe have highly confident thermal components (significance $>0.95$, see boldface in Tab.~\ref{tab:grbTemperature}), which means that the addition of a blackbody spectrum improves a single power-law fit (which is, conversely, excluded at $2$--$\sigma$ of confidence level). These blackbodies have fluxes in a range from $1\%$ to $30\%$ of the total flux and share similar order of magnitude radii, i.e., $\sim10^{11}$--$10^{12}$~cm. In order to have a highly significant thermal component, the blackbody radiation itself should be prominent, as well as its ratio to the non-thermal part. Another critical reason is that the observable temperature must be compatible with the satellite bandpass. For example, {Swift}-XRT observes in the $0.3$--$10$~keV photon energy band, but the hydrogen absorption affects the lower energy part ($\sim 0.5$~keV), and data is always not adequate beyond $5$~keV, due to the low effective area of satellite for high energy photons. The reliable temperature only ranges from $0.15$~keV to $1.5$~keV (since peak photon energy is equal to the temperature times 2.82), so the remaining $9$ GRBs may contain a thermal component in the flare but outside the satellite bandpass. 

We now attempt to perform a more refined analysis to infer the value of $\beta$ from the observations. We assume that during the flare the black body emitter has spherical symmetry and expands with a constant Lorentz gamma factor. Therefore, the expansion velocity $\beta$ is also constant during the flare. The relations between the comoving time $t_{com}$, the laboratory time $t$, the arrival time $t_a$, and the arrival time $t_a^d$ at the detector, given in Eq.~(\ref{tadef}), in this case become:
\begin{equation}
t_a^d = t_a(1+z) = t (1-\beta\cos\vartheta)(1+z) = \Gamma t_{com} (1-\beta\cos\vartheta)(1+z)\ .
\label{times}
\end{equation}
We can infer an effective radius $R$ of the black body emitter from: 1) the observed black body temperature $T_\mathrm{obs}$, which comes from the spectral fit of the data during the flare; 2) the observed bolometric black body flux $F_\mathrm{bb,obs}$, computed from $T_\mathrm{obs}$ and the normalization of the black body spectral fit; and 3) the cosmological redshift $z$ of the source \citep[see also][]{2012A&A...543A..10I}. We recall that $F_\mathrm{bb,obs}$ by definition is given by:
\begin{equation}
F_\mathrm{bb,obs} = \frac{L}{4\pi D_L(z)^2}\ ,
\label{fbbobs0}
\end{equation}
where $D_L(z)$ is the luminosity distance of the source, which in turn is a function of the cosmological redshift $z$, and $L$ is the source bolometric luminosity (i.e., the total emitted energy per unit time). $L$ is Lorentz invariant, so we can compute it in the co-moving frame of the emitter using the usual black body expression:
\begin{equation}
L=4\pi {R_\mathrm{com}}^2 \sigma {T_\mathrm{com}}^4\ ,
\label{lum}
\end{equation}
where $R_\mathrm{com}$ and $T_\mathrm{com}$ are the comoving radius and the comoving temperature of the emitter, respectively, and $\sigma$ is the Stefan-Boltzmann constant. We recall that $T_\mathrm{com}$ is constant over the entire shell due to our assumption of spherical symmetry. From Eq.~(\ref{fbbobs0}) and Eq.~(\ref{lum}) we then have:
\begin{equation}
F_\mathrm{bb,obs}= \frac{{R_\mathrm{com}}^2 \sigma {T_\mathrm{com}}^4}{D_L(z)^2}\ .
\label{fbbobs1}
\end{equation}

We now need the relation between $T_\mathrm{com}$ and the observed black body temperature $T_\mathrm{obs}$. Considering both the cosmological redshift and the Doppler effect due to the velocity of the emitting surface, we have:
\begin{equation}
T_\mathrm{obs} (T_\mathrm{com},z,\Gamma,\cos\vartheta)= \frac{T_\mathrm{com}}{\left(1+z\right)\Gamma\left(1-\beta\cos\vartheta\right)} = \frac{T_\mathrm{com}\mathcal{D}(\cos\vartheta)}{1+z}\, ,
\label{tdef}
\end{equation}
where we have defined the Doppler factor $\mathcal{D}(\cos\vartheta)$ as:
\begin{equation}
\mathcal{D}(\cos\vartheta)\equiv\frac{1}{\Gamma\left(1-\beta\cos\vartheta\right)}\, .
\label{defd}
\end{equation}
Eq.~(\ref{tdef}) gives us the observed black body temperature of the radiation coming from different points of the emitter surface, corresponding to different values of $\cos\vartheta$. However, since the emitter is at a cosmological distance, we are not able to resolve spatially the source with our detectors. Therefore, the temperature that we actually observe corresponds to an average of Eq.~(\ref{tdef}) computed over the emitter surface:\footnote{From the point of view of the observer the spectrum is not a perfect black body, coming from a convolution of black body spectra at different temperatures. The black body component we obtain from the spectral fit of the observed data is an effective black body of temperature $T_\mathrm{obs}$, analogously to other cases of effective temperatures in cosmology \citep[see, e.g.,][]{1983A&A...125..265R}.}
\begin{align}
\nonumber T_\mathrm{obs}(T_\mathrm{com},z,\Gamma)=&\,\displaystyle\frac{1}{1+z}\frac{\int^{1}_{\beta}{\mathcal{D}(\cos\vartheta)T_\mathrm{com}\cos\vartheta d\cos\vartheta}}{\int^{1}_{\beta}{\cos\vartheta d\cos\vartheta}} \\[6pt]
\nonumber=&\, \displaystyle\frac{2}{1+z}\frac{\beta\left(\beta-1\right)+\ln\left(1+\beta\right)}{\Gamma\beta^2\left(1-\beta^2\right)}T_\mathrm{com}\\[6pt]
\label{tobsbb}=&\,\Theta(\beta)\frac{\Gamma}{1+z}T_\mathrm{com}
\end{align}
where we defined
\begin{equation}
\Theta(\beta) \equiv 2\, \frac{\beta\left(\beta-1\right)+\ln\left(1+\beta\right)}{\beta^2}\, ,
\end{equation}
we have used the fact that due to relativistic beaming, we observe only a portion of the surface of the emitter defined by:
\begin{equation}
\beta \leq \cos\vartheta \leq 1\, ,
\label{visible}
\end{equation}
and we used the definition of $\Gamma$ given in Sec.~\ref{sec:Theory}. Therefore, inverting Eq.~(\ref{tobsbb}), the comoving black body temperature $T_\mathrm{com}$ can be computed from the observed black body temperature $T_\mathrm{obs}$, the source cosmological redshift $z$ and the emitter Lorentz gamma factor in the following way:
\begin{equation}
T_\mathrm{com} (T_\mathrm{obs},z,\Gamma) = \frac{1+z}{\Theta(\beta)\Gamma}T_\mathrm{obs}\, .
\label{tcomdef}
\end{equation}

We can now insert Eq.~(\ref{tcomdef}) into Eq.~(\ref{fbbobs1}) to obtain:
\begin{equation}
F_\mathrm{bb,obs} = \frac{{R_\mathrm{com}}^2}{D_L(z)^2} \sigma T_\mathrm{com}^4 = \frac{{R_\mathrm{com}}^2}{D_L(z)^2} \sigma \left[\frac{1+z}{\Theta(\beta)\Gamma}T_\mathrm{obs}\right]^4\, .
\label{fbbobs2}
\end{equation}
Since the radius $R_\mathrm{lab}$ of the emitter in the laboratory frame is related to $R_\mathrm{com}$ by:
\begin{equation}
R_\mathrm{com} = \Gamma R_\mathrm{lab}\, ,
\label{rcomdef}
\end{equation}
we can insert Eq.~(\ref{rcomdef}) into Eq.~(\ref{fbbobs2}) and obtain:
\begin{equation}
\label{fbbobs} F_\mathrm{bb,obs}=\frac{\left(1+z\right)^4}{\Gamma^2}\left(\frac{R_\mathrm{lab}}{D_L(z)}\right)^2\sigma \left[\frac{T_\mathrm{obs}}{\Theta(\beta)}\right]^4\ .
\end{equation}
Solving Eq.~(\ref{fbbobs}) for $R_\mathrm{lab}$ we finally obtain the thermal emitter effective radius in the laboratory frame:
\begin{equation}
\label{raggiorel} R_\mathrm{lab}=\Theta(\beta)^2\Gamma\frac{D_L(z)}{(1+z)^2}\sqrt{\frac{F_\mathrm{bb,obs}}{\sigma T_\mathrm{obs}^4}}=\Theta(\beta)^2\Gamma \phi_0\ ,
\end{equation}
where we have defined $\phi_0$:
\begin{equation}
\phi_0 \equiv \frac{D_L(z)}{(1+z)^2}\sqrt{\frac{F_\mathrm{bb,obs}}{\sigma T_\mathrm{obs}^4}}\, .
\label{rclass}
\end{equation}
In astronomy the quantity $\phi_0$ is usually identified with the radius of the emitter. However, in relativistic astrophysics this identity cannot be straightforwardly applied, because the estimate of the effective emitter radius $R_\mathrm{lab}$ in Eq.~\ref{raggiorel} crucially depends on the knowledge of its expansion velocity $\beta$ (and, correspondingly, of $\Gamma$).

It must be noted that Eq.~(\ref{raggiorel}) above gives the correct value of $R_\mathrm{lab}$ for all values of $0 \leq \beta \leq 1$ by taking all the relativistic transformations properly into account. In the non-relativistic limit ($\beta \rightarrow 0$, $\Gamma \rightarrow 1$) we have respectively:
\begin{align}
&\Theta\xrightarrow[\beta\rightarrow 0]{} 1\, , &\Theta^2\xrightarrow[\beta\rightarrow 0]{} 1\, , \\
&T_\mathrm{com}\xrightarrow[\beta\rightarrow 0]{}T_\mathrm{obs}(1+z)\, , &R_\mathrm{lab}\xrightarrow[\beta\rightarrow 0]{}\phi_0\, ,
\end{align}
as expected.

\section{Implications on the dynamics of the flares from their thermal emission} \label{sec:thermalflare2}

An estimate of the expansion velocity $\beta$ can be deduced from the ratio between the variation of the emitter effective radius $\Delta R_\mathrm{lab}$ and the emission duration in laboratory frame $\Delta t$, i.e., 
\begin{equation}
\beta=\frac{\Delta R_\mathrm{lab}}{c \Delta t}=\Theta(\beta)^2\Gamma(1-\beta\cos\vartheta)(1+z)\frac{\Delta \phi_0}{c \Delta t_a^d}\, ,
\label{beta1}
\end{equation}
where we have used Eq.~(\ref{raggiorel}) and the relation between $\Delta t$ and $\Delta t_a^d$ given in Eq.~(\ref{times}). We then have:
\begin{equation}
\beta=\Theta(\beta)^2\frac{1-\beta\cos\vartheta}{\sqrt{1-\beta^2}}(1+z)\frac{\Delta \phi_0}{c \Delta t_a^d}\, ,
\label{beta}
\end{equation}
where we used the definition of $\Gamma$ given in Sec.~\ref{sec:Theory}.

\begin{figure}
\centering
\includegraphics[width=0.8\hsize,clip]{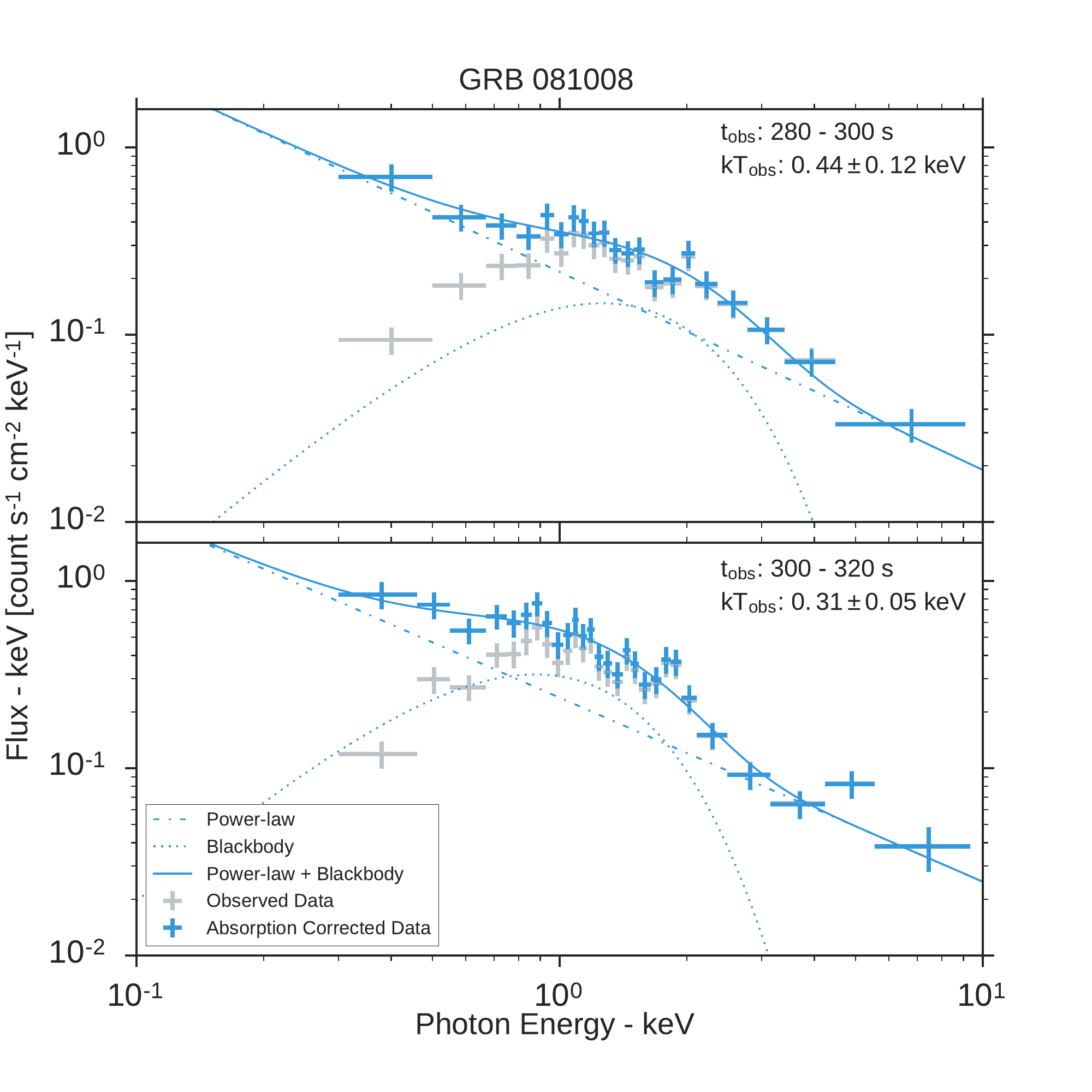}
\caption{Thermal evolution of GRB 081008 ($z=1.967$) in the observer frame. The X-ray flare of this GRB peaks at $304(\pm 17)$~s. \textbf{Upper}: Swift-{XRT} spectrum from $280$~s to $300$~s. \textbf{Lower}: Swift-\emph{XRT} spectrum from $300$~s to $320$~s. The grey points are the observed data markedly absorbed at low energies, while the blue points are absorption corrected ones. The data is fit with a combination of power-law (dot-dashed lines) and blackbody (dotted line curves) spectra. The power-law $+$ blackbody spectra are shown as solid curves. Clearly, the temperature decreases with time from $\sim 0.44$~keV to $\sim 0.31$~keV, but the ratio of thermal component goes up from $\sim 20\%$ to $\sim 30\%$. This is a remarkable high percentage among our sample.}
\label{fig:081008Spec}
\end{figure}

For example, in GRB 081008 we observe a temperature of $T_\mathrm{obs}=(0.44\pm0.12)$~keV between $t_a^d=280$ s and $t_a^d=300$ s (i.e.,~$20$~s before the flare peak time), and a temperature of $T_\mathrm{obs}=(0.31\pm0.05)$~keV between $t_a^d=300$~s and $t_a^d=320$~s (i.e., $20$~s after the flare peak time, see the corresponding spectra in Fig.~\ref{fig:081008Spec}). In these two time intervals we can infer $\phi_0$ and by solving Eq.~(\ref{beta}) and taking the errors of the parameters properly into account, get the value of $\langle\beta\rangle$ corresponding to the average expansion speed of the emitter from the beginning of its expansion up to the upper bound of the time interval considered. The results so obtained are listed in Tab.~\ref{tab:thermList}.
\begin{table}
\begin{center}
\begin{tabular}{|l|c|c|}
\hline
Time interval & $280\, \mathrm{s} \leq t_a^d \leq 300\, \mathrm{s}$ & $300\, \mathrm{s} \leq t_a^d \leq 320\, \mathrm{s}$ \\
\hline
$T_\mathrm{obs}$ [keV]& $0.44\pm0.12$ & $0.31\pm0.05$ \\[6pt]
$\phi_0$ [cm]& $(5.6\pm3.2)\times10^{11}$ & $(1.44\pm0.48)\times10^{12}$ \\[6pt]
$\langle\beta\rangle_{(\cos\vartheta=1)}$ & $0.19^{+0.10}_{-0.11}$ & $0.42^{+0.10}_{-0.12}$ \\[6pt]
$\langle\Gamma\rangle$ & $1.02^{+0.03}_{-0.02}$ & $1.10^{+0.07}_{-0.05}$ \\[6pt]
$R_\mathrm{lab}$ [cm]& $(7.1\pm4.1)\times10^{11}$ & $(2.34\pm0.78)\times10^{12}$ \\[6pt]
\hline
\end{tabular}
\end{center}
\caption{List of the physical quantities inferred from the thermal components observed during the flare of GRB 081008. For each time interval we summarize: the observed temperature $T_\mathrm{obs}$, $\phi_0$, the average expansion speed $\langle\beta\rangle$ computed from the beginning up to the upper bound of the considered time interval, and the corresponding average Lorentz factor $\langle\Gamma\rangle$ and laboratory radius $R_\mathrm{lab}$.}
\label{tab:thermList}
\end{table}
Moreover, we can also compute the value of $\langle\beta\rangle$ between the two time intervals considered above.
For $\cos\vartheta=1$, namely along the line of sight, we obtain $\langle\beta\rangle=0.90^{+0.06}_{-0.31}$ and $\langle\Gamma\rangle=2.34^{+1.29}_{-1.10}$. In conclusion, no matter what the details of the approximation adopted, the Lorentz gamma factor is always moderate, i.e., $\Gamma\lesssim 4$.

\section{The electron-positron plasma as the common origin of the prompt emission and the X-ray flares}
\label{sec:originprFPA}

\subsection{Necessity for a new hydrodynamic code for $10\leq B\leq 10^2$}

As stated above,  there are many different components in BdHNe: following episode 1 of the hypercritical accretion of the SN ejecta onto the NS, the prompt emission occurs with $\Gamma\approx 10^2$--$10^3$ which represents the most energetic component accelerated by the $e^+e^-$ plasma; a third component which encompasses the X-ray flare with $\Gamma\lesssim 4$, which represents only a fraction of $E_{e^+e^-}$ ranging from $2$ to $20$\% (see Fig.~\ref{fig:percentage}); finally, there are in addition the gamma-ray flare and the late X-ray flares which will be addressed in a forthcoming publication as well as the late afterglow phases, which have been already addressed in \citet{Pisani2013,2016ApJ...833..159P} but whose dynamics will be discussed elsewhere. As already mentioned, for definiteness we address here the case of X-ray flares.

In \ref{sec:dynamicse+e-} we have shown that our model successfully explains the entire prompt emission as originating from the transparency of an initially optically thick $e^+e^-$ plasma with a baryon load $B<10^{-2}$ reaching $\Gamma\approx10^2$--$10^3$ and the accelerated baryons interacting with the clouds of the CBM. The fundamental equations describing the dynamics of the optically thick plasma, its self-acceleration to ultra-relativistic velocities, and its interaction with the baryon load have been described in \citet{RSWX2,RSWX}. A semi-analytic approximate numerical code was developed, which assumed that the plasma expanded as a shell with a constant thickness in the laboratory frame \citep[the so called ``slab" approximation, see][]{RSWX2}. This semi-analytic approximate code was validated by comparing its results with the ones obtained by numerically integrating the complete system of equations, for selected values of the initial conditions. It turns out that the semi-analytic code is an excellent approximation to the complete system of equations for $B<10^{-2}$, which is the relevant regime for the prompt emission, but this approximation is not valid beyond this limit \citep[see][for details]{RSWX2,RSWX}.

We examine here the possibility that also the energy of the X-ray flare component originates from a fraction of the $e^+e^-$ plasma energy (see Fig.~\ref{fig:percentage}) interacting with the much denser medium of the SN ejecta with $10\lesssim B\lesssim10^2$. The above-mentioned semi-analytic approximate code cannot be used for this purpose, since it is valid only for $B<10^{-2}$, and therefore, thanks to the more powerful computers we have at present, we move on here to a new numerical code to integrate the complete system of equations.

We investigate if indeed the dynamics to be expected from an initially pure $e^+e^-$ plasma with a negligible baryon load relativistically expanding in the fireshell model, with an initial Lorentz factor $\Gamma \sim 100$, and then impacting onto such a SN ejecta can lead, reaching transparency, to the Lorentz factor $\Gamma\lesssim 4$ inferred from the thermal emission observed in the flares (see Tabs.~\ref{tab:grbTemperature} and \ref{tab:thermList}, and Fig.~\ref{fig:081008Spec}). 

We have performed hydrodynamical simulations of such a process using the one-dimensional relativistic hydrodynamical (RHD) module included in the freely available PLUTO\footnote{http://plutocode.ph.unito.it/} code \citep{PLUTO}. In the spherically symmetric case considered here, only the radial coordinate is used and the code integrates partial differential equations with two variables: radius and time. This permits the study of the evolution of the plasma along one selected radial direction at a time. The code integrates the equations of an ideal relativistic fluid in the absence of gravity, which can be written as follows:
\begin{align}
\frac{\partial(\rho \Gamma)}{\partial t} +\nabla.\left(\rho\Gamma
\mathbf{v}\right)=0, \label{consmass}\\
\frac{\partial m_r}{\partial t} +\nabla.\left(m_r
\mathbf{v}\right)+\frac{\partial p}{\partial r}=0,\label{consmomentum}\\
\frac{\partial \mathcal{E}}{\partial t}+ \nabla .\,\left(\mathbf{m}-\rho\Gamma\mathbf{v}\right)=0,\label{consenergy}
\end{align}
where $\rho$ and $p$ are, respectively, the comoving fluid density and pressure, $\mathbf{v}$ is the coordinate velocity in natural units ($c=1$), $\Gamma=(1-\mathbf{v}^2)^{-\frac{1}{2}}$ is the Lorentz gamma factor, $\mathbf{m}=h\Gamma^2\mathbf{v}$ is the fluid momentum, $m_r$ its radial component, $\mathcal{E}$ is the internal energy density, and $h$ is the comoving enthalpy density which is defined by $h=\rho+\epsilon+p$. In this last definition $\epsilon$ is equal to $\mathcal{E}$ measured in the comoving frame. We define $\mathcal{E}$ as follows:
\begin{equation}
\mathcal{E}=h\Gamma^2-p-\rho\Gamma.
\end{equation}
The first two terms on the right hand side of this equation coincide with the $T^{00}$ component of the fluid energy-momentum tensor $T^{\mu\nu}$, and the last one is the mass density in the  laboratory frame.

Under the conditions discussed in Appendix~\ref{app:gamma}, the plasma satisfies the equation of state of an ideal relativistic gas, which can be expressed in terms of its enthalpy as:
\begin{equation}
h=\rho+\frac{\gamma p}{\gamma-1},
\label{eq:eos}
\end{equation}
with $\gamma=4/3$. Fixing this equation of state completely defines the system, leaving the choice of the boundary conditions as the only remaining freedom. To compute the evolution of these quantities in the chosen setup, the code uses the Harten-Lax-van Leer-Contact Riemann solver. Time integration is performed by means of a second-order Runge-Kutta algorithm, and a second-order total variation diminishing scheme is used for spatial reconstruction \citep{PLUTO}. Before each integration step, the grid is updated according to an adaptive mesh refinement algorithm, provided by the CHOMBO library \citep{CHOMBO}.

\begin{figure}
\centering
\includegraphics[width=0.49\hsize,clip]{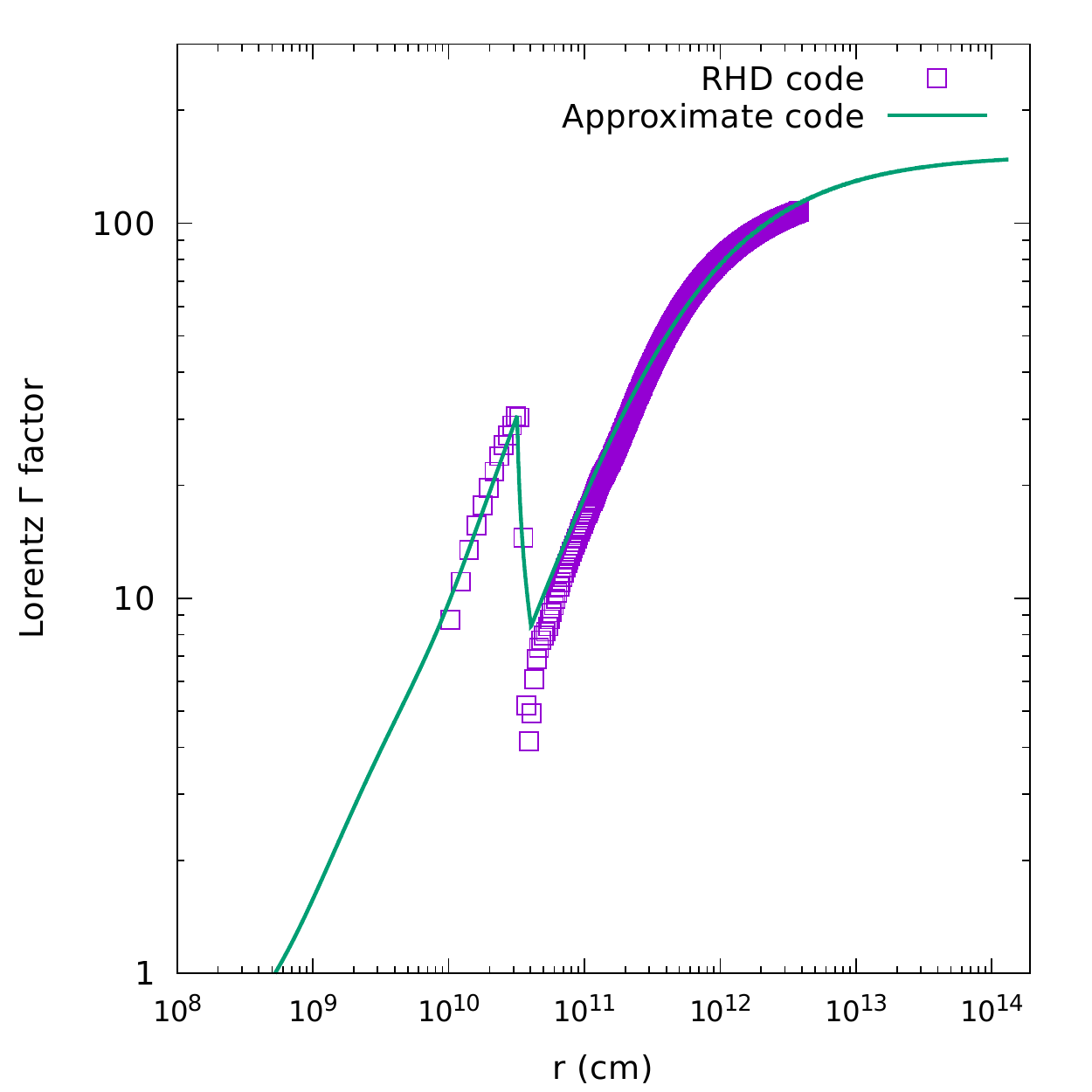}
\caption{Lorentz gamma factor computed with the new RHD code compared with the one computed with the old semi-analytic approximate code. This plot is for $E_{e^+e^-}=1.0\times 10^{53}$ erg and $B=6.61\times 10^{-3}$. Similar  agreement is found for other values of $E_{e^+e^-}$ and $B$ as long as $B<10^{-2}$.}
\label{fig:model3}
\end{figure}

It must be emphasized that the above equations are equivalent (although written in a different form) to the complete system of equations used in \citet{RSWX2,RSWX}. To validate this new numerical code, we compare its results with the ones obtained with the old semi-analytic ``slab" approximate code in the domain of its validity (i.e., for $B< 10^{-2}$), finding excellent agreement. As an example, in Fig.~\ref{fig:model3} we show the comparison between the Lorentz gamma factors computed with the two codes for one particular value of $E_{e^+e^-}$ and $B$.
We can then conclude that for $B<10^{-2}$ the new RHD code is consistent with the old semi-analytic ``slab" approximate one, which in turn is consistent with the treatment done in  \citet{RSWX2,RSWX}. This is not surprising, since we already stated that the above system of equations is equivalent to the one considered in \citet{RSWX2,RSWX}.

Having validated the new RHD code in the region of the parameter space where the old semi-analytic one can also be used, we now explore the region of $B>10^{-2}$ which is relevant for the interaction of the plasma with the SN ejecta.

\subsection{Inference from the IGC scenario for the ejecta mass profile}

We start with the shape of the SN ejecta, following the results of the numerical simulations in \citet{2016ApJ...833..107B}.

The first simulations of the IGC process were presented in \citet{2014ApJ...793L..36F} including: 1) detailed SN explosions of the CO$_{\rm core}$ obtained from a 1D core-collapse SN code code of Los Alamos \citep{1999ApJ...516..892F}; 2) the hydrodynamic details of the hypercritical accretion process; 3) the evolution of the SN ejecta material entering the Bondi-Hoyle region all the way up to its incorporation into the NS in a spherically symmetric approximation. Then in \cite{2015ApJ...812..100B} estimates were presented of the angular momentum carried by the SN ejecta and transferred to the NS via accretion. The effects of such angular momentum transfer into the evolution and fate of the system were examined there. These calculations followed the following procedure: first the accretion rate onto the NS is computed adopting an homologous expansion of the SN ejecta and introducing the pre-SN density profile of the CO$_{\rm core}$ envelope from numerical simulations. Then the angular momentum that the SN material might transfer to the NS is estimated: it turns out that the ejecta have enough angular momentum to circularize for a short time and form a disc-like structure around the NS. Then the evolution of the NS central density and rotation angular velocity is followed computing the equilibrium configurations from the numerical solution of the axisymmetric Einstein equations in full rotation, until the critical point of collapse of the NS to a BH is reached, accounting for the stability limits given by mass-shedding and the secular axisymmetric instability.
In \citet{2016ApJ...833..107B} an improved simulation of all the above processes leading to a BdHN was recently presented. In particular:
\begin{enumerate}
\item The accretion rate estimate includes effects of the finite size/thickness of the ejecta density profile.
\item Different CO$_{\rm core}$ progenitors leading to different SN ejecta masses were also considered.
\item The maximum orbital period, $P_{\rm max}$, up to which the accretion onto the NS companion is high enough to bring it to the critical mass for gravitational collapse to a BH, first estimated in \citet{2015ApJ...812..100B}, was computed for all the possible initial values of the mass of the NS companion. Various values of the angular momentum transfer efficiency parameter were also explored there.
\item It was shown there how the presence of the NS companion gives rise to large asymmetries in the SN ejecta. As we show here such a density of the SN ejecta modified by the presence of the NS companion plays a crucial role in the physical explanation for the occurrence of X-ray flares.
\item The evolution of the SN material and its consequent accretion onto the NS companion is followed via a smoothed-particle-hydrodynamic-like code in which point-like particles describe the SN ejecta. The trajectory of each particle is computed by solving the Newtonian equations of motion including the effects of the gravitational field of the NS on the SN ejecta including the orbital motion as well as the changes in the NS gravitational mass owing to the accretion process via the Bondi-Hoyle formalism. The initial conditions of the SN are obtained from the Los Alamos core-collapse SN code \citep{1999ApJ...516..892F}. The initial power-law density profile of the CO envelope is simulated by populating the inner layers with more particles. The particles crossing the Bondi-Hoyle radius are captured and accreted by the NS so we remove them from the system. We adopted a total number of 16 million particles in this simulation.
\end{enumerate}

For further details we refer the reader to \citet{2016ApJ...833..107B} and references therein.

\subsection{The density profile of the ejecta and the reaching of transparency}

We now use the results of a simulation with the following binary parameters: the NS has an initial mass of $2.0~M_\odot$; the CO$_{\rm core}$ obtained from a progenitor with a zero-age-main-sequence mass $M_{\rm ZAMS}=30~M_\odot$ leads to a total ejecta mass $7.94~M_\odot$, and follows an approximate power-law profile $\rho_{\rm ej}^0 \approx 3.1\times 10^8 (8.3\times 10^7/r)^{2.8}$~g~cm$^{-3}$. The orbital period is $P\approx 5$~min, i.e., a binary separation $a\approx 1.5\times 10^{10}$~cm. For these parameters the NS reaches the critical mass and collapses to form a BH.

\begin{figure}
\centering
\includegraphics[width=0.49\hsize,clip]{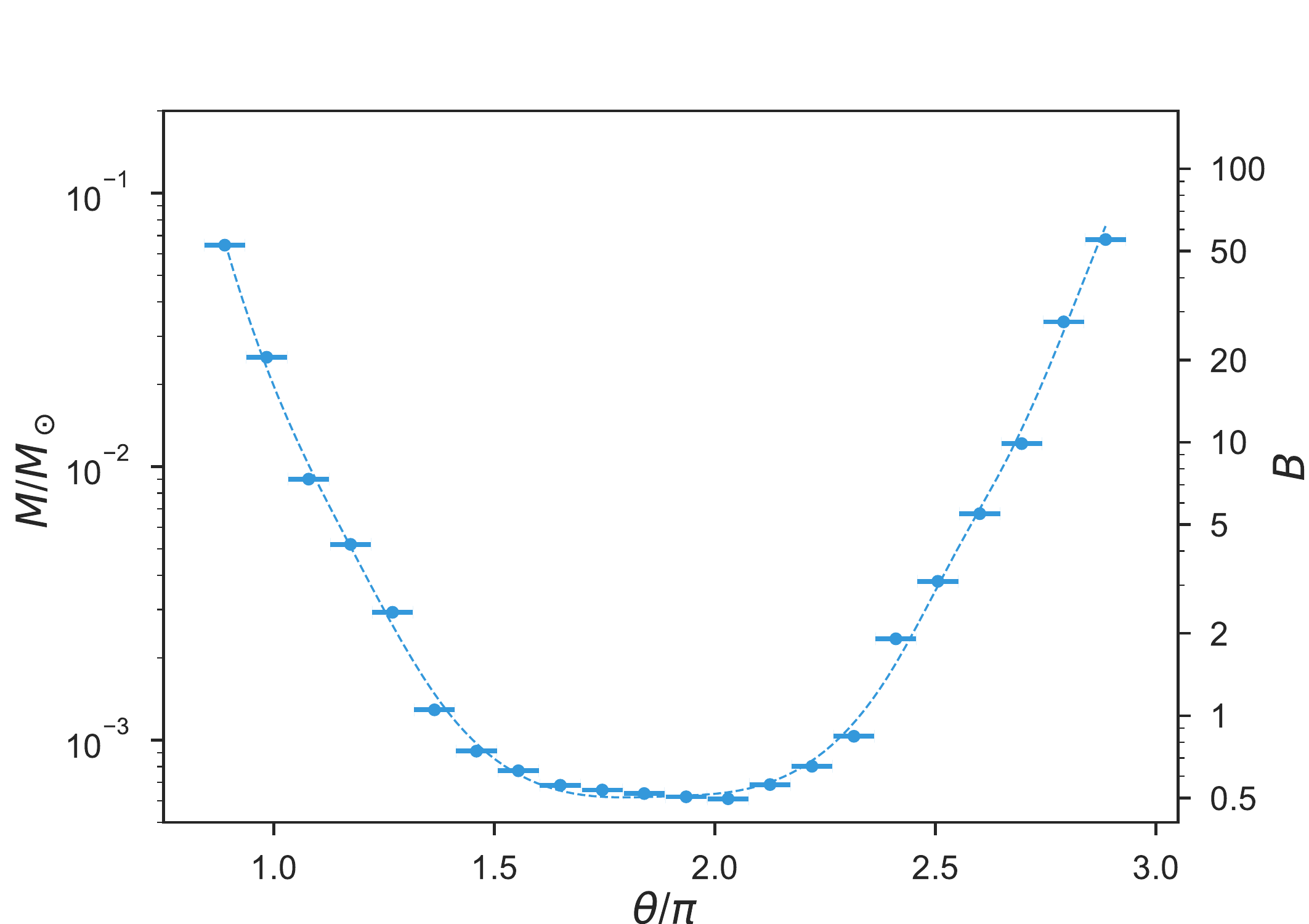}
\caption{The SN ejecta mass enclosed within a cone of 5 degrees of semi-aperture angle, whose vertex is at the position of the BH at the moment of its formation \citep[see the lower left panel of Fig.~6 in][]{2016ApJ...833..107B}, and whose axis is along various directions measured counterclockwise with respect to the line of sight. The binary parameters of this simulations are: the NS has an initial mass of $2.0~M_\odot$; the CO$_{\rm core}$ obtained from a progenitor with a zero-age-main-sequence mass $M_{\rm ZAMS}=30~M_\odot$ leads to a total ejecta mass $7.94~M_\odot$, and the orbital period is $P\approx 5$~min, i.e., a binary separation $a\approx 1.5\times 10^{10}$~cm. The vertical axis on the right side gives, as an example, the corresponding value of the baryon loading $B$ assuming a plasma energy of $E_{e^+e^-}=3.16\times10^{53}$ erg.}
\label{fig:model1}
\end{figure}

\begin{figure}
\centering
\includegraphics[width=0.49\hsize,clip]{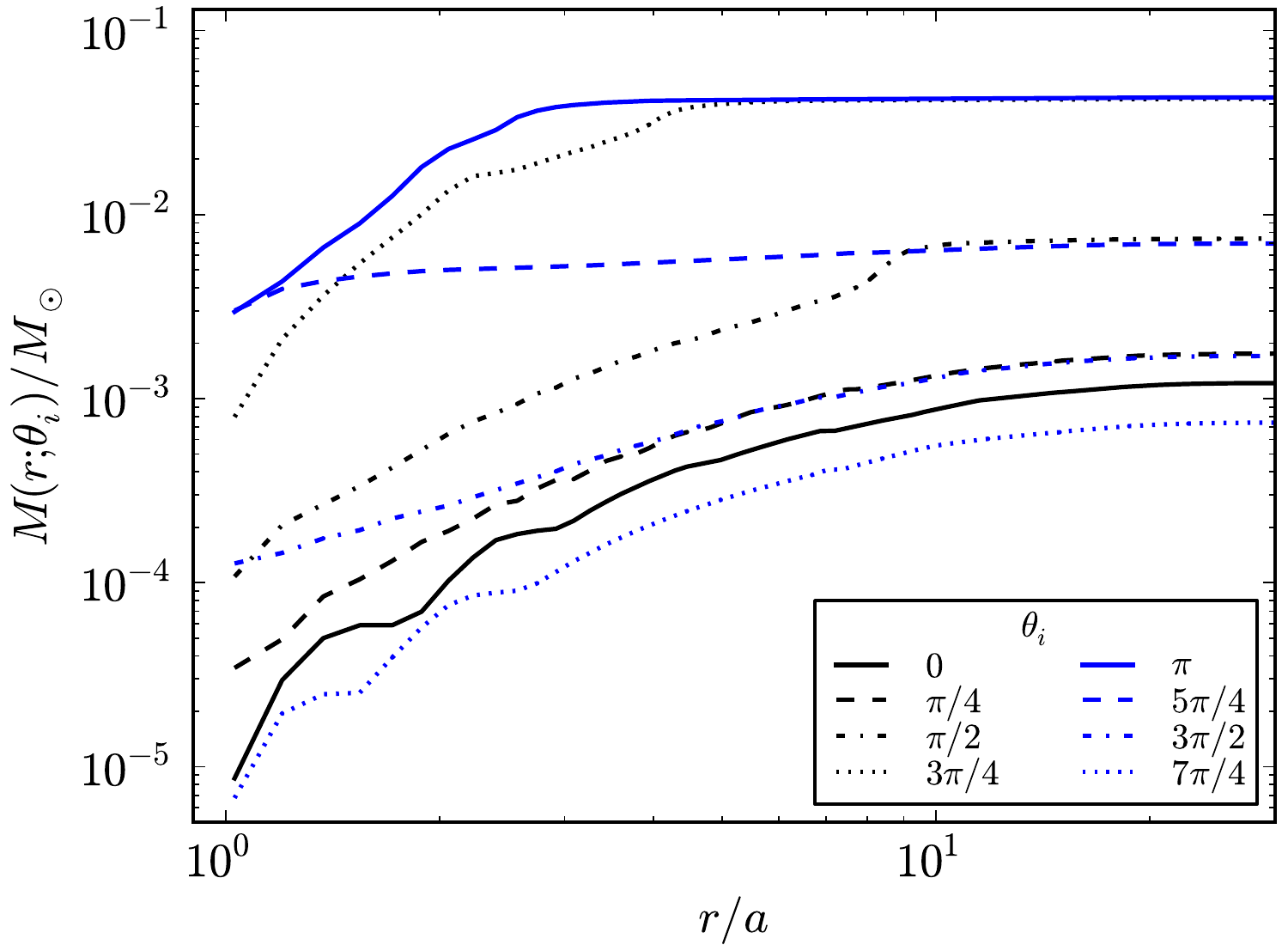}
\caption{Cumulative radial mass profiles within selected cones among the ones used in Fig.~\ref{fig:model1}. We note that the final value for the cumulative mass reached at the end of each direction, namely the value when each curve flattens, is consistent with the total integrated mass value of the corresponding direction shown in Fig.~\ref{fig:model1}. The binary parameters of these simulations are: the NS has an initial mass of $2.0~M_\odot$; the CO$_{\rm core}$ obtained from a progenitor with a zero-age-main-sequence mass $M_{\rm ZAMS}=30~M_\odot$ leads to a total ejecta mass $7.94~M_\odot$, and the orbital period is $P\approx 5$~min, i.e., a binary separation $a\approx 1.5\times 10^{10}$~cm.}
\label{fig:model2}
\end{figure}

Fig.~\ref{fig:model1} shows the SN ejecta mass that is enclosed within a cone of $5$ degrees of semi-aperture angle, whose vertex is at the position of the BH at the moment of its formation \citep[see the lower left panel of Fig.~6 in][]{2016ApJ...833..107B}, and whose axis is along various directions measured counterclockwise with respect to the line of sight. Fig.~\ref{fig:model2} shows instead the cumulative radial mass profiles within a selected number of the aforementioned cones. We can see from these plots how the $e^+e^-$ plasma engulfs different amounts of baryonic mass along different directions due to the asymmetry of the SN ejecta created by the presence of the NS binary companion and the accretion process onto it \citep[see][]{2016ApJ...833..107B}.

In these calculations, we have chosen initial conditions consistent with those of the BdHNe. At the initial time, the $e^+e^-$ plasma has $E_{e^+e^-}=3.16\times10^{53}$ erg, a negligible baryon load and is distributed homogeneously within a region of radii on the order of $10^8$--$10^9$~cm. The surrounding SN ejecta, whose pressure has been assumed to be negligible, has a mass density radial profile given by:
\begin{equation}
\rho\propto(R_0-r)^\alpha\, ,
\label{eq.profilo}
\end{equation}
where the parameters $R_0$ and $\alpha$, with $2<\alpha<3$, as well as the normalization constant, are chosen to fit the profiles obtained in \citet{2016ApJ...833..107B} and represented in Fig.~\ref{fig:model2}. The initial radial velocity is taken to be $v_r\propto r$ in order to reproduce the homologous expansion of the SN ejecta before its interaction with the plasma. Every choice of these parameters corresponds to studying the evolution along a single given direction. 

\begin{figure}
\centering
\includegraphics[width=0.49\hsize,clip]{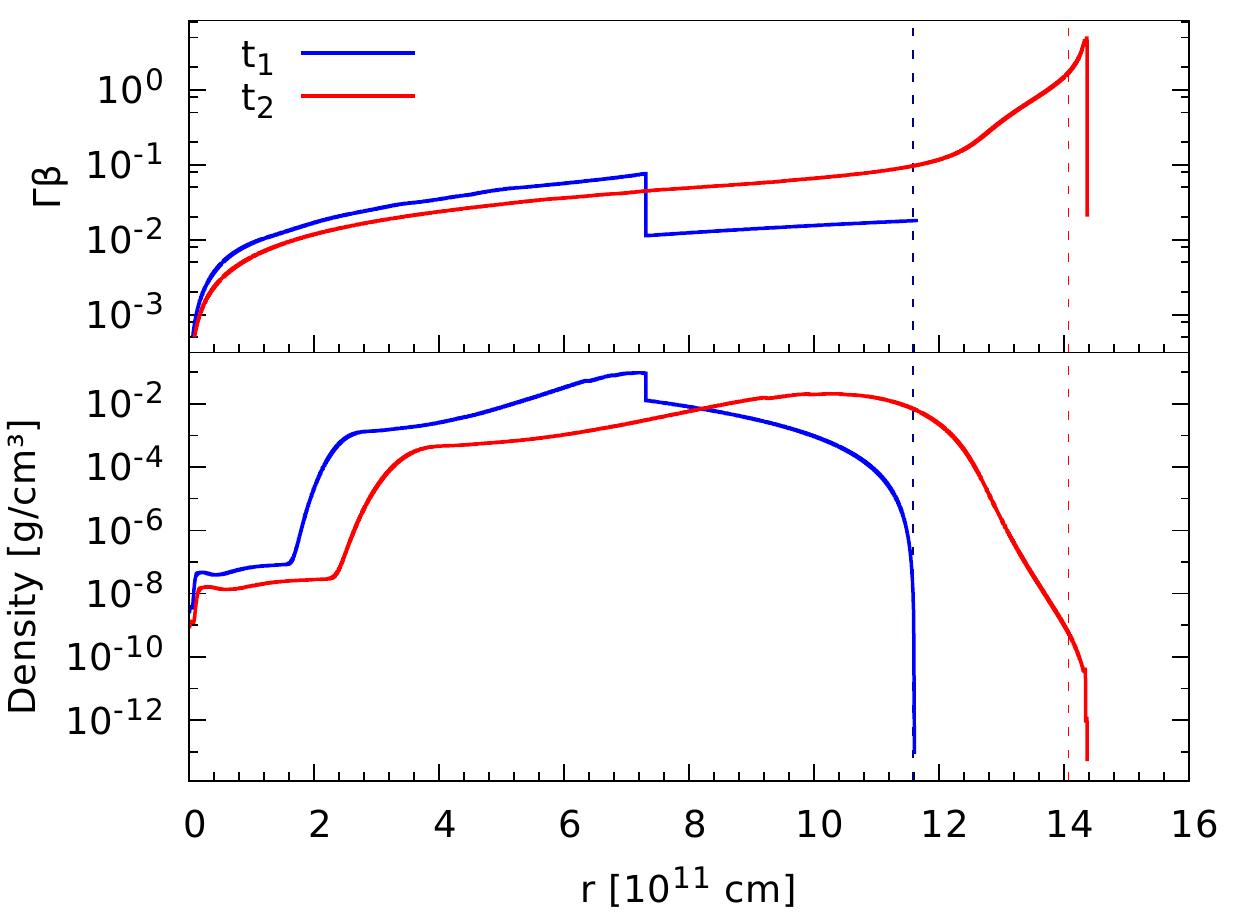}
\caption{\textbf{Above:} Distribution of the velocity inside the SN ejecta at the two fixed values of the laboratory time $t_1$ (before the plasma reaches the external surface of the ejecta) and $t_2$ (the moment at which the plasma, after having crossed the entire SN ejecta, reaches the external surface). We plotted the quantity $\Gamma\beta$, recalling that we have $\Gamma\beta \sim \beta$ when $\beta < 1$ and $\Gamma\beta \sim \Gamma$ when $\beta \sim 1$. \textbf{Below:} Corresponding distribution of the mass density of the SN ejecta in the laboratory frame $\rho_{lab}$. These particular profiles are made using a baryon load $B=200$. The dashed vertical lines corresponds to the two values of the transparency radius $R_{ph}$, see Fig.~\ref{fig:model4} and Eq.(\ref{eqrph}). In particular, we see that at $t_1$ the shock front did not reach $R_{ph}$ yet and the system is optically thick.}
\label{fig:model5}
\end{figure}

The evolution from these initial conditions leads to the formation of a shock and to its subsequent expansion until reaching the outermost part of the SN. In Fig.~\ref{fig:model5} we show the radial distribution profiles of the velocity and mass density $\rho_{lab}$ in the laboratory frame inside the SN ejecta as a function of $r$ for $B=200$ at two selected values of the laboratory time. The velocity distribution peaks at the shock front (with a Lorentz gamma factor $\Gamma\lesssim 4$), and behind the front it is formed a broad tail of accelerated material with $0.1 \lesssim \beta \lesssim 1$.

\begin{figure}
\centering
\includegraphics[width=0.49\hsize,clip]{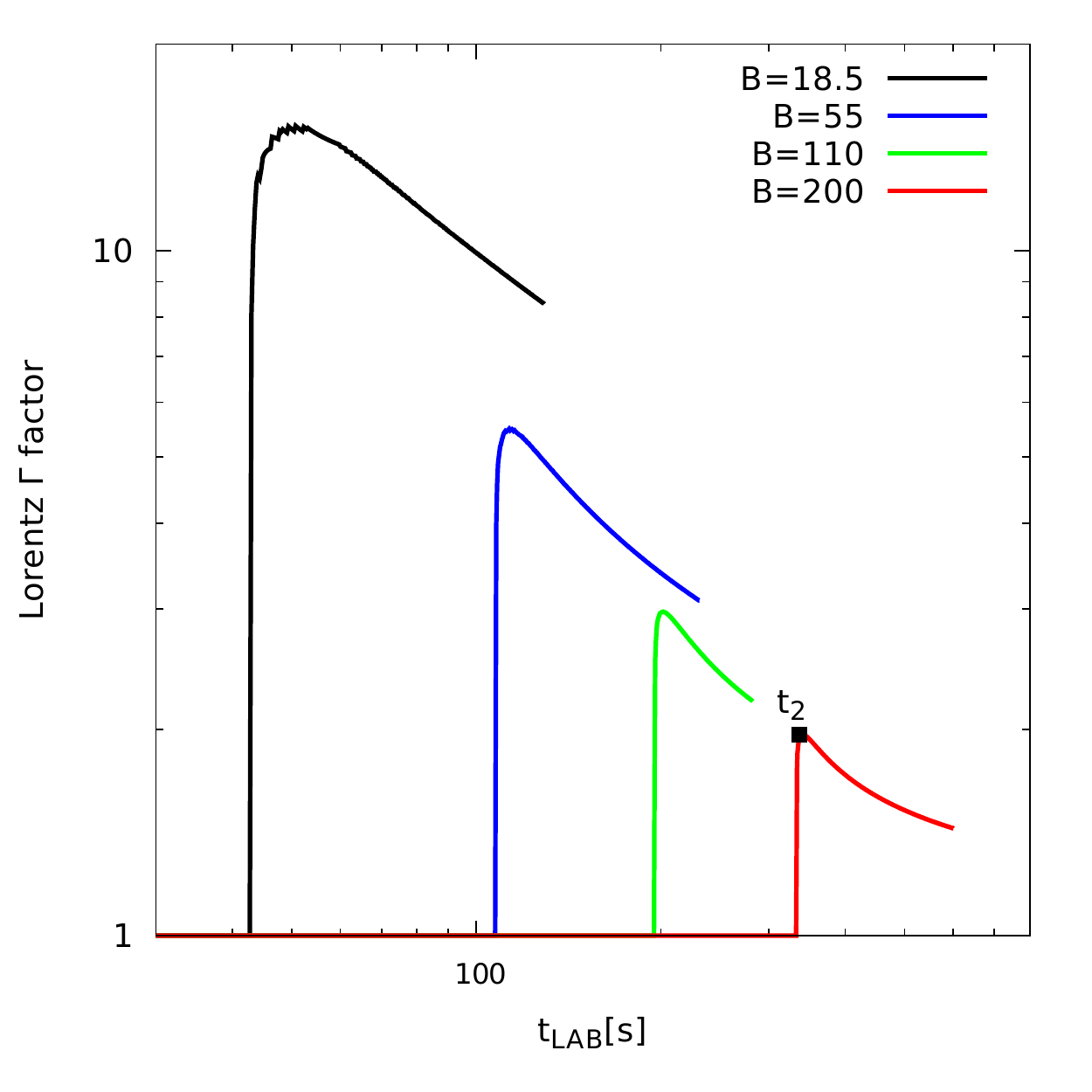}
\caption{Lorentz $\Gamma$ factor at the transparency radius $R_{ph}$ as a function of the laboratory time for $E_{e^+e^-}=3.16\times10^{53}$ erg and various selected values of the $B$ parameter. Such $B$ values correspond to the expansion of the $e^+e^-$ plasma along various selected directions inside the remnant (see Figs.~\ref{fig:model1} and \ref{fig:model2}). Along the red curve, corresponding to $B=200$, there is marked the laboratory time instant $t_2$ represented in Fig.~\ref{fig:model5} (at $t_1$ the plasma did not reach $R_{ph}$ yet). We see that these results are in agreement with the Lorentz gamma factor $\Gamma\lesssim 4$ inferred from the thermal emission observed in the flare (see Sec.~\ref{sec:thermalflare2}).}
\label{fig:model4}
\end{figure}

Fig.~\ref{fig:model4} shows the Lorentz $\Gamma$ factor at the transparency radius $R_{ph}$, namely the radius at which the optical depth $\tau$, calculated from the observer's line of sight, is equal to $1$. If we assume to have a constant cross section, $\tau$ becomes Lorentz invariant, and therefore we can compute it in laboratory coordinates in the following way:
\begin{equation}
\tau=\int_{R_{ph}}^\infty\mathrm{d}r \,\sigma_T \,n_{e^-}(r),
\label{eqrph}
\end{equation}
where $\sigma_T=6.65\times10^{-25}$ cm$^2$ is the Thomson cross section, and the electron density is related to the baryon mass density by means of the formula $n_{e^-}=\rho\,\Gamma/m_P$, where $m_P$ is the proton mass, the mass of the electrons and positrons is considered to be negligible with respect to that of the baryons, and we have assumed to have one electron per nucleon on average. The values of $\Gamma$ at $r=R_{ph}$ computed in this way are shown in Fig.~\ref{fig:model4}, as a function of the time measured in the laboratory frame, for several values of $B>10^{-2}$ corresponding to the expansion of the $e^+e^-$ plasma along several different directions inside the SN ejecta (see Figs.~\ref{fig:model1} and \ref{fig:model2}).

We conclude that the relativistic expansion of an initially pure $e^+e^-$ plasma (see Fig.~\ref{fig:model3}), interacting with a SN ejecta with the above-described induced asymmetries (see Figs.~\ref{fig:Carlo}--\ref{fig:Carlo2}), leads to the formation of a shock that reaches the outermost part of the ejecta with Lorentz gamma factors at the transparency radius $\Gamma(R_{ph})\lesssim 4$. This is in striking agreement with the one inferred from the thermal component observed in the flares (see Sec.~\ref{sec:thermalflare2}). The space-time diagram of the global scenario is represented in Fig.~\ref{fig:Carlo}. Clearly in this approach neither ultra-relativistic jetted emission nor synchrotron or inverse Compton processes play any role.

\section{Summary, Conclusions and Perspectives}
\label{sec:conclusions}

\subsection{Summary}

In the last twenty five years the number of observed GRBs has exponentially increased thanks to unprecedented technological developments in all ranges of wavelengths, going from the X-ray, to the gamma ray to the GeV radiation, as well as to the radio and the optical. In spite of this progress, the traditional GRB approach has continued to follow the paradigm of a single system \citep[the ``collapsar'' paradigm, see][]{1993ApJ...405..273W}, where accretion into an already formed BH occurs \citep[see, e.g.,][and references therein]{2004RvMP...76.1143P}. Following the fireball model, synchrotron and inverse Compton emission processes, related to an ultra-relativistic jetted emission described by the \citet{1976PhFl...19.1130B} solution, have been assumed to occur \citep[see, e.g.,][for one of the latest example where this approach is further extended to the GeV emission component]{2015ApJ...803...10T}. The quest for a ``standard'' GRB model has been pursued even lately \citep[see, e.g.,][]{2007ApJ...671.1903C,2010MNRAS.406.2149M} neglecting differences among GRB subclasses and/or neglecting all relativistic corrections in the time parameterizations presented in Sec.~\ref{sec:Theory}. Under these conditions, it is not surprising that the correlations we have found here have been missed.

It is appropriate to recall that a ``standard'' GRB energy of $10^{51}$~erg \citep{2001ApJ...562L..55F} was considered, assuming the collimation of GRBs and the existence of a light-curve break in the GRB afterglows. This possibility followed from the traditional approach expecting the ultra-relativistic component to extend all the way from the prompt emission to the last phases of the afterglow \citep{1994ApJ...424L.131M,1999ApJ...519L..17S,1999ApJ...526..707P}. This ``traditional" approach to GRBs has appeared in a  large number of papers over recent decades and is well summarized in a series of review papers \citep[see, e.g.,][]{Piran1999,2004RvMP...76.1143P,Meszaros2002,Meszaros2006,2014ARA&A..52...43B,2015PhR...561....1K}, which are disproved by the data presented here in which the upper limit for the Lorentz factor $\Gamma\lesssim 4$ is established in the FPA phase.

Since 2001 we have followed an alternative approach introducing three paradigms: the space-time parametrization of GRBs \citep{Ruffini2001a}, the field equations of the prompt emission phase \citep{2002ApJ...581L..19R}, and the IGC paradigm \citep{Rueda2012,2013A&A...551A.133P,2015ApJ...798...10R}, see Sec.~\ref{sec:Theory}. Since then:

a) we demonstrated  that all GRBs originate in binary systems: the short GRBs in binary NSs or in binaries composed of a NS and a BH \citep{2015PhRvL.115w1102F,2016ApJ...832..136R}; the long GRBs in binary systems composed of CO$_{\rm core}$ and a NS, or alternatively a BH and a CO$_{\rm core}$, or also a white dwarf and a NS;

b) we distinguish GRBs into seven different subclasses \citep{2016ApJ...832..136R}, each characterized by specific signatures in their spectra and luminosities in the various energy bands;

c) we address the new physical and astrophysical processes in the  ultra-relativistic regimes made possible by the vast amount of gravitational and rotational energies in such binaries.

As recalled in Secs.~\ref{sec:intro}--\ref{sec:Theory}, we have confirmed the binary nature of the GRB progenitors \citep[see, e.g.,][]{2014ApJ...793L..36F,2015ApJ...812..100B,2015PhRvL.115w1102F,2016ApJ...833..107B,2016ApJ...831..178R,2017ApJ...844...83A}. We have obtained the first evidence of the formation of a BH in the hypercritical accretion process of the SN ejecta onto the binary NS companion: the BdHN  \citep{2014A&A...565L..10R, 2015ApJ...798...10R,2016ApJ...832..136R}, clearly different from the single star collapsar model. Finally, in this paper we have addressed the interaction which occurs in a BdHN of the GRB on the SN ejecta considered as the origin of the X-ray flares. We use this process and the mildly-relativistic region in which it occurs as a discriminant between the traditional approach and our binary system approach: we use the X-ray flare properties as a discriminant between our BdHN and the ``fireball" GRB models.

\subsection{Conclusions}

We have reached three major results.

1) We have searched X-ray flares in all GRBs and identified 16 of them with excellent data. After examining the seven GRB subclasses
\citep{2016ApJ...832..136R}, we conclude that they all occur in BdHNe and no X-ray flares are observed in other GRB sources. This indicates a link between the occurrence of the flare and the formation of a Black Hole in long GRBs. In Sec.~\ref{sec:sample} we have shown how the previously proposed association of X-ray flares with short GRBs 050724 and 050709 has been superseded.

By a statistical analysis we correlate the time of occurrence of their peak luminosity in the cosmological  rest frame, their duration, their energy, their X-ray luminosity to the corresponding GRB $E_{iso}$. We also correlate the energy of the FPA phase, $E_{FPA}$, as well as the relative ratio $E_{FPA}/E_{iso}$, to the $E_{iso}$.

2) Using the data from the associated thermal emission, the relativistic relation between the co-moving time, the arrival time at the detector, the cosmological and Doppler corrections, we determine the thermal emitter effective radii as a function of the rest frame time. We determine the expansion velocity of the emitter $\beta$ as the ratio between the variation of the emitter effective radius $\Delta R_{\rm lab}$ and the emission duration in the Laboratory time; see Eq.~(\ref{beta}). We obtain a radius $10^{12}$~cm for the effective radius of the emitter, moving with $\Gamma\lesssim 4$ at a time $\sim 100$~s in the rest frame (see Tab.~\ref{tab:thermList}). These results show the clear rupture between the processes in the prompt emission phase, occurring prior to the flares at radii of the order of $10^{16}$~cm and $\Gamma = 10^2$--$10^3$, and the ones of the X-ray flares.

3) We have modeled the X-ray flares by considering the impact of the GRB on the SN ejecta introducing a new set of relativistic hydrodynamic equations for the expansion of the optically thick $e^+e^-$ plasma into a medium with baryon load in the range $10$--$10^2$. The matter density and velocity profiles of the ejecta are obtained from the 1D core-collapse code developed at Los Alamos \citep{1999ApJ...516..892F}. With this we generate initial conditions for our smoothed-particle-hydrodynamics-like simulation \citep{2016ApJ...833..107B} which follows the evolution of the ejecta matter and the accretion rate at the position of the Bondi-Hoyle surface of the NS binary companion. In our simulations we have adopted 16 million particles (see Section~\ref{sec:originprFPA} for further details). We start the simulation of the interaction of the $e^+e^-$ plasma with such ejecta at $10^{10}$~cm and continue all the way to $10^{12}$~cm where transparency is reached. We found full agreement between the radius of the emitter at transparency and the one derived from the observations, as well as between the time of the peak energy emission and the observed time of arrival of the flare, derived following Eq.~(\ref{tadef}) using the computed Lorentz $\Gamma$ factor of the worldline of the process.

We can now conclude that:

The existence of such mildly relativistic Lorentz gamma factors in the FPA phase rules out the traditional GRB model, including  the claims of the existence of GRB beaming, collimation and break in the luminosity \citep[see, e.g.,][]{Piran1999,2001ApJ...562L..55F,Meszaros2002,2004RvMP...76.1143P,Meszaros2006,2014ARA&A..52...43B,2015PhR...561....1K}. In these models the common underlying assumption is the existence of a single ultra-relativistic component extending from the prompt radiation, through the FPA phase, all the way to the late afterglow and to the GeV emission assuming a common dynamics solely described by the \citet{1976PhFl...19.1130B} solution, see however \citet{2005ApJ...633L..13B,2006ApJ...644L.105B}. These assumptions were made without ever looking for observational support. It is not surprising that all GRB models in the current literature purport the existence of an ultra-relativistic Lorentz gamma factor extending into the afterglow, among many others, see e.g., \citet{2010ApJ...724..861J,2015ApJ...807...92Y}. All these claims have been disproven by the present article, where a drastic break from the ultra-relativistic physics with $\Gamma\sim 10^2$--$10^3$, occurring in the prompt emission, is indicated already at times $\sim 100$~s, when the Lorentz gamma factor is limited to $\Gamma\leq 4$.

In our approach a multi-episode structure for each GRB is necessary. Each episode, being characterized by a different physical process, leads to a different worldline with a specific Lorentz gamma factor at each event. The knowledge of the worldline is essential, following Eq.~(\ref{tadef}) in Sec.~\ref{sec:Theory}, to compute the arrival time of the signals in the observer frame and compare with the observations. This procedure, previously routinely adopted in the prompt emission phase of BdHN, has for the first time been introduced here for the X-ray flares. As a byproduct we have confirmed both the binarity and the nature of the progenitors of the BdHNe, composed of a CO$_{\rm core}$ undergoing a SN explosion and accreting onto a close-by binary NS and impact of the GRB on the hypernova ejecta.

\subsection{Perspectives}

Far from representing solely a criticism of the traditional approach, in this paper: 1) we exemplify new procedures in data analysis, see Sec.~\ref{sec:sample} to Sec.~\ref{sec:partition}, 2) we open up the topic to an alternative style of conceptual analysis which adopts procedures well tested in high energy physics and not yet appreciated in the astrophysical community, see Sec.~\ref{sec:thermalflare} to Sec.~\ref{sec:originprFPA}, and 3) we introduce new tools for simulation techniques affordable with present day large computer facilities, see figures in Sec.~\ref{sec:conclusions}, which, if properly guided by a correct theoretical understanding, can be particularly helpful in the visualization of these phenomena.

We give three specific examples of our new approach and indicate as well, when necessary, some disagreements with current approaches:

\begin{itemize}
\item[A)]
The first step in any research on GRBs is to represent the  histogram of the $T_{90}$ for the GRB subclasses. We report in Fig.~\ref{fig:t90Histogram} the $T_{90}$ values for all the GRB subclasses we have introduced \citep[see][]{2016ApJ...832..136R}. The values reported are both in the observer frame \citep[left panel; see, e.g.,][]{Koveliotou1993,2013ApJ...764..179B} and properly converted to the cosmological rest frame of the sources (right panel). The large majority of papers on GRBs have been neglecting the cosmological corrections and subdivision in the subclasses, making impossible the comparison of the $T_{90}$ among different GRBs \citep[see e.g.][]{2007ApJ...671.1921F,2010MNRAS.406.2113C}.

\begin{figure}
\centering
\includegraphics[width=0.48\hsize,clip]{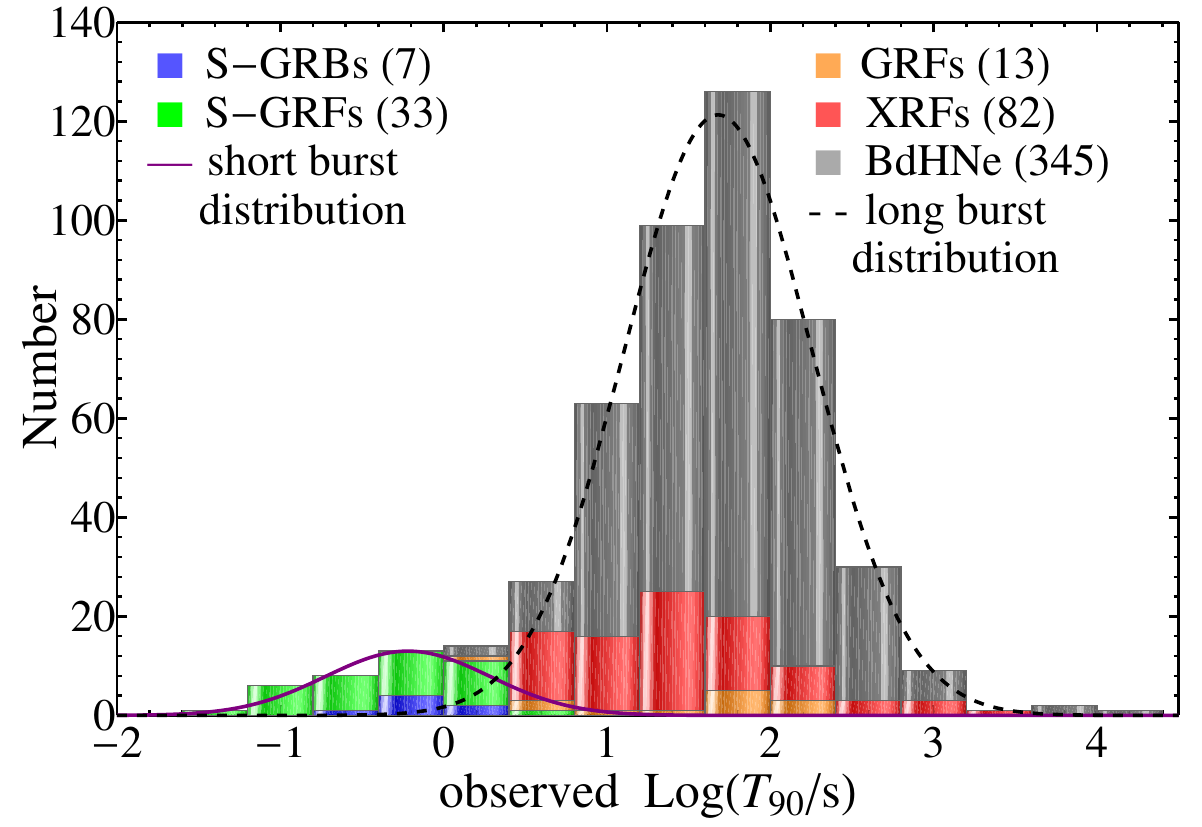}
\includegraphics[width=0.48\hsize,clip]{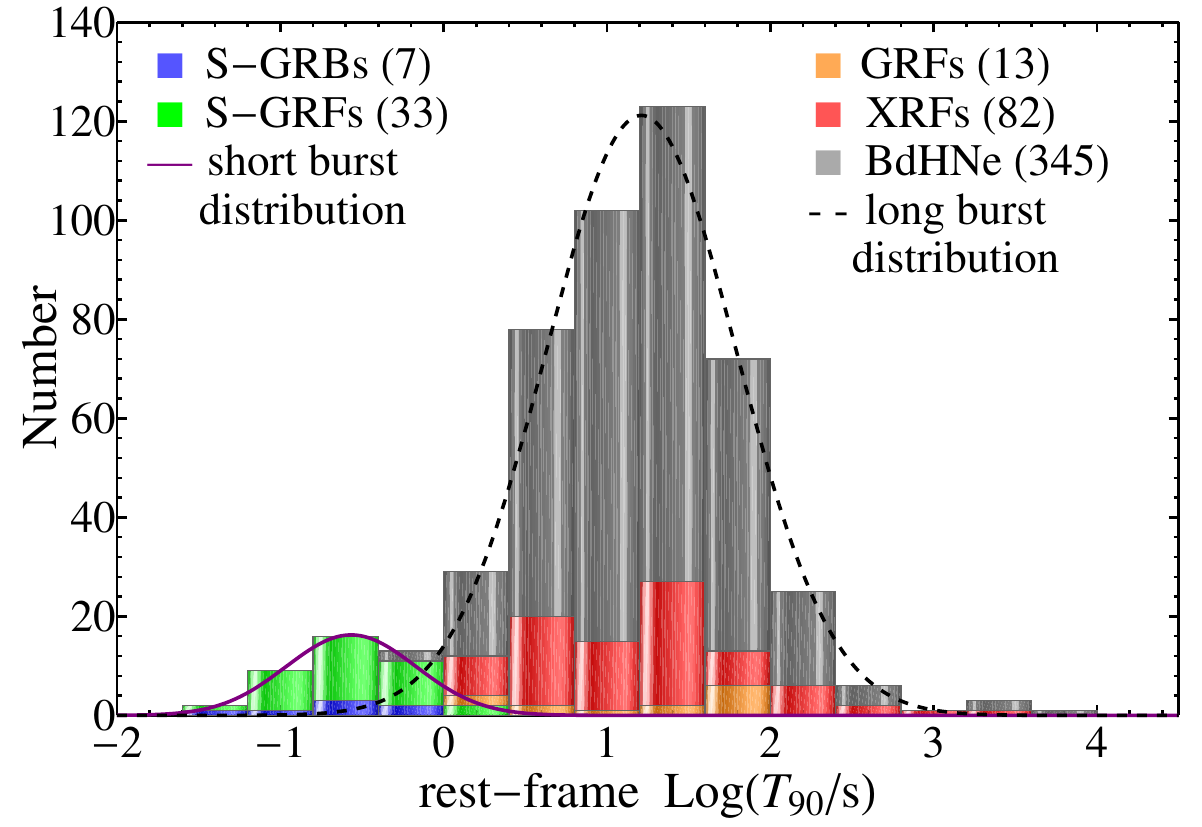}
\caption{Histograms of $T_{90}$ distributions in the observer frame \citep[left panel, which is the traditional treatment widely adopted in many previous articles, see e.g.][and references therein]{Koveliotou1993,2013ApJ...764..179B} and in the cosmological rest frame (right panel, which is the approach adopted in the present paper). Both histograms are built using the total number of GRBs with known redshift. The contribution to the total distributions and the number of sources of each subclass are highlighted in the legend (the choice of the colors is the same as in Fig.~\ref{fig:EpEiso}). The short burst (solid purple curve) and the long burst (dashed black curve) distributions are also shown.
In the observer frame we obtain $T_{90}^{\rm short}=0.60^{+1.31}_{-0.41}$~s and $T_{90}^{\rm long}=48^{+133}_{-35}$~s; in the cosmological rest frame we have $T_{90}^{\rm short}=0.27^{+0.41}_{-0.16}$~s and $T_{90}^{\rm long}=16^{+46}_{-12}$~s.
The $T_{90}$ value discriminating between short and long bursts shifts from $\approx2$~s in the observer frame, to $\approx0.75$~s in the cosmological rest frame.
The existence of BdHNe with $T_{90} \gtrsim 10^2$~s indicates the origin of the possible contamination between the prompt emission spikes and the X-ray flares, which is indeed observed in some cases (see Sec.~\ref{sec:sample} for details).}
\label{fig:t90Histogram}
\end{figure}

\item[B)]
For the first time, we present a simplified space-time diagram of the BdHNe (see Fig.~\ref{fig:Carlo}). This space-time diagram emphasizes the many different emission episodes, each one with distinct corresponding Lorentz gamma factors and consequently leading through Eq.~(\ref{tadef}) to a specific value of their distinct times of occurrence in the cosmological rest frame of the GRB (see Fig.~\ref{fig:Carlo}). In all Episodes we analyzed for the X-ray flares, and more generally for the entire FPA phase, there is no need for collapsar related concepts. Nevertheless, in view of the richness of the new scenario in Fig.~\ref{fig:Carlo}, we have been examining the possibility that such concepts can play a role in additional Episodes, either in BdHNe or in any of the additional six GRB sub-classes, e.g. in S-GRBs. These results are being submitted for publication. The use of space-time diagrams in the description of GRBs is indeed essential in order to illustrate the causal relation between the source in each episode, the place of occurrence and the time at detection. Those procedures have been introduced long ago in the study of high energy particle physics processes and codified in textbooks. Our group, since the basic papers \citep{Ruffini2001a,Ruffini2001b,Ruffini2001c} has widely shared these space-time formulations \citep[see e.g. in][]{1992spai.book.....T} and also extended the concept of the quantum S-Matrix \citep{1943ZPhy..120..513H,1937PhRv...52.1107W} to the classic astrophysical regime of the many components of a BdHN introducing the concept of the cosmic matrix \citep{2015ApJ...798...10R}. The majority of astrophysicists today make wide use of the results of nuclear physics in the study of stellar evolution \citep{1991rla..book.....B} and of Fermi statistics also in general relativity \citep{1939PhRv...55..374O}. They have not yet been ready, however, to approach these additional concepts more typical of relativistic astrophysics and relativistic field theories which are necessary for the study of GRBs and active galactic nuclei.

\begin{figure}
\centering
\includegraphics[width=0.64\hsize,clip]{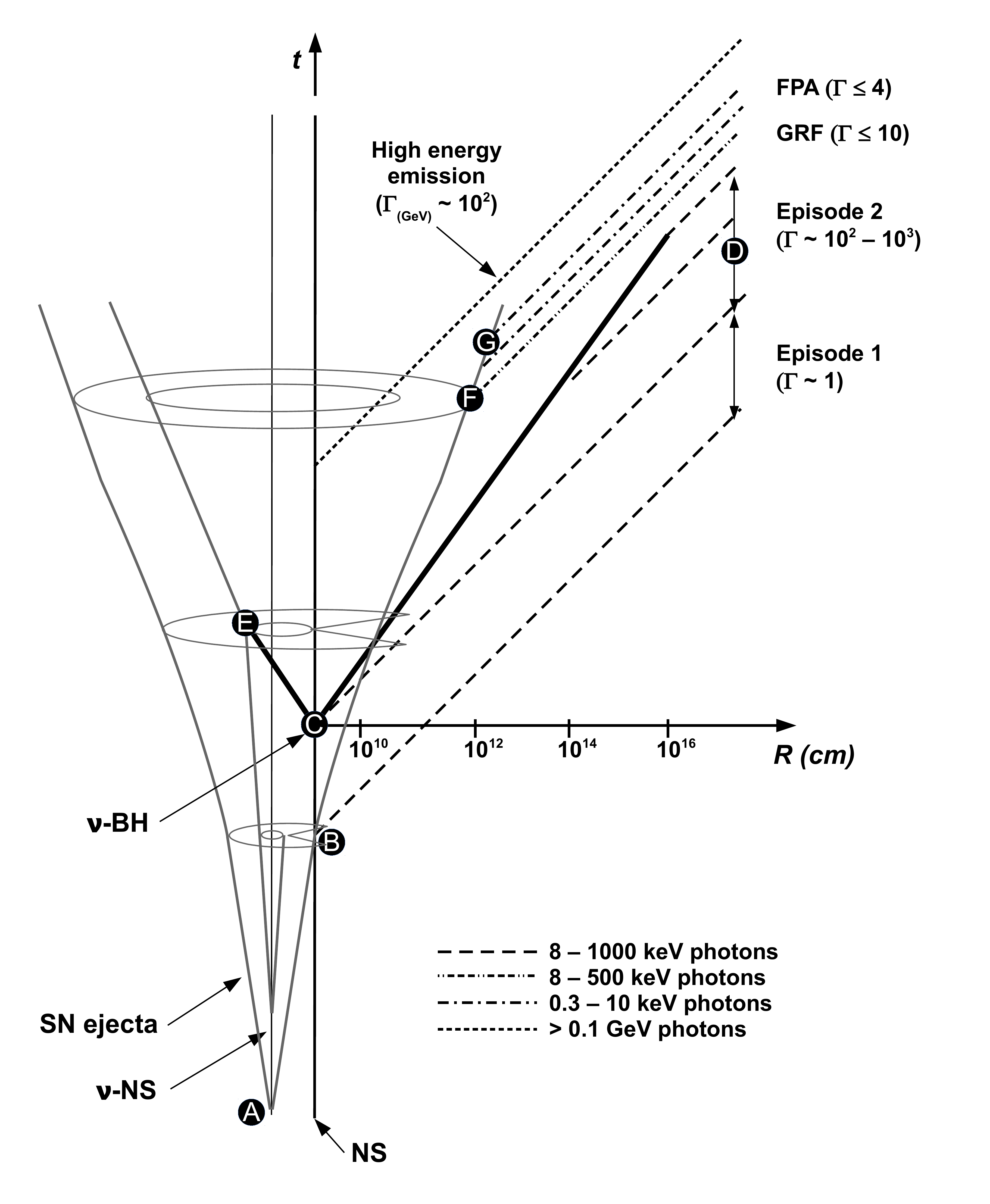}
\caption{Space-time diagram (not in scale) of BdHNe. The CO$_\mathrm{core}$ explodes as a SN at point A and forms a new NS ($\nu$NS). The companion NS (bottom right line) accretes the SN ejecta starting from point B, giving rise to the non-relativistic Episode 1 emission (with Lorentz factor $\Gamma\approx 1$). At the point C the NS companion collapses into a BH, and an $e^+e^-$ plasma --- the dyadosphere --- is formed \citep{RSWX2}. The following self-acceleration process occurs in a spherically symmetric manner (thick black lines). A large portion of plasma propagates in the direction of the line of sight, where the environment is cleared by the previous accretion into the NS companion, finding a baryon load $B \lesssim 10^{-2}$ and leading to the GRB prompt gamma-ray spikes (GRS, Episode 2, point D) with $\Gamma \sim 10^2$--$10^3$. The remaining part of the plasma impacts with the high density portion  of the SN ejecta (point E), propagates inside the ejecta encountering a baryon load $B \sim 10^{1}-10^2$, and finally reaches transparency, leading to the gamma-ray flare emission (point F) in gamma rays with an effective Lorentz factor  $\Gamma \lesssim 10$ and to the FPA emission (point G) corresponding to the X-ray Flares  with an effective $\Gamma \lesssim 4$ (see Secs.~\ref{sec:thermalflare2} and \ref{sec:originprFPA}). In the meantime, accretion over the newly formed BH produces the high energy GeV emission with $\Gamma \sim 10^2$. For simplicity, this diagram is 2D and static and does not attempt to show the 3D rotation of the ejecta.}
\label{fig:Carlo}
\end{figure}

\item[C)]
The visual representation of our result (see Fig.~\ref{fig:Carlo2}) have been made possible thanks to the simulations of SN explosions with the core-collapse SN code developed at Los Alamos \citep[see e.g.][]{2041-8205-773-1-L7,1999ApJ...516..892F,2014ApJ...793L..36F}, the smoothed-particle-hydrodynamics-like simulations of the evolution of the SN ejecta accounting for the presence of a NS companion \citep{2016ApJ...832..136R}, and the possibility to vary the parameters of the NS, of the SN, and of the distance between the two to explore all possibilities \citep{2015ApJ...812..100B,2016ApJ...832..136R}. We recall that these signals occur in each galaxy every $\sim$ hundred million years, but with their luminosity $\sim 10^{54}$~erg they can be detected in all $10^9$ galaxies. The product of these two factors give the ``once per day'' rate. They are not visualizable in any other way, but analyzing the spectra and time of arrival of the photons now, and simulating these data on the computer: they indeed already occurred billions of years ago in our past light cone, and they are revived by scientific procedures today.

\end{itemize}

\begin{figure}
\centering
\includegraphics[width=\hsize,clip]{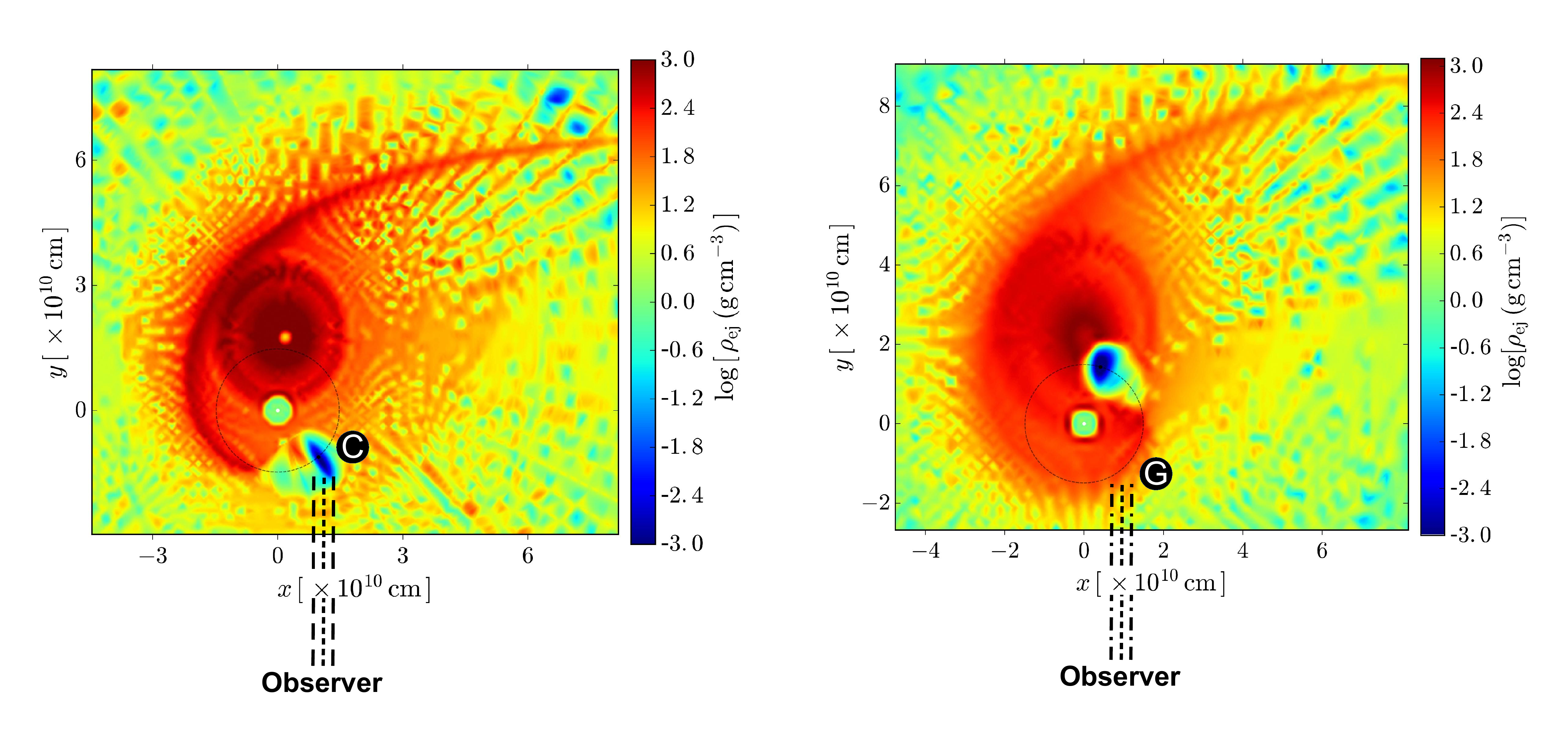}
\caption{Two snapshots of the distribution of matter in the equatorial plane of the progenitors binary system. The one on the right side corresponds to point C, when the BH is formed and a large portion of the $e^+e^-$ plasma starts to self-accelerate in a low density environment ($B\lesssim 10^{-2}$) toward the observer producing the GRB prompt emission. The one on the left side corresponds to point G, when the remaining part of the plasma, after the propagation inside the high density SN ejecta ($B \sim 10^{2}$--$10^3$), reaches transparency and produces the FPA emission in the X-rays which is directed toward the observer due to the rotation of the ejecta in the equatorial plane. The simulations of the matter distributions in the three snapshots are from \citet{2016ApJ...833..107B}.}
\label{fig:Carlo2}
\end{figure}

\acknowledgments
We are happy to acknowledge fruitful discussions with Fulvio Melia, Tsvi Piran and Bing Zhang, in our attempt to make these new results clearer for a broader audience, and with Roy Kerr on the implication of GRB GeV emission for the Kerr-Newman solution. We also thank the second referee for important suggestions. We acknowledge the continuous support of the MAECI. This work made use of data supplied by the UK Swift Science Data Center at the University of Leicester. M.~Ka. and Y.A. are supported by the Erasmus Mundus Joint Doctorate Program Grant N. 2014--0707 from EACEA of the European Commission. M.M. and J.A.R. acknowledge the partial support of the project N 3101/GF4 IPC-11, and the target program F.0679  0073-6/PTsF of the Ministry of Education and Science of the Republic of Kazakhstan.

\appendix

\section{The complete list of the BdHNe}
\label{sec:bdhne}

We present here in Tab.~\ref{tab:bdhne} the complete list of the $345$ BdHNe observed up through the end of 2016, which includes the $161$ BdHNe already presented in \citet{2016ApJ...833..159P}.

\begin{longtable}{lcccccccl}
\caption{List of the BdHNe considered in this work. It is composed by $345$ sources spanning 12 years of Swift/XRT observation activity. In the table we report important observational features: the redshift $z$, the isotropic energy $E_{iso}$, the observing instrument in the gamma-ray band, and the correspondent reference from which we take the gamma-ray spectral parameters in order to estimate $E_{iso}$.\\
$^{(a)}$: in units of $10^{52}$ erg.\\
$^{(b)}$: ``LX'' marks the sources with {Swift}/XRT data observed up to times larger than $10^4$ s in the rest-frame after the initial explosion.\\
$^{(c)}$: ``C'' and ``E'' mark the sources showing an early flare in their {Swift}/XRT, and they stay for ``confirmed'' and ``excluded'' respectively. The 16 C sources compose the sample considered in the present paper.\\
$^{(d)}$: ``UL'' stays for ultra-long, indicating sources with $T_{90} \gtrsim 1000$ s.\\
$^{(e)}$: observed $T_{90}$ (s).\\
$^{(f)}$: ``B-SAX'' stays for {Beppo-SAX}/GRBM; ``BATSE'' stays for {Compton-GRO}/BATSE; ``Ulysses'' stays for {Ulysses}/GRB; ``KW'' stays for Konus-WIND; ``HETE'' stays for {HETE-2}/FREGATE; ``Swift'' stays for {Swift}/BAT; ``Fermi'' stays for {Fermi}/GBM.\\
$^{(g)}$: (1) \citet{1998ApJ...493L..67F}; (2) \citet{2015ARep...59..626R}; (3) \citet{1998ApJ...505L.119I}; (4) \citet{2000Sci...290..953A}; (5) \citet{2000ApJ...534L..23H}; (6) \citet{2001ApJ...559..710I}; (7) \citet{2003A&A...400.1021B}; (8) \citet{2008PASJ...60..919S}; (9) \citet{2006ApJ...652..490C}.
\label{tab:bdhne}}\\
\hline\hline
GRB & z & $E_{iso}^{(a)} $  & LX$^{(b)}$ & Early flare$^{(c)}$  & UL$^{(d)}$ & $T_{90}^{(e)}$ & Instrument$^{(f)}$ & Reference$^{(g)}$ \\
\hline
\endfirsthead
\caption{continued.}\\
\hline\hline
GRB & z & $E_{iso}^{(a)} $  & LX$^{(b)}$ & Early flare$^{(c)}$  & UL$^{(d)}$ & $T_{90}^{(e)}$ & Instrument$^{(f)}$ & Reference$^{(g)}$ \\
\hline
\endhead
\hline
\endfoot
970228	&	$0.695$	&	$1.65\pm0.16$	&		&		&		&	$80$	&	B-SAX	&	(1) 	\\
970828	&	$0.958$	&	$30.4\pm3.6$	&		&		&		&	$90$	&	BATSE	&	(2) 	\\
971214	&	$3.42$	&	$22.1\pm2.7$	&		&		&		&	$40$	&	BATSE	&	IAUC 6789 	\\
980329	&	$3.5$	&	$267\pm53$	&		&		&		&	$54$	&	B-SAX	&	(3) 	\\
980703	&	$0.966$	&	$7.42\pm0.74$	&		&		&		&	$400$	&	BATSE	&	GCN 143	\\
990123	&	$1.6$	&	$241\pm39$	&		&		&		&	$63.3$	&	BATSE	&	GCN 224	\\
990506	&	$1.3$	&	$98.1\pm9.9$	&		&		&		&	$131.33$	&	BATSE	&	GCN 306	\\
990510	&	$1.619$	&	$18.1\pm2.7$	&		&		&		&	$75$	&	BATSE	&	GCN 322	\\
990705	&	$0.842$	&	$18.7\pm2.7$	&		&		&		&	$42$	&	B-SAX	&	(4)	\\
991208	&	$0.706$	&	$23.0\pm2.3$	&		&		&		&	$68$	&	Ulysses	&	(5)	\\
991216	&	$1.02$	&	$69.8\pm7.2$	&		&		&		&	$15.17$	&	BATSE	&	GCN 504 	\\
000131	&	$4.5$	&	$184\pm32$	&		&		&		&	$50$	&	KW+Ulysses	&	GCN 529	\\
000210	&	$0.846$	&	$15.4\pm1.7$	&		&		&		&	$12.3$	&	BATSE	&	GCN 540	\\
000301C	&	$2.0335$	&	$4.96\pm0.50$	&		&		&		&	$10$	&	Ulysses	&	GCN 568	\\
000418	&	$1.12$	&	$9.5\pm1.8$	&		&		&		&	$30$	&	KW+Ulysses	&	GCN 642 	\\
000911	&	$1.06$	&	$70\pm14$	&		&		&		&	$500$	&	KW+Ulysses	&	GCN 791	\\
000926	&	$2.07$	&	$28.6\pm6.2$	&		&		&		&	$25$	&	KW+Ulysses	&	GCN 801	\\
010222	&	$1.48$	&	$84.9\pm9.0$	&		&		&		&	$170$	&	B-SAX	&	(6)	\\
010921	&	$0.45$	&	$0.97\pm0.10$	&		&		&		&	$12$	&	HETE	&	GCN 1096	\\
011121	&	$0.36$	&	$8.0\pm2.2$	&		&		&		&	$28$	&	Ulysses	&	GCN 1148	\\
011211	&	$2.14$	&	$5.74\pm0.64$	&		&		&		&	$270$	&	B-SAX	&	GCN 1215	\\
020124	&	$3.2$	&	$28.5\pm2.8$	&		&		&		&	$78.6$	&	HETE	&	(7)	\\
020127	&	$1.9$	&	$3.73\pm0.37$	&		&		&		&	$9.3$	&	HETE	&	(7)	\\
020405	&	$0.69$	&	$10.6\pm1.1$	&		&		&		&	$40$	&	KW+Ulysses	&	GCN 1325	\\
020813	&	$1.25$	&	$68\pm17$	&		&		&		&	$90$	&	HETE	&	(7)	\\
021004	&	$2.3$	&	$3.47\pm0.46$	&		&		&		&	$57.7$	&	HETE	&	(7)	\\
021211	&	$1.01$	&	$1.16\pm0.13$	&		&		&		&	$5.7$	&	HETE	&	GCN 1734	\\
030226	&	$1.98$	&	$12.7\pm1.4$	&		&		&		&	$100$	&	HETE	&	GCN 1888	\\
030323	&	$3.37$	&	$2.94\pm0.92$	&		&		&		&	$26$	&	HETE	&	GCN 1956	\\
030328	&	$1.52$	&	$38.9\pm3.9$	&		&		&		&	$100$	&	HETE	&	GCN 1978 	\\
030329	&	$0.169$	&	$1.62\pm0.16$	&		&		&		&	$50$	&	HETE	&	IAUC 8101	\\
030429	&	$2.65$	&	$2.29\pm0.27$	&		&		&		&	$14$	&	HETE	&	GCN 2211	\\
030528	&	$0.78$	&	$2.22\pm0.27$	&		&		&		&	$21.6$	&	HETE	&	GCN 2256 	\\
040912	&	$1.563$	&	$1.36\pm0.36$	&		&		&		&	$122$	&	HETE	&	GCN 2723	\\
040924	&	$0.859$	&	$0.98\pm0.10$	&		&		&		&	$2.4$	&	KW	&	GCN 2754	\\
041006	&	$0.716$	&	$3.11\pm0.89$	&		&		&		&	$27.3$	&	HETE	&	(8) 	\\
041219A	&	$0.31$	&	$10.0\pm1.0$	&		&		&		&	$520$	&	Swift	&	GCN 2874	\\
050126	&	$1.29$ 	&	$2.47\pm0.25$	&		&		&		&	$26$	&	Swift	&	GCN 2987 	\\
050315	&	$1.95$	&	$6.15\pm0.30$	&	LX	&		&		&	$96$	&	Swift	&	GCN 3099	\\
050318	&	$1.444$	&	$2.30\pm0.23$	&	LX	&		&		&	$32$	&	Swift	&	GCN 3134	\\
050319	&	$3.243$	&	$4.63\pm.0.56$	&	LX	&		&		&	$10$	&	Swift	&	GCN 3119	\\
050401	&	$2.898$	&	$37.6\pm7.3$	&	LX	&		&		&	$33$	&	KW	&	GCN 3179	\\
050408	&	$1.2357$	&	$2.48\pm0.25$	&	LX	&		&		&	$34$	&	HETE	&	GCN 3188	\\
050502B	&	$5.2$	&	$2.66\pm0.22$	&		&		&		&	$17.5$	&	Swift	&	GCN 3339 	\\
050505	&	$4.27$	&	$16.0\pm1.1$	&	LX	&		&		&	$60$	&	Swift	&	GCN 3364 	\\
050525A	&	$0.606$	&	$2.30\pm0.49$	&	LX	&		&		&	$5.2$	&	KW	&	GCN 3479	\\
050603	&	$2.821$	&	$64.1\pm6.4$	&		&		&		&	$6$	&	KW	&	GCN 3518	\\
050714B	&	$2.4383$	&	$4.99\pm0.85$	&		&		&		&	$46.7$	&	Swift	&	GCN 3615	\\
050730	&	$3.969$	&	$11.8\pm0.8$	&	LX	&		&		&	$155$	&	Swift	&	GCN 3715	\\
050802	&	$1.71$	&	$5.66\pm0.47$ 	&	LX	&		&		&	$13$	&	Swift	&	GCN 3737	\\
050803	&	$4.3$	&	$1.16\pm0.12$	&	LX	&	E	&		&	$85$	&	Swift	&	GCN 3757	\\
050814	&	$5.3$	&	$9.9\pm1.1$	&	LX	&		&		&	$65$	&	Swift	&	GCN 3803	\\
050819	&	$2.5043$	&	$3.60\pm0.55$	&		&		&		&	$36$	&	Swift	&	GCN 3828	\\
050820	&	$2.615$	&	$103\pm10$	&	LX	&		&		&	$549.2$	&	KW	&	(9)	\\
050822	&	$1.434$	&	$10.8\pm1.1$	&	LX	&	E	&		&	$102$	&	Swift	&	GCN 3856	\\
050904	&	$6.295$	&	$133\pm14$	&		&		&		&	$225$	&	Swift	&	GCN 3938	\\
050908	&	$3.347$	&	$1.54\pm0.16$	&		&		&		&	$20$	&	Swift	&	GCN 3951	\\
050915	&	$2.5273$	&	$1.8\pm1.3$	&		&		&		&	$53$	&	Swift	&	GCN 3982	\\
050922B	&	$4.9$	&	$46.4\pm4.6$	&	LX	&	E	&		&	$980$	&	Swift	&	GCN 4019 	\\
050922C	&	$2.199$	&	$5.6\pm1.8$	&	LX	&		&		&	$8.4$	&	KW	&	GCN 4030	\\
051001	&	$2.4296$	&	$2.3\pm1.7$	&		&		&		&	$190$	&	Swift	&	GCN 4052	\\
051006	&	$1.059$	&	$1.02\pm0.56$	&		&		&		&	$26$	&	Swift	&	GCN 4063	\\
051008	&	$2.77$	&	$115\pm20$	&		&		&		&	$280$	&	KW	&	GCN 4078	\\
051022	&	$0.8$	&	$56.0\pm5.6$	&		&		&		&	$200$	&	KW	&	GCN 4150	\\
051109A	&	$2.346$	&	$6.85\pm0.73$	&	LX	&		&		&	$130$	&	KW	&	GCN 4238	\\
051111	&	$1.55$	&	$15.4\pm1.9$	&		&		&		&	$47$	&	KW	&	GCN 4260	\\
060108	&	$2.03$	&	$1.51\pm1.33$	&	LX	&		&		&	$14.4$	&	Swift	&	GCN 4445	\\
060111	&	$2.32$	&	$1.62\pm0.08$	&	LX	&	E	&		&	$13$	&	Swift	&	GCN 4486	\\
060115	&	$3.533$	&	$5.9\pm3.8$	&	LX	&		&		&	$142$	&	Swift	&	GCN 4518	\\
060124	&	$2.296$	&	$43.8\pm6.4$	&	LX	&		&		&	$300$	&	KW	&	GCN 4599	\\
060202	&	$0.785$	&	$1.20\pm0.09$	&	LX	&		&		&	$203.7$	&	Swift	&	GCN 4635	\\
060204B	&	$2.3393$	&	$29.3\pm6.0$	&	LX	&	C	&		&	$134$	&	Swift	&	GCN 4671	\\
060206	&	$4.056$	&	$4.1\pm1.9$	&	LX	&		&		&	$7$	&	Swift	&	GCN 4697	\\
060210	&	$3.91$	&	$32.2\pm3.2$	&	LX	&		&		&	$255$	&	Swift	&	GCN 4748	\\
060223	&	$4.41$	&	$9.73\pm0.72$	&		&		&		&	$11$	&	Swift	&	GCN 4820	\\
060306	&	$3.5$	&	$7.6\pm1.0$	&		&		&		&	$61$	&	Swift	&	GCN 4851	\\
060418	&	$1.489$	&	$13.5\pm2.7$	&	LX	&		&		&	$52$	&	Swift	&	GCN 4975	\\
060502A	&	$1.51$	&	$10.57\pm0.48$	&	LX	&		&		&	$33$	&	Swift	&	GCN 5053	\\
060510B	&	$4.9$	&	$19.1\pm0.8$	&	LX	&		&		&	$276$	&	Swift	&	GCN 5107	\\
060512	&	$2.1$	&	$2.38\pm2.70$	&	LX	&		&		&	$8.6$	&	Swift	&	GCN 5124	\\
060522	&	$5.11$	&	$6.47\pm0.63$	&		&		&		&	$69$	&	Swift	&	GCN 5153	\\
060526	&	$3.22$	&	$2.75\pm0.37$	&	LX	&		&		&	$298$	&	Swift	&	GCN 5174	\\
060602A	&	$0.787$	&	$6.63\pm0.41$	&		&		&		&	$60$	&	Swift	&	GCN 5206	\\
060605	&	$3.773$	&	$4.23\pm0.61$	&	LX	&		&		&	$15$	&	Swift	&	GCN 5231	\\
060607A	&	$3.082$	&	$21.4\pm11.9$	&	LX	&	C	&		&	$100$	&	Swift	&	GCN 5242	\\
060707	&	$3.424$	&	$4.3\pm1.1$	&	LX	&		&		&	$68$	&	Swift	&	GCN 5289	\\
060708	&	$1.92$	&	$1.06\pm0.08$	&	LX	&		&		&	$9.8$	&	Swift	&	GCN 5295	\\
060714	&	$2.7108$	&	$7.67\pm0.44$	&	LX	&		&		&	$115$	&	Swift	&	GCN 5334	\\
060719	&	$1.532$	&	$1.4\pm1.3$	&		&		&		&	$55$	&	Swift	&	GCN 5349	\\
060729	&	$0.54$	&	$1.20\pm0.53$	&	LX	&	E	&		&	$116$	&	Swift	&	GCN 5370	\\
060814	&	$1.923$	&	$56.7\pm5.7$	&	LX	&		&		&	$40$	&	KW	&	GCN 5460	\\
060906	&	$3.6856$	&	$7.81\pm0.51$	&	LX	&		&		&	$43.6$	&	Swift	&	GCN 5538	\\
060908	&	$1.884$	&	$7.2\pm1.9$	&		&		&		&	$19.3$	&	Swift	&	GCN 5551	\\
060923B	&	$1.5094$	&	$2.71\pm0.34$	&		&		&		&	$8.8$	&	Swift	&	GCN 5595 	\\
060926	&	$3.2086$	&	$2.29\pm0.37$	&		&		&		&	$8$	&	Swift	&	GCN 5621	\\
060927	&	$5.46$	&	$12.0\pm2.8$	&		&		&		&	$22.6$	&	Swift	&	GCN 5639	\\
061007	&	$1.262$	&	$90.0\pm9.0$	&	LX	&		&		&	$75$	&	KW	&	GCN 5722	\\
061110B	&	$3.4344$	&	$17.9\pm1.6$	&		&		&		&	$128$	&	Swift	&	GCN 5810 	\\
061121	&	$1.314$	&	$23.5\pm2.7$	&	LX	&		&		&	$81$	&	Swift	&	GCN 5831	\\
061126	&	$1.1588$	&	$31.4\pm3.6$	&	LX	&		&		&	$191$	&	Swift	&	GCN 5860	\\
061202	&	$2.2543$	&	$21.99\pm0.63$	&		&		&		&	$91$	&	Swift	&	GCN 5887	\\
061222A	&	$2.088$	&	$30.0\pm6.4$	&	LX	&		&		&	$72$	&	Swift	&	GCN 5964	\\
061222B	&	$3.355$	&	$8.1\pm1.5$	&		&		&		&	$40$	&	Swift	&	GCN 5974	\\
070110	&	$2.3521$	&	$4.98\pm0.30$	&	LX	&		&		&	$85$	&	Swift	&	GCN 6007	\\
070125	&	$1.547$	&	$84.1\pm8.4$	&		&		&		&	$75$	&	Swift	&	GCN 6049	\\
070129	&	$2.3384$	&	$16.8\pm1.7$	&	LX	&	E	&		&	$460$	&	Swift	&	GCN 6058 	\\
070223	&	$1.6295$	&	$4.73\pm0.28$	&		&		&		&	$89$	&	Swift	&	GCN 6132	\\
070224	&	$1.9922$	&	$2.37\pm0.28$	&		&		&		&	$34$	&	Swift	&	GCN 6141	\\
070306	&	$1.4959$	&	$8.26\pm0.41$	&	LX	&		&		&	$210$	&	Swift	&	GCN 6173	\\
070318	&	$0.84$	&	$3.41\pm2.14$	&	LX	&	C	&		&	$63$	&	Swift	&	GCN 6212 	\\
070328	&	$2.0627$	&	$56.7\pm7.7$	&		&		&		&	$45$	&	KW	&	GCN 6230	\\
070411	&	$2.954$	&	$8.31\pm0.45$	&		&		&		&	$101$	&	Swift	&	GCN 6274 	\\
070419B	&	$1.959$	&	$12.1\pm1.7$	&		&		&		&	$236.5$	&	Swift	&	GCN 6327	\\
070508	&	$0.82$	&	$7.74\pm0.29$	&	LX	&		&		&	$40$	&	KW	&	GCN 6403	\\
070521	&	$1.35$	&	$10.8\pm1.8$	&		&		&		&	$55$	&	KW	&	GCN 6459	\\
070529	&	$2.4996$	&	$12.8\pm1.1$	&	LX	&		&		&	$109$	&	Swift	&	GCN 6468	\\
070611	&	$2.0394$	&	$0.92\pm0.13$	&		&		&		&	$12$	&	Swift	&	GCN 6502	\\
070612A	&	$0.617$	&	$1.96\pm0.40$	&		&		&		&	$370$	&	Swift	&	GCN 6522	\\
070721B	&	$3.6298$	&	$24.2\pm1.4$	&		&		&		&	$340$	&	Swift	&	GCN 6649	\\
070802A	&	$2.45$	&	$1.65\pm2.78$	&	LX	&		&		&	$16.4$	&	Swift	&	GCN 6699 	\\
070810A	&	$2.17$	&	$91.5\pm1.1$	&		&		&		&	$11$	&	Swift	&	GCN 6748	\\
071003	&	$1.604$	&	$38.3\pm4.5$	&	LX	&		&		&	$30$	&	KW	&	GCN 6849	\\
071010B	&	$0.947$	&	$2.32\pm0.40$	&		&		&		&	$16.6$	&	KW	&	GCN 6879	\\
071020	&	$2.145$	&	$10.0\pm4.6$	&		&		&		&	$8.45$	&	KW	&	GCN 6960	\\
071021	&	$2.452$	&	$8.18\pm0.82$	&	LX	&	E	&		&	$225$	&	Swift	&	GCN 6966	\\
071025	&	$5.2$	&	$115\pm4$	&		&		&		&	$109$	&	Swift	&	GCN 6996	\\
071031	&	$2.6918$	&	$4.99\pm0.97$	&		&		&		&	$180$	&	Swift	&	GCN 7029	\\
071112C	&	$0.823$	&	$15.7\pm2.1$	&		&		&		&	$15$	&	Swift	&	GCN 7081	\\
071117	&	$1.331$	&	$5.86\pm2.7$	&		&		&		&	$5$	&	KW	&	GCN 7114	\\
080129	&	$4.349$	&	$7.7\pm3.5$	&		&		&		&	$48$	&	Swift	&	GCN 7235	\\
080205	&	$2.72$	&	$15.21\pm0.72$	&		&		&		&	$106.5$	&	Swift	&	GCN 7257 	\\
080207	&	$2.0858$	&	$16.4\pm1.8$	&		&		&		&	$340$	&	Swift	&	GCN 7272	\\
080210	&	$2.6419$	&	$4.77\pm0.29$	&	LX	&		&		&	$45$	&	Swift	&	GCN 7289 	\\
080310	&	$2.4274$	&	$20.9\pm2.1$	&	LX	&	E	&		&	$365$	&	Swift	&	GCN 7402 	\\
080319A	&	$2.0265$	&	$27.0\pm2.2$	&		&		&		&	$64$	&	Swift	&	GCN 7447	\\
080319B	&	$0.937$	&	$118\pm12$	&	LX	&		&		&	$50$	&	KW	&	GCN 7482	\\
080319C	&	$1.95$	&	$14.9\pm3.0$	&	LX	&		&		&	$15$	&	KW	&	GCN 7487	\\
080325	&	$1.78$	&	$9.55\pm0.84$	&		&		&		&	$128.4$	&	Swift	&	GCN 7531	\\
080411	&	$1.03$	&	$16.2\pm1.6$	&		&		&		&	$70$	&	KW	&	GCN 7589	\\
080413A	&	$2.433$	&	$8.6\pm2.1$	&		&		&		&	$46$	&	Swift	&	GCN 7604	\\
080413B	&	$1.1$	&	$1.61\pm0.27$	&		&		&		&	$8$	&	Swift	&	GCN 7606	\\
080514B	&	$1.8$	&	$18.1\pm3.6$	&		&		&		&	$7$	&	KW	&	GCN 7751	\\
080515	&	$2.47$	&	$5.11\pm0.77$	&		&		&		&	$21$	&	Swift	&	GCN 7726	\\
080602	&	$1.8204$	&	$6.08\pm0.38$	&		&		&		&	$74$	&	Swift	&	GCN 7786	\\
080603B	&	$2.69$	&	$6.0\pm3.1$	&		&		&		&	$70$	&	KW	&	GCN 7812 	\\
080604	&	$1.4171$	&	$1.05\pm0.12$	&		&		&		&	$82$	&	Swift	&	GCN 7817	\\
080605	&	$1.64$	&	$28\pm14$	&	LX	&		&		&	$20$	&	Swift	&	GCN 7854	\\
080607	&	$3.04$	&	$187\pm11$	&	LX	&	C	&		&	$85$	&	KW	&	GCN 7862	\\
080710	&	$0.8454$	&	$1.68\pm0.22$	&		&		&		&	$120$	&	Swift	&	GCN 7969	\\
080721	&	$2.591$	&	$134\pm23$	&	LX	&		&		&	$30$	&	KW	&	GCN 7995	\\
080804	&	$2.205$	&	$12.0\pm1.2$	&	LX	&		&		&	$34$	&	Swift	&	GCN 8067	\\
080805	&	$1.51$	&	$7.16\pm1.90$	&	LX	&	C	&		&	$78$	&	Swift	&	GCN 8068	\\
080810	&	$3.35$	&	$50.0\pm4.4$	&	LX	&	C	&		&	$79.4$	&	KW	&	GCN 8101	\\
080825B	&	$4.3$	&	$38.4\pm3.8$	&		&		&		&	$110$	&	KW	&	GCN 8142 	\\
080905B	&	$2.3739$	&	$4.55\pm0.37$	&	LX	&		&		&	$128$	&	Swift	&	GCN 8188	\\
080906	&	$2.1$	&	$21.2\pm1.2$	&		&		&		&	$147$	&	Swift	&	GCN 8196	\\
080913	&	$6.695$	&	$9.2\pm2.7$	&		&		&		&	$8.8$	&	KW	&	GCN 8280	\\
080916A	&	$0.689$	&	$0.98\pm0.10$	&		&		&		&	$40$	&	KW	&	GCN 8259	\\
080916C	&	$4.35$	&	$407\pm86$	&	LX	&		&		&	$60$	&	Fermi	&	GCN 8263	\\
080928	&	$1.692$	&	$3.99\pm0.91$	&	LX	&		&		&	$66$	&	Fermi	&	GCN 8278	\\
081008	&	$1.967$	&	$13.5\pm6.6$	&	LX	&	C	&		&	$185.5$	&	Swift	&	GCN 8351	\\
081028	&	$3.038$	&	$18.3\pm1.8$	&	LX	&		&		&	$260$	&	Swift	&	GCN 8428	\\
081029	&	$3.8479$	&	$12.1\pm1.4$	&		&		&		&	$270$	&	Swift	&	GCN 8447 	\\
081109	&	$0.9787$	&	$1.81\pm0.12$	&	LX	&		&		&	$45$	&	Fermi	&	GCN 8505	\\
081118	&	$2.58$	&	$12.2\pm1.2$	&		&		&		&	$20$	&	Fermi	&	GCN 8550	\\
081121	&	$2.512$	&	$32.4\pm3.7$	&	LX	&		&		&	$18$	&	KW	&	GCN 8548	\\
081203A	&	$2.05$	&	$32\pm12$	&	LX	&		&		&	$213$	&	KW	&	GCN 8611	\\
081210	&	$2.0631$	&	$15.6\pm5.4$	&	LX	&	C	&		&	$146$	&	Swift	&	GCN 8649 	\\
081221	&	$2.26$	&	$31.9\pm3.2$	&	LX	&		&		&	$40$	&	Fermi	&	GCN 8704	\\
081222	&	$2.77$	&	$27.4\pm2.7$	&	LX	&		&		&	$30$	&	Fermi	&	GCN 8715	\\
081228	&	$3.44$	&	$9.9\pm2.0$	&		&		&		&	$3$	&	Swift	&	GCN 8749 	\\
081230	&	$2.0$	&	$3.21\pm0.31$	&		&		&		&	$60.7$	&	Swift	&	GCN 8759	\\
090102A	&	$1.547$	&	$22.6\pm2.7$	&	LX	&		&		&	$30$	&	KW	&	GCN 8776	\\
090113A	&	$1.7493$	&	$1.00\pm0.17$	&		&		&		&	$9.1$	&	Swift	&	GCN 8808	\\
090201A	&	$2.1$	&	$93.4\pm8.1$	&		&		&		&	$110$	&	KW	&	GCN 8878	\\
090205A	&	$4.6497$	&	$1.12\pm0.16$	&		&		&		&	$8.8$	&	Swift	&	GCN 8886	\\
090313A	&	$3.375$	&	$4.42\pm0.79$	&	LX	&		&		&	$78$	&	Swift	&	GCN 8986	\\
090323A	&	$3.57$	&	$438\pm53$	&		&		&		&	$150$	&	Fermi	&	GCN 9035	\\
090328A	&	$0.736$	&	$14.2\pm1.4$	&	LX	&		&		&	$80$	&	Fermi	&	GCN 9057	\\
090404A	&	$3$	&	$59.2\pm6.1$	&	LX	&	E	&		&	$84$	&	Swift	&	GCN 9089	\\
090418A	&	$1.608$	&	$17.2\pm2.7$	&	LX	&		&		&	$64.8$	&	KW+Swift	&	GCN 9196	\\
090423A	&	$8.26$	&	$8.8\pm2.1$	&	LX	&		&		&	$12$	&	Fermi	&	GCN 9229	\\
090424A	&	$0.544$	&	$4.07\pm0.41$	&	LX	&		&		&	$52$	&	Fermi	&	GCN 9230	\\
090429B	&	$9.3$	&	$6.7\pm1.3$	&		&		&		&	$5.5$	&	Swift	&	GCN 9290	\\
090516A	&	$4.109$	&	$99.6\pm16.7$	&	LX	&	C	&		&	$350$	&	Fermi	&	GCN 9415	\\
090519A	&	$3.85$	&	$24.7\pm2.8$	&		&		&		&	$64$	&	Swift	&	GCN 9406	\\
090529A	&	$2.625$	&	$2.56\pm0.30$	&		&		&		&	$100$	&	Swift	&	GCN 9434	\\
090530A	&	$1.266$	&	$1.73\pm0.19$	&		&		&		&	$48$	&	Swift	&	GCN 9443	\\
090618A	&	$0.54$	&	$28.6\pm2.9$	&	LX	&		&		&	$113.2$	&	Fermi	&	GCN 9535	\\
090715B	&	$3.$	&	$63.9\pm3.7$	&	LX	&	E	&		&	$100$	&	KW	&	GCN 9679	\\
090726A	&	$2.71$	&	$1.82\pm0.40$	&		&		&		&	$67$	&	Swift	&	GCN 9716	\\
090809A	&	$2.737$	&	$1.88\pm0.26$	&	LX	&		&		&	$5.4$	&	Swift	&	GCN 9756	\\
090812A	&	$2.452$	&	$44.0\pm6.5$	&	LX	&	C	&		&	$64.8$	&	KW+Swift	&	GCN 9821	\\
090902B	&	$1.822$	&	$292\pm29.2$	&	LX	&		&		&	$21$	&	Fermi	&	GCN 9866	\\
090926A	&	$2.106$	&	$228\pm23$	&	LX	&		&		&	$20$	&	Fermi	&	GCN 9933	\\
090926B	&	$1.24$	&	$4.14\pm0.45$	&		&		&		&	$81$	&	Fermi	&	GCN 9957	\\
091003A	&	$0.897$	&	$10.7\pm1.8$	&	LX	&		&		&	$21.1$	&	Fermi	&	GCN 9983	\\
091020A	&	$1.71$	&	$8.4\pm1.1$	&	LX	&		&		&	$37$	&	Fermi	&	GCN 10095	\\
091024A	&	$1.092$	&	$18.4\pm2.0$	&		&		&	UL	&	$1250$	&	KW	&	GCN 10083	\\
091029A	&	$2.752$	&	$7.97\pm0.82$	&	LX	&		&		&	$39.2$	&	Swift	&	GCN 10103	\\
091109A	&	$3.076$	&	$10.6\pm1.4$	&		&		&		&	$48$	&	Swift	&	GCN 10141	\\
091127A	&	$0.49$	&	$1.64\pm0.18$	&	LX	&		&		&	$9$	&	Fermi	&	GCN 10204	\\
091208B	&	$1.063$	&	$2.06\pm0.21$	&	LX	&		&		&	$15$	&	Fermi	&	GCN 10266	\\
100219A	&	$4.6667$	&	$3.93\pm0.61$	&		&		&		&	$18.8$	&	Swift	&	GCN 10434	\\
100302A	&	$4.813$	&	$1.33\pm0.17$	&	LX	&		&		&	$17.9$	&	Swift	&	GCN 10462	\\
100414A	&	$1.368$	&	$55.0\pm5.5$	&		&		&		&	$26.4$	&	Fermi	&	GCN 10595	\\
100424A	&	$2.465$	&	$3.05\pm0.53$	&		&		&		&	$104$	&	Swift	&	GCN 10670	\\
100425A	&	$1.755$	&	$2.76\pm3.45$	&	LX	&		&		&	$37$	&	Swift	&	GCN 10685	\\
100513A	&	$4.8$	&	$6.75\pm0.53$	&	LX	&		&		&	$84$	&	Swift	&	GCN 10753	\\
100615A	&	$1.398$	&	$5.81\pm0.11$	&		&		&		&	$37.7$	&	Fermi	&	GCN 10851	\\
100621A	&	$0.542$	&	$2.82\pm0.35$	&	LX	&		&		&	$80$	&	KW	&	GCN 10882	\\
100728A	&	$1.567$	&	$86.8\pm8.7$	&		&		&		&	$162.9$	&	Fermi	&	GCN 11006	\\
100728B	&	$2.106$	&	$3.55\pm0.36$	&		&		&		&	$11.8$	&	Fermi	&	GCN 11015	\\
100814A	&	$1.44$	&	$15.3\pm1.8$	&	LX	&		&		&	$149$	&	Fermi	&	GCN 11099	\\
100816A	&	$0.8049$	&	$0.75\pm0.10$	&	LX	&		&		&	$2$	&	Fermi	&	GCN 11124	\\
100901A	&	$1.408$	&	$4.22\pm0.50$	&	LX	&		&		&	$439$	&	Swift	&	GCN 11169	\\
100906A	&	$1.727$	&	$29.9\pm2.9$	&	LX	&		&		&	$105$	&	Fermi	&	GCN 11248	\\
101213A	&	$0.414$	&	$2.72\pm0.53$	&		&		&		&	$45$	&	Fermi	&	GCN 11454	\\
110128A	&	$2.339$	&	$1.58\pm0.21$	&	LX	&		&		&	$12$	&	Fermi	&	GCN 11628	\\
110205A	&	$2.22$	&	$48.3\pm6.4$	&	LX	&		&		&	$330$	&	KW	&	GCN 11659	\\
110213A	&	$1.46$	&	$5.78\pm0.81$	&	LX	&		&		&	$33$	&	Fermi	&	GCN 11727	\\
110213B	&	$1.083$	&	$8.3\pm1.3$	&		&		&		&	$50$	&	KW	&	GCN 11722	\\
110422A	&	$1.77$	&	$79.8\pm8.2$	&	LX	&		&		&	$40$	&	KW	&	GCN 11971	\\
110503A	&	$1.613$	&	$20.8\pm2.1$	&	LX	&		&		&	$12$	&	KW	&	GCN 12008	\\
110715A	&	$0.82$	&	$4.36\pm0.45$	&	LX	&		&		&	$20$	&	KW	&	GCN 12166	\\
110731A	&	$2.83$	&	$49.5\pm4.9$	&	LX	&		&		&	$7.3$	&	Fermi	&	GCN 12221	\\
110801A	&	$1.858$	&	$10.9\pm2.7$	&		&		&		&	$415.1$	&	KW+Swift	&	GCN 12276	\\
110808A	&	$1.348$	&	$6.09\pm4.83$	&	LX	&		&		&	$48$	&	KW	&	GCN 12270	\\
110818A	&	$3.36$	&	$26.6\pm2.8$	&		&		&		&	$75$	&	Fermi	&	GCN 12287	\\
110918A	&	$0.982$	&	$185\pm5$	&	LX	&		&		&	$22$	&	KW	&	GCN 12362	\\
111008A	&	$4.9898$	&	$24.7\pm1.2$	&	LX	&		&		&	$40$	&	KW	&	GCN 12433	\\
111107A	&	$2.893$	&	$3.76\pm0.55$	&		&		&		&	$12$	&	Fermi	&	GCN 12545	\\
111123A	&	$3.1516$	&	$24\pm14$	&	LX	&		&		&	$290$	&	Swift	&	GCN 12598	\\
111209A	&	$0.677$	&	$5.14\pm0.62$	&	LX	&		&	UL	&	$11900$	&	KW	&	GCN 12663	\\
111215A	&	$2.06$	&	$22.1\pm2.5$	&		&		&		&	$796$	&	Swift	&	GCN 12689	\\
111228A	&	$0.716$	&	$2.75\pm0.28$	&	LX	&		&		&	$101.2$	&	Fermi	&	GCN 12744	\\
120118B	&	$2.943$	&	$6.24\pm0.55$	&		&		&		&	$23.26$	&	Swift	&	GCN 12873	\\
120119A	&	$1.728$	&	$27.2\pm3.6$	&	LX	&		&		&	$55$	&	Fermi	&	GCN 12874	\\
120211A	&	$2.4$	&	$7.1\pm1.0$	&		&		&		&	$61.7$	&	Swift	&	GCN 12924	\\
120326A	&	$1.798$	&	$3.27\pm0.33$	&	LX	&		&		&	$12$	&	Fermi	&	GCN 13145	\\
120327A	&	$2.813$	&	$14.42\pm0.46$	&	LX	&		&		&	$62.9$	&	Swift	&	GCN 13137	\\
120404A	&	$2.876$	&	$4.18\pm0.34$	&		&		&		&	$38.7$	&	Swift	&	GCN 13220	\\
120521C	&	$6.01$	&	$11.9\pm1.9$	&		&		&		&	$26.7$	&	Swift	&	GCN 13333	\\
120624B	&	$2.197$	&	$319\pm32$	&		&		&		&	$271$	&	Fermi	&	GCN 13377	\\
120711A	&	$1.405$	&	$180\pm18$	&	LX	&		&		&	$44$	&	Fermi	&	GCN 13437	\\
120712A	&	$4.175$	&	$21.2\pm2.1$	&	LX	&		&		&	$23$	&	Fermi	&	GCN 13469	\\
120716A	&	$2.486$	&	$30.2\pm3.0$	&		&		&		&	$234$	&	Fermi	&	GCN 13498	\\
120802A	&	$3.796$	&	$12.9\pm2.8$	&		&		&		&	$50$	&	Swift	&	GCN 13559	\\
120805A	&	$3.1$	&	$19.0\pm3.2$	&		&		&		&	$48$	&	Swift	&	GCN 13594	\\
120811C	&	$2.671$	&	$6.41\pm0.64$	&		&		&		&	$26.8$	&	Swift	&	GCN 13634	\\
120815A	&	$2.358$	&	$1.65\pm0.27$	&		&		&		&	$9.7$	&	Swift	&	GCN 13652	\\
120909A	&	$3.93$	&	$87\pm10$	&	LX	&		&		&	$112$	&	Fermi	&	GCN 13737	\\
120922A	&	$3.1$	&	$22.4\pm1.4$	&	LX	&		&		&	$180$	&	Fermi	&	GCN 13809	\\
121024A	&	$2.298$	&	$4.61\pm0.55$	&	LX	&		&		&	$69$	&	Swift	&	GCN 13899	\\
121027A	&	$1.773$	&	$1.50\pm0.17$	&	LX	&	E	&	UL	&	$6000$	&	Swift	&	GCN 13910	\\
121128A	&	$2.2$	&	$8.66\pm0.87$	&	LX	&		&		&	$17$	&	Fermi	&	GCN 14012	\\
121201A	&	$3.385$	&	$2.52\pm0.34$	&		&		&		&	$85$	&	Swift	&	GCN 14028	\\
121209A	&	$2.1$	&	$24.31\pm0.84$	&		&		&		&	$42.7$	&	Swift	&	GCN 14052	\\
121217A	&	$3.1$	&	$25.9\pm19.7$	&	LX	&		&		&	$780$	&	Fermi	&	GCN 14094	\\
121229A	&	$2.707$	&	$3.7\pm1.1$	&		&		&		&	$100$	&	Swift	&	GCN 14123	\\
130408A	&	$3.758$	&	$35.0\pm6.4$	&		&		&		&	$15$	&	KW	&	GCN 14368	\\
130131B	&	$2.539$	&	$7.15\pm0.84$	&		&		&		&	$4.3$	&	Swift	&	GCN 14164	\\
130215A	&	$0.597$	&	$4.45\pm0.11$	&		&		&		&	$140$	&	Fermi	&	GCN 14219	\\
130408A	&	$3.757$	&	$35.4\pm5.9$	&		&		&		&	$15$	&	KW	&	GCN 14368	\\
130418A	&	$1.218$	&	$9.9\pm1.6$	&	LX	&		&		&	$120$	&	KW	&	GCN 14417	\\
130420A	&	$1.297$	&	$7.74\pm0.77$	&	LX	&		&		&	$102$	&	Fermi	&	GCN 14429	\\
130427A	&	$0.334$	&	$92\pm13$	&	LX	&		&		&	$162.8$	&	Fermi	&	GCN 14473	\\
130427B	&	$2.78$	&	$13.3\pm0.5$	&	LX	&	E	&		&	$27$	&	Swift	&	GCN 14469	\\
130505A	&	$2.27$	&	$347\pm35$	&	LX	&		&		&	$21$	&	KW	&	GCN 14575	\\
130514A	&	$3.6$	&	$49.5\pm9.2$	&	LX	&	E	&		&	$204$	&	Swift	&	GCN 14636	\\
130518A	&	$2.488$	&	$193\pm19$	&		&		&		&	$48$	&	Fermi	&	GCN 14674	\\
130528A	&	$1.25$	&	$18.0\pm2.3$	&	LX	&	E	&		&	$55$	&	Fermi	&	GCN 14729	\\
130606A	&	$5.91$	&	$28.3\pm5.1$	&	LX	&	E	&		&	$165$	&	KW	&	GCN 14808	\\
130610A	&	$2.092$	&	$6.99\pm0.46$	&	LX	&		&		&	$28$	&	Fermi	&	GCN 14858	\\
130701A	&	$1.155$	&	$2.60\pm0.09$	&	LX	&		&		&	$5.5$	&	KW	&	GCN 14958	\\
130907A	&	$1.238$	&	$304\pm19$	&	LX	&		&		&	$214$	&	KW	&	GCN 15203	\\
130925A	&	$0.347$	&	$3.23\pm0.37$	&	LX	&	E	&	UL	&	$4500$	&	Fermi	&	GCN 15261	\\
131011A	&	$1.874$	&	$86.67\pm0.39$	&		&		&		&	$77$	&	Fermi	&	GCN 15331	\\
131030A	&	$1.293$	&	$30.0\pm2.0$	&	LX	&	C	&		&	$28$	&	KW	&	GCN 15413	\\
131105A	&	$1.686$	&	$34.7\pm1.2$	&	LX	&		&		&	$112$	&	Fermi	&	GCN 15455	\\
131108A	&	$2.4$	&	$70.87\pm0.97$	&	LX	&		&		&	$19$	&	Fermi	&	GCN 15477	\\
131117A	&	$4.042$	&	$1.02\pm0.16$	&	LX	&		&		&	$11$	&	Swift	&	GCN 15499	\\
131227A	&	$5.3$	&	$24.2\pm1.7$	&		&		&		&	$18$	&	Swift	&	GCN 15620	\\
140114A	&	$3.0$	&	$27.6\pm0.8$	&	LX	&	E	&		&	$139.7$	&	Swift	&	GCN 15738	\\
140206A	&	$2.73$	&	$35.8\pm7.9$	&	LX	&	C	&		&	$27$	&	Fermi	&	GCN 15796	\\
140213A	&	$1.2076$	&	$9.93\pm0.15$	&	LX	&		&		&	$18.6$	&	Fermi	&	GCN 15833	\\
140226A	&	$1.98$	&	$5.8\pm1.1$	&	LX	&		&		&	$15$	&	KW	&	GCN 15889	\\
140301A	&	$1.416$	&	$0.95\pm0.18$	&	LX	&	C	&		&	$31$	&	Swift	&	GCN 15906	\\
140304A	&	$5.283$	&	$15.3\pm1.1$	&	LX	&	E	&		&	$32$	&	Fermi	&	GCN 15923	\\
140311A	&	$4.954$	&	$11.6\pm1.5$	&	LX	&		&		&	$71.4$	&	Swift	&	GCN 15962	\\
140419A	&	$3.956$	&	$185\pm77$	&	LX	&	C	&		&	$80$	&	KW	&	GCN 16134	\\
140423A	&	$3.26$	&	$65.3\pm3.3$	&	LX	&		&		&	$95$	&	Fermi	&	GCN 16152	\\
140428A	&	$4.7$	&	$1.88\pm0.31$	&		&		&		&	$17.42$	&	Swift	&	GCN 16186	\\
140430A	&	$1.6$	&	$1.54\pm0.23$	&		&		&		&	$173.6$	&	Swift	&	GCN 16200	\\
140506A	&	$0.889$	&	$7.75\pm0.80$	&	LX	&	E	&		&	$64$	&	Fermi	&	GCN 16220	\\
140508A	&	$1.027$	&	$23.24\pm0.26$	&	LX	&		&		&	$44.3$	&	Fermi	&	GCN 16224	\\
140509A	&	$2.4$	&	$3.77\pm0.44$	&	LX	&		&		&	$23.2$	&	Swift	&	GCN 16240	\\
140512A	&	$0.725$	&	$7.76\pm0.18$	&	LX	&		&		&	$148$	&	Fermi	&	GCN 16262	\\
140515A	&	$6.32$	&	$5.41\pm0.55$	&		&		&		&	$23.4$	&	Swift	&	GCN 16284	\\
140518A	&	$4.707$	&	$5.89\pm0.59$	&		&		&		&	$60.5$	&	Swift	&	GCN 16306	\\
140614A	&	$4.233$	&	$7.3\pm2.1$	&	LX	&		&		&	$720$	&	Swift	&	GCN 16402	\\
140620A	&	$2.04$	&	$6.28\pm0.24$	&	LX	&		&		&	$46$	&	Fermi	&	GCN 16426	\\
140623A	&	$1.92$	&	$7.69\pm0.68$	&		&		&		&	$110$	&	Fermi	&	GCN 16450	\\
140629A	&	$2.275$	&	$6.15\pm0.90$	&	LX	&		&		&	$26$	&	KW	&	GCN 16495	\\
140703A	&	$3.14$	&	$1.72\pm0.09$	&	LX	&		&		&	$84$	&	Fermi	&	GCN 16512	\\
140801A	&	$1.32$	&	$5.69\pm0.05$	&		&		&		&	$7$	&	Fermi	&	GCN 16658	\\
140808A	&	$3.29$	&	$11.93\pm0.75$	&		&		&		&	$4.7$	&	Fermi	&	GCN 16669	\\
140907A	&	$1.21$	&	$2.29\pm0.08$	&	LX	&		&		&	$35$	&	Fermi	&	GCN 16798	\\
141026A	&	$3.35$	&	$7.17\pm0.90$	&	LX	&		&		&	$146$	&	Swift	&	GCN 16960	\\
141028A	&	$2.33$	&	$68.9\pm0.02$	&		&		&		&	$31.5$	&	Fermi	&	GCN 16971	\\
141109A	&	$2.993$	&	$33.1\pm6.9$	&	LX	&		&		&	$94$	&	KW	&	GCN 17055	\\
141121A	&	$1.47$	&	$14.2\pm1.1$	&	LX	&		&	UL	&	$1200$	&	KW	&	GCN 17108	\\
141220A	&	$1.3195$	&	$2.44\pm0.07$	&		&		&		&	$7.6$	&	Fermi	&	GCN 17205	\\
141221A	&	$1.47$	&	$6.99\pm1.98$	&	LX	&	C	&		&	$23.8$	&	Fermi	&	GCN 17216	\\
141225A	&	$0.915$	&	$2.29\pm0.11$	&		&		&		&	$56$	&	Fermi	&	GCN 17241	\\
150120B	&	$3.5$	&	$7.37\pm1.09$	&	LX	&		&		&	$24.3$	&	Swift	&	GCN 17330	\\
150206A	&	$2.087$	&	$55.6\pm20.1$	&	LX	&		&		&	$60$	&	KW	&	GCN 17427	\\
150301B	&	$1.5169$	&	$2.87\pm0.42$	&	LX	&		&		&	$13$	&	Fermi	&	GCN 17525	\\
150314A	&	$1.758$	&	$95.2\pm3.1$	&	LX	&		&		&	$10.7$	&	Fermi	&	GCN 17579	\\
150323A	&	$0.593$	&	$1.30\pm0.30$	&		&		&		&	$38$	&	KW	&	GCN 17640	\\
150403A	&	$2.06$	&	$98.1\pm6.3$	&	LX	&		&		&	$22.3$	&	Fermi	&	GCN 17674	\\
150413A	&	$3.139$	&	$49.80\pm7.01$	&		&		&		&	$263.6$	&	KW+Swift	&	GCN 17731	\\
150821A	&	$0.755$	&	$14.7\pm1.1$	&	LX	&		&		&	$103$	&	Fermi	&	GCN 18190	\\
150910A	&	$1.359$	&	$21.6\pm1.8$	&	LX	&		&		&	$112.2$	&	Swift	&	GCN 18268	\\
151021A	&	$2.33$	&	$112.2\pm35$	&	LX	&		&		&	$100$	&	KW	&	GCN 18433	\\
151027A	&	$0.81$	&	$3.94\pm1.33$	&	LX	&	C	&		&	$124$	&	Fermi	&	GCN 18492	\\
151027B	&	$4.063$	&	$18.6\pm3.7$	&	LX	&		&		&	$80$	&	Swift	&	GCN 18514	\\
151111A	&	$3.5$	&	$3.43\pm1.19$	&	LX	&	E	&		&	$40$	&	Fermi	&	GCN 18582	\\
151112A	&	$4.1$	&	$12.1\pm1.5$	&	LX	&		&		&	$19.32$	&	Swift	&	GCN 18593	\\
151215A	&	$2.59$	&	$1.89\pm0.43$	&	LX	&		&		&	$17.8$	&	Swift	&	GCN 18699	\\
160121A	&	$1.96$	&	$2.54\pm0.21$	&	LX	&		&		&	$12$	&	Swift	&	GCN 18919	\\
160131A	&	$0.972$	&	$58.7\pm32.7$	&	LX	&		&		&	$200$	&	KW	&	GCN 18974	\\
160203A	&	$3.52$	&	$12.0\pm1.0$	&	LX	&		&		&	$20.2$	&	Swift	&	GCN 18998	\\
160227A	&	$2.38$	&	$5.52\pm2.38$	&	LX	&		&		&	$316.5$	&	Swift	&	GCN 19106	\\
160228A	&	$1.64$	&	$15.98\pm0.80$	&		&		&		&	$98.36$	&	Swift	&	GCN 19113	\\
160509A	&	$1.17$	&	$84.5\pm2.3$	&	LX	&		&		&	$371$	&	Fermi	&	GCN 19411	\\
160623A	&	$0.367$	&	$22.4\pm1.5$	&	LX	&		&		&	$38.9$	&	KW	&	GCN 19554	\\
160625B	&	$1.406$	&	$419.0\pm4.8$	&	LX	&		&		&	$460$	&	Fermi	&	GCN 19587	\\
160629A	&	$3.332$	&	$48.8\pm9.9$	&		&		&		&	$66.6$	&	Fermi	&	GCN 19628	\\
160804A	&	$0.736$	&	$2.46\pm0.51$	&	LX	&		&		&	$130$	&	Fermi	&	GCN 19769	\\
161014A	&	$2.823$	&	$10.1\pm1.7$	&		&		&		&	$37$	&	Fermi	&	GCN 20051	\\
161017A	&	$2.013$	&	$7.56\pm1.55$	&	LX	&		&		&	$32$	&	Fermi	&	GCN 20068	\\
161023A	&	$2.708$	&	$73.9\pm27.5$	&	LX	&		&		&	$50$	&	KW	&	GCN 20111	\\
161108A	&	$1.159$	&	$1.66\pm0.15$	&	LX	&		&		&	$105.1$	&	Swift	&	GCN 20151	\\
161117A	&	$1.549$	&	$31.2\pm5.5$	&	LX	&		&		&	$122$	&	Fermi	&	GCN 20192
\end{longtable}

\section{Parameters of the equation of state}\label{app:gamma}

We give here details concerning the determination of the value of the index $\gamma$ and verify the accuracy of our assumption $\gamma=4/3$ adopted in the equation of state of the plasma (\ref{eq:eos}). This index is defined as:
\begin{equation}
\gamma\equiv1+\frac{p}{\epsilon}\, .
\label{gammathdef}
\end{equation}
The total internal energy density and pressure are computed as
\begin{align}
\epsilon&=\epsilon_{e^-}+\epsilon_{e^+}+\epsilon_{\gamma}+\epsilon_{B}\\
p&=p_{e^-}+p_{e^+}+p_{\gamma}+p_{B}\, ,
\end{align}
where the subscript $B$ indicates the contributions of the baryons in the fluid. The number and energy densities, as well as the pressure of the different particles, can be computed in natural units ($c=\hbar=k_B=1$) using the following expressions \citep[see, e.g.,][]{Landau5}:
\begin{align}
n_{e^-}&=A\,T^3\int_0^\infty f(z,T,m_e,\mu_{e^-})\,z^2\,\mathrm{d}z\\
n_{e^+}&=A\,T^3\int_0^\infty f(z,T,m_e,\mu_{e^+})\,z^2\,\mathrm{d}z\\
\epsilon_{e^-}&=A\,T^4\int_0^\infty f(z,T,m_e,\mu_{e^-})\,\sqrt{z^2+(m_e/T)^2}\,z^2\,\mathrm{d}z\,-\,m_e\,n_{e^-}\\
\epsilon_{e^+}&=A\,T^4\int_0^\infty f(z,T,m_e,\mu_{e^+})\,\sqrt{z^2+(m_e/T)^2}\,z^2\,\mathrm{d}z\,-\,m_e\,n_{e^+}\\
p_{e^-}&=A\,\frac{T^4}{3}\int_0^\infty f(z,T,m_e,\mu_{e^-})\,\frac{z^4}{\sqrt{z^2+(m_e/T)^2}}\,\mathrm{d}z\\
p_{e^+}&=A\,\frac{T^4}{3}\int_0^\infty f(z,T,m_e,\mu_{e^+})\,\frac{z^4}{\sqrt{z^2+(m_e/T)^2}}\,\mathrm{d}z\\
\epsilon_\gamma&=a\,T^4\\
p_\gamma&=\frac{a\,T^4}{3}\\
\epsilon_B&=\frac{3}{2}n_N\,T\\
p_B&=n_N\,T\, ,\\
\end{align}
where
\begin{equation}
f(z,T,m,\mu)=\frac{1}{e^{\sqrt{z^2+(m/T)^2}-\mu/T}+1}
\end{equation}
is the Fermi-Dirac distribution, $m_e$ is the electron mass, $n_N$ the nuclei number density, $a=8\pi^5k_B^4/15h^3c^3=7.5657\times 10^{-15}$erg cm$^{-3}$ K$^{-4}$ is the radiation constant, and $A=15a/\pi^4$. If the pair annihilation rate is zero, i.e., if the reaction $e^-+e^+\leftrightarrows 2\gamma$ is in equilibrium, then the equality $\mu_{e^-}=-\mu_{e^+}\equiv\mu$ holds, since the equilibrium photons have zero chemical potential. Besides, charge neutrality implies that the difference in the number of electrons and positrons is equal to the number of protons in the baryonic matter, which can be expressed as:
\begin{equation}
n_{e^-}(\mu,T)-n_{e^+}(\mu,T)=Z\,n_B,
\label{eq:ConsCharge}
\end{equation}
where $n_B$ is the baryon number density and $1/2<Z<1$ is the average number of electrons per nucleon. The number density $n_B$ is related to  the other quantities as
\begin{equation}
\rho=m_p\,n_B+m_e\,(n_{e^-}+n_{e^+}),
\end{equation}
where $m_p$ is the proton mass. If the baryons are only protons, then $Z=1$ and $n_N=n_B$. Together with Eq.~(\ref{eq:ConsCharge}), this completely defines the mass density as a function of ($\mu$,$T$). The equation of state that relates the pressure with the mass and internal energy densities is thus defined implicitly as the parametric surface
\begin{equation}
\{(\rho(\mu,T),\epsilon(\mu,T),p(\mu,T))\,:\,T>0\,,\,\mu\geq0\}
\end{equation}
that satisfies all of the above relations.

In the cases relevant for the simulations performed in Sec.~\ref{sec:originprFPA}, we have that indeed the index $\gamma$ in the equation of state of the plasma (\ref{eq:eos}) satisfies $\gamma=4/3$ with a maximum error of $0.2\%$.

\end{document}